\documentclass[lettersize,twocolumn]{IEEEtran}
\usepackage{amsmath,amsfonts}
\usepackage{algorithmic}
\usepackage{algorithm}
\usepackage{array}
\usepackage[caption=false,font=normalsize,labelfont=sf,textfont=sf]{subfig}
\usepackage{stfloats}
\usepackage{verbatim}
\usepackage{graphicx}
\usepackage{cite}
\usepackage{tikz}
\usepackage{placeins}

\usepackage{multicol}
\usepackage{multirow}
\usepackage{mathtools}
\usepackage{amssymb}
\usepackage{indentfirst}
\usepackage{nomencl}
\usepackage{booktabs}
\usepackage{bm}
\usepackage[utf8]{inputenc}
\usepackage[numbers,sort&compress]{natbib}
\usepackage{makecell}
\usepackage{scalerel}

\usepackage{newtxtext,newtxmath} 
\usepackage{color, colortbl}
\usepackage[hyphens]{url}
\usepackage[hidelinks]{hyperref}

\newcommand{\Lag}{\mathcal{L}}
\newcommand{\x}{\mathbf{x}}

\newcommand{\hx}{\bar{\mathbf{x}}}
\newcommand{\z}{\mathbf{z}}
\newcommand{\w}{\mathbf{w}}

\newcommand{\bb}{\mathbf{b}}
\newcommand{\uu}{\mathbf{u}}

\newcommand{\vbold}{\mathbf{v}}

\newcommand{\y}{\mathbf{y}}
\newcommand{\X}{\mathbf{\mathcal{X}}}
\newcommand{\A}{\mathbf{A}}
\newcommand{\B}{\mathbf{B}}
\newcommand{\Q}{\mathbf{Q}}

\newcommand{\PP}{\mathbf{P}}

\newcommand{\R}{\mathbf{R}}

\newcommand{\Y}{\mathbf{\mathcal{Y}}}
\newcommand{\boldlambda}{\boldsymbol{\lambda}}
\newcommand{\kmid}{k + \frac{1}{2}}
\newcommand{\tbold}{\mathbf{t}}

\DeclareMathOperator{\Lips}{\text{Lipschitz}}
\DeclareMathOperator{\gauss}{Gauss-Seidel}
\DeclareMathOperator{\Jacobian}{Jacobian}

\newtheorem{definition}{Definition}

\definecolor{Gray}{gray}{0.9}

\begin{document}
	
	\title{A Survey of ADMM Variants for Distributed Optimization: Problems, Algorithms and Features}
	
	
	
	
	\author{Yu Yang,~\IEEEmembership{Member,~IEEE}, Xiaohong Guan,~\IEEEmembership{Fellow,~IEEE}, Qing-Shan Jia,~\IEEEmembership{Senior Member,~IEEE}, Liang Yu,~\IEEEmembership{Member,~IEEE}, Bolun Xu,~\IEEEmembership{Member,~IEEE}, Costas J. Spanos,~\IEEEmembership{Fellow,~IEEE}
		\thanks{ This work is  supported  by National Natural Science Foundation of China (62192752, 62192750, 62125304, 62073182), 111 International Collaboration Project (BP2018006),  Tsinghua University Initiative Scientific Research Program, and Columbia University Data Science Institute Seed Grant UR010067.  		
	}
		\thanks{Y. Yang is with School of Automation Science and Engineering, Xi'an Jiaotong University,  Xi’an, Shaanxi 710049, China (email: \texttt{yangyu21@xjtu.edu.cn).}  	\texttt{Y. Yang is the corresponding author. }}
		\thanks{X. Guan is with the School of Automation Science and Engineering,
			Xi’an Jiaotong University, Xi’an, Shaanxi 710049, China and is also with Center for
			Intelligent and Networked Systems, Department of Automation, Tsinghua University, Beijing 100084, China (email: \texttt{xhguan@xjtu.edu.cn}).}
		
		\thanks{Q. -S. Jia is with the Center 	for Intelligent and Networked Systems, Department of Automation, BNRist, Tsinghua University, Beijing 100084, China (e-mail: \texttt{jiaqs@tsinghua.edu.cn}).}
		
				\thanks{L. Yu is with College of Automation and College of Artificial Intelligence, Nanjing University of Posts and Telecommunications,
					Nanjing 210003, China (email: \texttt{liang.yu@njupt.edu.cn}).}
				
				\thanks{B. Xu is with Earth and Environmental Engineering, Columbia University, New York, NY 10027, United States (email: \texttt{bx2177@columbia.edu}).}
	\thanks{C. J. Spanos is with the Department of Electrical Engineering and Computer Sciences, University of California, Berkeley, CA, 94720 USA (email:
	\texttt{spanos@berkeley.edu}).}
	}
	\markboth{Manuscript 2022}%
	{Shell \MakeLowercase{\textit{et al.}}: A Sample Article Using IEEEtran.cls for IEEE Journals}
	
	
	\maketitle
	
	\begin{abstract}
By coordinating terminal smart devices or microprocessors to engage in cooperative computation to achieve system-level targets, distributed optimization is incrementally favored by both engineering and computer science. The well-known alternating direction method of multipliers (ADMM) has turned out to be one of the most popular tools for distributed optimization due to many advantages, such as modular structure, superior convergence, easy implementation and high flexibility. In the past decade, ADMM has experienced widespread developments. The developments manifest in both handling more general problems and enabling more effective implementation. Specifically, the method has been generalized to broad classes of problems (i.e., multi-block, coupled objective, nonconvex, etc.). Besides, it has been extensively reinforced for more effective implementation, such as improved convergence rate, easier subproblems, higher computation efficiency, flexible communication, compatible with inaccurate information, robust to communication delays, etc. These developments lead to a plentiful of ADMM variants  to be celebrated by broad areas ranging from  smart grids, smart buildings, wireless communications, machine learning and beyond. However, there lacks a survey to document those developments and discern the results.  To achieve such a goal, this paper provides a comprehensive survey on  ADMM variants. Particularly, we discern the five major classes of problems that have been mostly concerned and discuss the related ADMM variants in terms of main ideas, main assumptions, convergence behaviors and main features. In addition, we figure out several important future research directions to be addressed. This survey is expected to work as a tutorial for both developing distributed optimization in broad areas and identifying existing research gaps.
	\end{abstract}
	
	
	
	%
	%
	%
	%
	
%
%
%

\section{Introduction}


\IEEEPARstart{D}{istributed} optimization is gaining  importance and popularity in both  engineering  and computer science for  decision making and data processing  with the intensifying computation demands \cite{zhong2019admm, qiu2016survey, sun2019survey}. 
The notion of distributed optimization is to engage dispersed smart devices or microprocessors in collaborative computation to  fulfill a certain system-level target. When it comes to an engineering system, distributed optimization is often used to empower subsystems to make decisions locally while  interacting with each other to pursue desirable  system performance.  In the context of computer science, distributed optimization is often utilized to distributed a heavy training task across  multiple microprocessors and coordinate them to fulfill a coherent training target. 
Though the scenarios are diverse, the philosophy of distributed optimization essentially corresponds to breaking a comprehensive mathematical optimization problem into several small-sized subproblems and empowering multiple computing agents to solve the subproblems in a coordinated manner so as to approach an optimal or near-optimal solution of original mathematical optimization.

When the system or problem is in large scale and many practical issues are considered, distributed optimization is often preferred over its centralized counterpart due to many advantages \cite{yang2019survey}.  Specifically, distributed optimization often shows high computation efficiency and favorable scaling property since the computation is distributed to multiple computing agents.  In contrast,  a  centralized method usually relies on a central  unit to solve a comprehensive mathematical optimization independently. 
This  is often computationally intensive or intractable considering the time constraints. 
Besides,  distributed optimization can directly utilize locally available information generated by distributed sensing and monitoring.    Whereas  a centralized counterpart has to collect information of an entire system (often geographically dispersed) and store massive data in an unified memory for computation.  
These highlight the multiplied advantages of distributed optimization,  such as high computation efficiency, low communication overheads,  low footprint memory, and high data privacy.


The well-known alternating direction method of multipliers (ADMM)  has  emerged as one of the most popular tools  for distributed optimization. It has found massive applications in broad areas ranging from  statistical learning  \cite{boyd2011distributed, gong2021research}, multi-agent reinforcement learning \cite{wai2018multi}, imaging processing \cite{afonso2010augmented, almeida2013deconvolving}, data mining \cite{tan2019learning, zhang2016dynamic}, power system control  \cite{molzahn2017survey, patari2021distributed, chen2022towards, yang2021optimal}, smart grid operation \cite{maneesha2021survey, yang2022optimal, yang2017distributed, yang2018decentralized, long2021efficient, yang2017stochastic, yang2016joint}, smart building management \cite{yang2020hvac, yang2021distributed, yang2021stochastic}, multi-robot coordination \cite{halsted2021survey}, wireless communication control \cite{shen2012distributed, liu2021admm},  autonomous vehicle routing \cite{yao2019admm, zhang2021semi}   and beyond. 
The popularity of  ADMM  can be attributed to its many distinguishing advantages, such as \emph{modular structure}, \emph{superior convergence}, \emph{easy implementation} and \emph{high flexibility}.  
The \emph{modular structure} characterizes that ADMM generally explores the decomposition of a large-scale optimization across objective components (often called features). This results in a fixed number of subproblems.  Each subproblem corresponds to one objective function (one feature) and one disjoint  block of decision variables. Afterwards, the subproblems can be handled separately by individual agents using customized solvers. 
The \emph{superiors convergence} describes both the less restrictive convergence conditions and  faster convergence rates of ADMM over many other distributed methods. For example, ADMM does not entail any smoothness for convex optimization.   Besides,  ADMM ensures an $\mathcal{O}(1/k)$ convergence rate for convex optimization whereas subgradient  methods only promise $\mathcal{O}(1/\sqrt{k})$  convergence rate \cite{yang2010distributed}.  
ADMM  is often  more reliable and robust in convergence  compared with  dual ascent methods  at the lack of strong convexity  \cite{boyd2011distributed}.   
In addition, ADMM  often requires  less iterations to approach an optimal or near-optimal solution  than distributed gradient methods \cite{ling2015dlm}.  Note that the number of iterations often determines the communication overhead of a distributed method. 
Overall, ADMM has been recognized at least comparable to very specialized algorithms \cite{boyd2011distributed}.
ADMM enables  \emph{easy implementation} due to the rather small dependence on parameter settings compared with many other distributed methods. This is largely attributed to the quadratic penalty terms that enhance problem convexity.  Besides, the subproblems of ADMM often admit closed-form solutions, yielding low per-iteration complexity. 
The \emph{high flexibility} can be perceived from  the broad classes of problems that can be handled by ADMM either naturally or by means of reformulations. This will be clear from the rest of this paper.

%

\begin{figure}[h]
	\centering
	\includegraphics[width = 3.2 in]{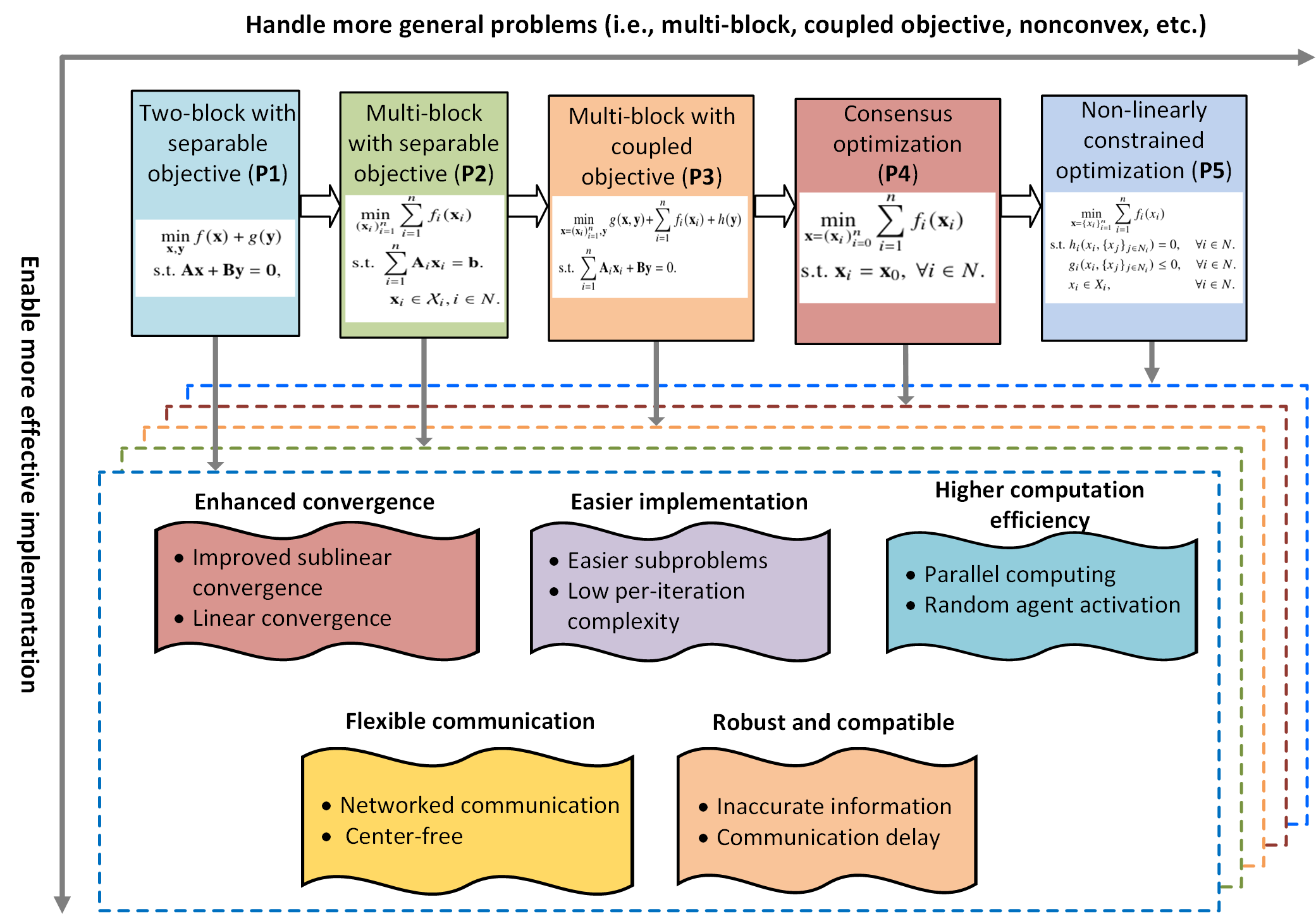}
	\caption{Developments of ADMM  for both handling more general problems and enabling more effective implementation.}
	\label{fig:ADMM_extension}
\end{figure}

With the growing demand for distributed optimization, ADMM has experienced widespread developments in the past decade. An overview of the developments is shown in Fig. \ref{fig:ADMM_extension}. Specifically, the developments manifest in both  \textbf{handling more general problems} (i.e., multi-block, coupled objective, nonconvex, etc.) and \textbf{enabling more effective implementation}. 
Primarily, the method was developed for convex optimization with a two-block separable structure. Whereas it has been generalized to diverse structured convex and nonconvex optimization, which includes  \emph{i)} two-block with separable objective \eqref{pp:p1}, \emph{ii)} multi-block with separable objective \eqref{pp:p2}, \emph{iii)} multi-block with coupled objective \eqref{pp:p3}, \emph{iv)} consensus optimization \eqref{pp:p4}, and \emph{v)} non-linearly constrained optimization \eqref{pp:p5}.  
In addition, the method has been extensively reinforced for more effective implementation, such as improved convergence rate, easier implementation, higher computation efficiency, flexible communication,  compatible with inaccurate information,  robust to communication delays, etc. 
These developments lead to a plentiful of ADMM variants that are suitable for different problems or situations. Though some of those ADMM variants have found successful applications, many of the them are still  limited to the theoretical research community and expected to enjoy broader applications and success.   This is mainly caused by the fairly large number of variants developed for different problems and under different conditions, making it rather difficult for one who lacks strong theoretical background to  identify an appropriate one for their problems.   
There lacks a survey to document those developments and discern the results. To fulfill the gap,  this paper provides a comprehensive survey on ADMM variants for distributed optimization.  Specifically, this paper makes the following main contributions.
\begin{itemize}
	\item [C1)] We survey ADMM and its variants developed throughout the decades broadly and comprehensively  in both convex and nonconvex settings.
	\item[C2)] We discern the  five  major classes of  problems that have been mainly concerned and discuss  the related ADMM variants in terms of  main assumptions, decomposition scheme, convergence properties and main features. 
	\item[C3)] Based on the existing results, we figure out the important future research directions to be addressed.  
\end{itemize}


This paper focuses on  ADMM and its variants for distributed optimization as they are being incrementally attractive and popular to account for the growing computation demand  of broad areas.   
 Though a number of celebrated reviews on distributed optimization have discussed ADMM, most of them only focused on classical ADMM for  two-block convex optimization and  the  significant developments that occurred in  very recent decade haven't been covered yet.   
We report those  reviews  in TABLE  \ref{tab:existing_review} by years. Particularly, we distinguish  this paper and those  reviews  by the convexity of concerned problems (\texttt{Convexity}), the presence of constraints (\texttt{Constraints}), the formulations of concerned problems (\texttt{Problems}),  the classification criterion (\texttt{Classification}),  the related distributed methods (\texttt{Methods}), the specialized applications (\texttt{Specialized applications}) and the year of publication (\texttt{Year}). Note that many of the reviews only involved classical ADMM as one distributed solution. Moreover, they were  mainly concerned with convex optimization.   As highlighted, there exist five exceptional reviews that are  specialized to ADMM like this paper. However, they are in quite different perspectives.  
\begin{itemize}
	\item Boyd \emph{et.al.} \cite{boyd2011distributed} (2011) gave the earliest tutorial on classical ADMM,  which documented the fundamental theory of ADMM for convex optimization followed by some applications arising from statistical and machine learning.  This tutorial exactly renewed  ADMM and raised the surge of interest in the method for distributed optimization. 
	\item Glowinski \cite{glowinski2014alternating} (2014) gave a introductory review on the origination of ADMM. Specifically, the method originated from  an inexact implementation of augmented Lagrangian method (ALM) for solving  partial differential equations (PDE). Afterwards, the relationship between the inexact ALM and Douglas-Rachford alternating direction method was discovered, leading to  the ADMM  that is well-known today.  	
	\item Eckstein \emph{et.al.} \cite{eckstein2015understanding} (2015) gave a thorough overview on  understanding  and establishing the convergence of ADMM from the perspective of operator splitting.  Specially, this paper argued that ADMM is actually not an approximate ALM as commonly recognized considering its quite different convergence behaviors from  the real approximate ALM variants observed in some numerical studies. 
	
	\item Maneesha \emph{et.al.} \cite{maneesha2021survey} (2021) conducted a survey on the applications of ADMM to smart grid operation. The classical ADMM was introduced in details, followed by its diverse applications  to smart grids (e.g., optimal power flow control, economic dispatch, demand response, etc.). 
	
	\item Han \emph{et.al.} \cite{han2022survey} (2021) gave  a comprehensive survey  on the recent developments of ADMM  and its variants from the perspective of parameter selecting,  easier subproblems, approximate iteration,  convergence rate characterizations,  multi-block  and nonconvex extensions. 
\end{itemize}

\begin{table*}[h] 
	\setlength\tabcolsep{2pt}
	\renewcommand\arraystretch{1.6}
	\centering
	\caption{Existing reviews on distributed optimization that involve ADMM.}
	\label{tab:existing_review}
	\begin{tabular}{lclcllll}   
		\toprule
		\textbf{References} & 	\textbf{Convexity}    & \textbf{Constraints}  & \textbf{Problems} & \textbf{Classification} &  ~~~~\textbf{Methods}  & \makecell[c]{ \textbf{Specialized} \\ \textbf{applications}} &  ~~\textbf{Year}  \\
		\hline
		\cite{yang2010distributed} & Convex  & \makecell[l]{Unconstrained\\Constrained}  & \ref{pp:p1}, \ref{pp:p2}, \ref{pp:p4} &\makecell[l]{Problems. \\ Methods.}  & \makecell[l]{Gradient method.  \\ Subgradient method. \\Incremental subgradient method. \\ Dual decomposition.\\  Primal decomposition. \\ Classical ADMM. } &\makecell[l]{Game theory. \\Networked system.} &  2010 \\
		\hline
				\rowcolor{Gray}
		&   &  & & & & Data analysis & \\ 
		\rowcolor{Gray}
		\multirow{-2}{*}{\cite{boyd2011distributed}} & \multirow{-2}{*}{Convex} &  \multirow{-2}{*}{Constrained}  & \multirow{-2}{*}{\ref{pp:p1}} & \multirow{-2}{*}{--} & \multirow{-2}{*}{Classical ADMM}  & Statistical learning.  & \multirow{-2}{*}{2011}\\
		\hline
		\rowcolor{Gray}
		\cite{glowinski2014alternating} & Convex &  Constrained & \ref{pp:p1} & --  & Classical ADMM.  & --  & 2014\\
		\hline
		\rowcolor{Gray}
		\cite{eckstein2015understanding} & Convex &  Constrained & \ref{pp:p1} & --  & Classical ADMM.  & --  & 2015\\
		\hline
		\cite{wang2017distributed} &	Convex &  \makecell[l]{Unconstrained\\Constrained}& -- & Applications.  & \makecell[l]{Classical ADMM.\\Dual Decomposition.\\ATC.} & \makecell[l]{Power system\\ operation} & 2017\\		
		\hline
		\cite{molzahn2017survey} &Convex & Constrained   & \ref{pp:p1}, \ref{pp:p2}   &  \makecell[l]{Methods.\\Applications.}   &\makecell[l]{Dual decomposition. \\ Classical ADMM.\\ ATC. \\ Proximal Message Passing. \\  Consensus+Innovation.}  & \makecell[l]{Electrical power \\system operation.} &   2017\\		
		\hline
		\cite{nedic2018distributed} &	Convex  & Unconstrained & \ref{pp:p4}  &  \makecell[c]{Problems. \\Methods.}     & \makecell[l]{Distributed average-weighting \\ algorithm. \\ Classical ADMM.} & \makecell[l]{--} &  2018 \\
		\hline
		\cite{notarstefano2019distributed}	& Convex &\makecell[l]{Unconstrained\\Constrained}   & \ref{pp:p2}, \ref{pp:p4} & \makecell[l]{Problems. \\Methods.\\Applications.} & \makecell[l]{Distributed subgradient methods.\\ Dual decomposition.\\ Classical ADMM.\\Distributed dual \\ subgradient methods. \\Constraints exchange.} & Cyber-physical Network  & 2019\\
		\hline 
		\cite{abeynanda2021study} &	Convex   &\makecell[l]{Unconstrained\\Constrained}  &\ref{pp:p3}     &  Methods.  & \makecell[l]{Gradient methods.\\Subgradient methods.\\Classical ADMM.}  & \makecell[l]{--}   & 2021 \\
		\hline
		\cite{patari2021distributed} &  Convex  & Constrained & \ref{pp:p2} & \makecell[l]{Methods.\\Applications.} &  \makecell[l]{Dual ascent.\\ Primal-dual method.\\Proximal Atomic Coordination. \\Classical ADMM.}& \makecell[l]{ Electric distribution \\ system control.} &2021\\
		\hline
		\rowcolor{Gray}
		\cite{maneesha2021survey} &  Convex  & Constrained &--  & Applications. & Classical ADMM  & Smart grid operation. &2021\\
		\hline
				\rowcolor{Gray}
		\cite{han2022survey} &  \makecell[c]{Convex\\ Nonconvex~~~~}  & \makecell[l]{Unconstrained~~ \\Constrained}& \ref{pp:p2} & ~~~-- &  \makecell[l]{Classical ADMM. \\ ADMM variants.~~~~~~~~~~~~~~~~~~~~~~~~~~~}&-- &2021\\
		\hline
			\rowcolor{Gray}
		\textbf{This work} &	\makecell[c]{ \textbf{Convex} \\ \textbf{Nonconvex~~~}}&  \makecell[l]{\textbf{Unconstrained} \\ \textbf{Constrained}} & \ref{pp:p1}, \ref{pp:p2}, \ref{pp:p3}, \ref{pp:p4}, \ref{pp:p5} &  \makecell[l]{\textbf{Problems}\\ \textbf{Methods.} \\ \textbf{Features}~~~~~~~~~}  & \makecell[l]{\textbf{Classical ADMM.}\\ \textbf{ADMM variants.}~~~~~~~~~~~~~~~~~~~~~~~~~}& -- &  \textbf{2022}\\
		\bottomrule
	\end{tabular}\\
\begin{tabular}{l}
Notes: we highlight the reviews that focused on ADMM and its variants in gray.  
\end{tabular}
\end{table*}

To the authors' best knowledge, \cite{han2022survey} has been the most updated and comprehensive survey  on ADMM variants. However, this paper differs from  \cite{han2022survey} in various aspects. First of all, we make an effort to  involve the  broad classes of problems (i.e., multi-block, coupled objective, non-linearly constrained, etc.) that have been studied  whereas \cite{han2022survey} only considered the standard linearly constrained multi-block  optimization that represents one class of this paper.   Besides, we discern the results from a quite different perspective including  \textbf{problems}, \textbf{methods} and \textbf{features} (i.e., parallel computation, low per-iteration complexity, fast convergence, etc.). 
Specifically, have been able to identify the five major classes of problems that have been mainly concerned in the literature.  We then comprehensively discuss the related ADMM variants for each class of problems in terms of main assumptions, decomposition scheme, convergence behaviors and main features.   This is relevant to help advance the transfer of ADMM related theory to practice considering that one often intends to search for  appropriate distributed solutions by their problems and requirements. 
In contrast, \cite{han2022survey} organized the results by  parameter selection, easier subproblems, approximate iteration, convergence rate characterization, multi-block and nonconvex extensions.  Based on our experience, this review is more suitable for those who have  quite solid theoretical backgrounds on ADMM and its variants. 
Last but importantly,  we comprehensively survey and discuss ADMM variants for  nonconvex optimization  whereas \cite{han2022survey} only gave  a short and simple discussion on that topic.

%
%

\begin{figure}[h]
	\centering
	\includegraphics[width = 3.5 in]{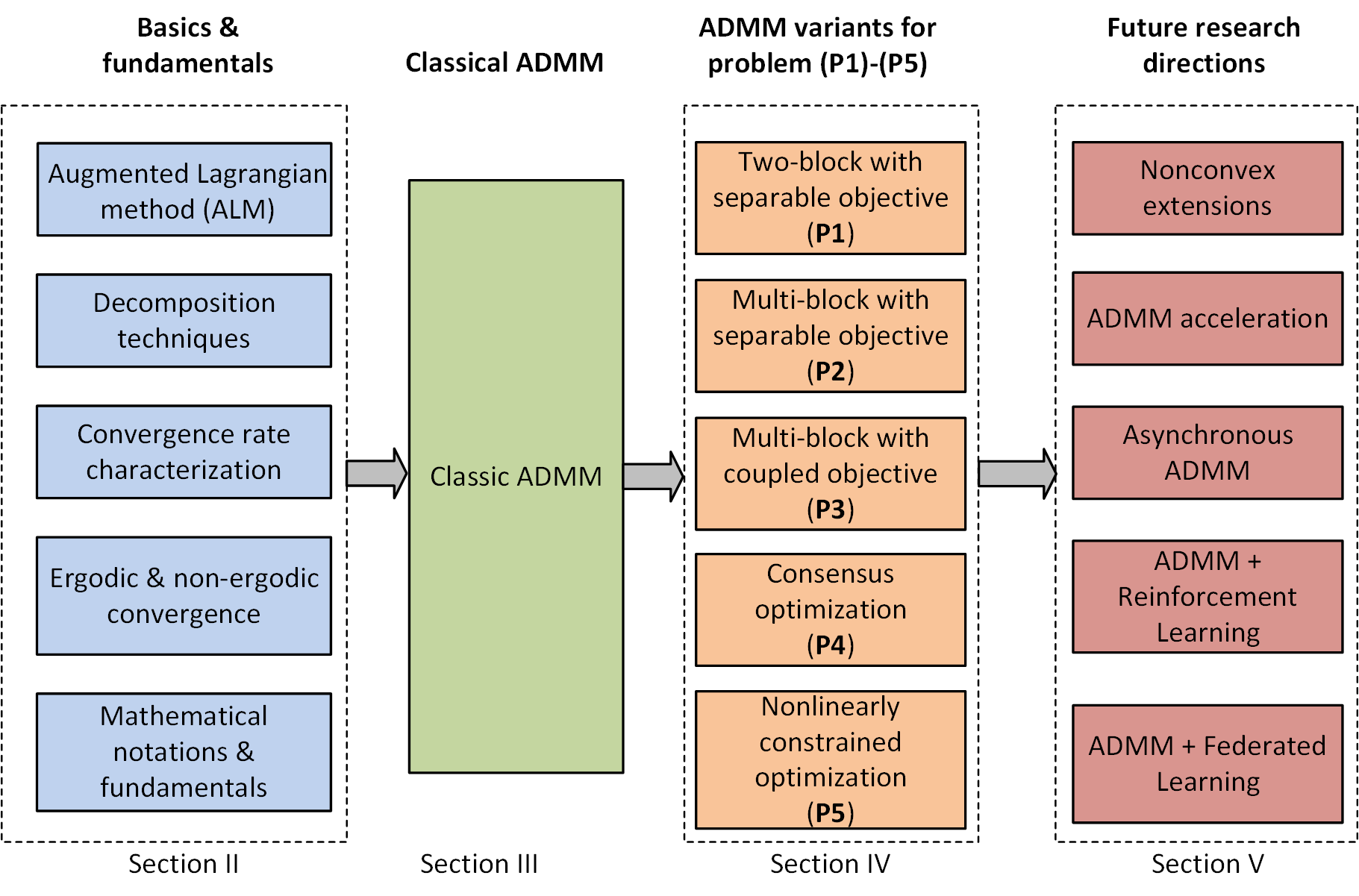}
	\caption{A roadmap for the major sections of this paper (Section II-Section V).}
	\label{fig:roadmap}
\end{figure}

The rest of this survey is as follows. In Section II, we  introduce some basic and fundamental knowledge related to ADMM and its variants, which include augmented Lagrangian method (ALM), decomposition techniques, convergence rate characterization, ergodic and nonergodic concepts of convergence. Afterwards, we clarify the mathematical notations  frequently used in this paper. 
In Section III, we introduce classical ADMM and its related theoretical results. 
In Section IV, we survey ADMM variants for solving the five major classes of problems \eqref{pp:p1} - \eqref{pp:p5}. For each class of problems, we discuss the related ADMM variants from the perspectives of main assumptions, decomposition schemes, convergence properties and main features. In Section V, we discuss several important and promising  future research directions.  A roadmap for the above major sections (Section  II-Section V) is shown in Fig. \ref{fig:roadmap}.  In Section VI, we conclude this paper.

\section{Basics and Fundamentals}
In this section, we first introduce  augmented Lagrangian method (ALM) and some decomposition techniques which are the basics of ADMM and its variants. We then introduce convergence rate characterization and ergodic/nonergodic  convergence. We finally define the mathematical notions. 
\subsection{Augmented Lagrangian Method}

\emph{Augmented Lagrangian method} (ALM), also  known as \emph{method of multipliers},  is a basic tool for constrained optimization. ALM is the precursor of ADMM. More specifically, ADMM was primarily developed as an approximate implementation of ALM  \cite{glowinski2014alternating}.

Central to ALM is to relax  some or all constraints of a constrained problem by Lagrangian multipliers and penalty functions (usually quadratic) and then solve a sequence of unconstrained or partially constrained relaxed problems to approach an optimal or near-optimal solution of the original problem. 
We use a simple linearly constrained optimization to illustrate the idea, i.e., 
\begin{align}
& \label{pp:p}~\min_{\x, \y} f(\x) + g(\y)~~\tag{$\mathbf{P1}$} \\
& {\rm s.t.~} \A \x + \B \y = \bm{0}, \notag
\end{align}
where $f: \R^n \rightarrow \R$ and $g: \R^n \rightarrow \R$ are given objective functions related to decision variables  $\x \in \R^n$ and $\y \in \R^n $ respectively;  $\A \in \R^{l  \times n}$  and $\B \in \R^{l \times n}$ are coefficient matrices encoding the linear constraints.  %

Considering the difficulty to solve the constrained optimization \eqref{pp:p} directly,   ALM proposes to relax the  constraints by Lagrangian multipliers and penalty functions. Specifically, associating the Lagrangian multipliers  $\boldlambda\in\R^l$ and  quadratic penalty parameter $\rho>0$ with the linear constraints, we have the augmented Lagrangian (AL) function
\begin{align*}
\Lag_{\rho}(\x, \y, \boldlambda) = f(\x) + g(\y) + \langle \boldlambda, \A \x + \B \y \rangle + \frac{\rho}{2}\Vert \A \x + \B \y \Vert^2. 
\end{align*}

ALM then performs the following primal-dual  updates to approach  an optimal or near-optimal solution of \eqref{pp:p}. 
\begin{align}
\label{eq:primal-update} & \textbf{Primal update:}~~~(\x^{k+1}, \y^{k+1}) = \arg \min_{\x, \y} \Lag_{\rho_k}(\x, \y, \boldlambda^k) \\
\label{eq:dual-update} & \textbf{Dual update:}~~~~~ \boldlambda^{k+1} = \boldlambda^k + \rho_k (\A \x^{k+1} + \B \y^{k+1})
\end{align} 
where $k$ denotes the iteration;  $\rho_k$ represents some varying penalty parameters which may be preselected or dynamically generated in the iterative process. ALM is composed of two alternative steps: \textbf{Primal update} solves AL problems  with  given Lagrangian multipliers $\boldlambda^k$, and \textbf{Dual update} updates  Lagrangian multipliers $\boldlambda^k$ based on the obtained solutions. The dual update formula \eqref{eq:dual-update} can be interpreted as a dual  gradient ascent step  with stepsize $\rho_k$. 
Specifically, we have the dual of AL problem $d(\boldlambda^k) = \min_{\x, \y}\Lag_{\rho_k} (\x, \y, \boldlambda^k)$ and its gradient $\nabla d(\boldlambda^k)  = \A \x^{k+1} + \B \y^{k+1}$  \cite{hong2017linear}, thereby a dual ascent step for maximizing $d(\boldlambda)$ reads as $\boldlambda^{k+1} = \boldlambda^k + \rho_k (\A \x^{k+1} + \B \y^{k+1})$.

ALM  was first proposed by Hestenes \cite{hestenes1969multiplier} and Powell \cite{powell1969method} in the 1969s as an alternative to  the penalty  method for constrained optimization. For the penalty method, the Lagrangian multipliers are absent ($\boldlambda^k =0$).  The motivation of developing  ALM  is that penalty method generally requires to  increase the penalty parameter $\rho_k$ to be very large (e.g., infinity), making the resulting relaxed problems ill-conditioning  and very difficult to solve. Moreover,  the method  was found to be very sensitive to the round-off error caused by the means of  analogy computing at that time.  In such context, ALM  proposes to add Lagrangian multipliers to the objective of penalty method. It was found that  when the Lagrangian multipliers  are close to its corresponding optima, one  does not require  quite large  penalty parameters, alleviating the difficulty of ill-conditioning.



ALM can be viewed as the  combination of penalty method ($\boldlambda^k=0$) and Lagrangian method ($\rho_k=0$). Lagrangian method was primarily proposed with the idea of obtaining optimal solutions by solving the equations of optimality conditions of a constrained optimization. Since the equations   involve primal-dual variables and  usually do not admit analytical solutions,   a primal-dual iterative update scheme is often used by Lagrangian method to approach the solutions gradually. The implementation of  Lagrangian method also falls into the general primal-dual framework \eqref{eq:primal-update}-\eqref{eq:dual-update} but with zero penalty parameter (i.e., $\rho_k =0$) in the primal update and general stepsize $\rho_k =\alpha_k$  in the dual update.  Despite both penalty method and Lagrangian method have found wide applications, they suffer from different drawbacks and limitations.  While penalty method  tends to face the ill-conditioning issue,  Lagrangian method is often very sensitive to the dual stepsize settings. Moreover, Lagrangian method  depends on fairly restrictive conditions to ensure convergence, such as  local or global  convexity over the constrained subsets.

As a combination, ALM   moderates the disadvantages of  penalty method and Lagrangian method. On one hand, ALM does not require to increase the penalty parameters to be very large and a  small fixed one often  works quite well, thereby alleviating the ill-conditioning problem with penalty method  \cite{bertsekas2014constrained}. On the other hand, ALM often shows  smaller dependence on the parameters, such as penalty parameters.  It was found that any penalty parameters over certain  threshold are admissible to ensure convergence of the method (Prop. 2.4, Ch2, \cite{bertsekas2014constrained}).  Besides, ALM often ensures the existence of minimizer of AL problems, which is not provided by  penalty method and Lagrangian method as often.  
Moreover, the Lagrangian multipliers of ALM often converge faster to the optima than that of Lagrangian method and the corresponding  terms of penalty method, implying a faster convergence rate with  ALM  over the other two. 

Because of those attractive features,  ALM has emerged as one of the most important and popular  tools for constrained optimization.  The  idea and basic theory   of ALM  have been comprehensively documented in the textbooks authored by Bertsekas \cite{bertsekas2014constrained} and Bergin \cite{birgin2014practical}. We refer the interested readers there for more details.

\subsection{Decomposition techniques}
Despite the benefits, one major drawback of ALM is its nondecomposable structure caused by  penalty functions. 
Consider \eqref{pp:p} as an example, though the objective functions are separable across the decision variables $\x$ and $\y$, we still require to deal with the joint optimization  \eqref{eq:primal-update} due to the quadratic penalties. 
The joint optimization is usually difficult,  at least not much easier than the original constrained optimization. To overcome such drawback, ALM is often combined with certain decomposition techniques to break the joint optimization into small subproblems. This idea exactly leads to ADMM and its variants to be discussed. 
There are  two widely used  decomposition techniques  for ALM. One is  $\gauss$ decomposition and the other one is $\Jacobian$ decomposition. 

\subsubsection{Gauss-Seidel decomposition} For a joint optimization, $\gauss$ decomposition (also known as alternating minimization)  proposes to update the decision variables one by one.  Specifically,  when $\gauss$ decomposition  is applied to the joint primal update \eqref{eq:primal-update},  we have 
{	\setlength{\abovedisplayskip}{3pt}
	\setlength{\belowdisplayskip}{-10pt}
\begin{align*}
& \textbf{$\x$-update:}~\x^{k+1} = \arg\min_{\x} \Lag_{\rho}(\x, \y^k, \boldlambda^k) \\
& \textbf{$\y$-update:}~\y^{k+1} = \arg\min_{\y} \Lag_{\rho}(\x^{k+1}, \y, \boldlambda^k) \\
\end{align*} }
Note that the joint optimization breaks into two serial block updates.   $\gauss$ decomposition resembles block coordinate method in the sense that multiple decision variables are optimized one by one by assuming the others with latest updates. 

\subsubsection{Jacobian decomposition}  For a joint optimization, 
$\Jacobian$ decomposition proposes to update the decision variables separately but in parallel by using the previous updates of the others.  Specifically, when the $\Jacobian$ decomposition is employed to the joint primal update \eqref{eq:primal-update}, we have
{	\setlength{\abovedisplayskip}{1pt}
	\setlength{\belowdisplayskip}{-10pt}
\begin{align*}
& \textbf{$\x$-update}:~\x^{k+1} = \arg\min_{\x} \Lag_{\rho}(\x, \y^k, \boldlambda^k) \\
& \textbf{$\y$-update}:~\y^{k+1} = \arg\min_{\y} \Lag_{\rho}(\x^{k}, \y, \boldlambda^k) \\
\end{align*}
}

Essentially,  both  $\gauss$ and $\Jacobian$ decomposition are expected  to run multiple rounds to approach an optimal or near-optimal solution of the joint optimization.  However, when combined with  ALM that already takes an iterative primal-dual scheme,  $\gauss$ and $\Jacobian$ decomposition are often performed only once to reduce computation. Clearly, this only provides an approximate  solution (may be very rough) to the joint primal update.  That is why  ADMM and its variants resulting from the combination of ALM and these decomposition techniques are often viewed as inexact or approximate ALM. 

One may note that one obvious advantage of $\Jacobian$ over $\gauss$ decomposition for distributed optimization is the parallelizable implementation.  However, this usually comes at a cost of being more likely to diverge. This is because a $\Jacobian$ decomposition  usually provides a less accurate approximation to a joint primal update of ALM  \cite{deng2017parallel, he2015full}.

We  has illustrated how  $\gauss$ and $\Jacobian$ decomposition are usually combined with ALM to enable distributed computation by a two-block example. This idea is  readily extended to general multi-block optimization, which will be commonly seen in the rest of this paper.

\subsection{Convergence rate characterization}
In distributed optimization where an iterative scheme is often used to achieve the coordination of multi-agent computation, convergence rate is an important metric for characterizing the computation efficiency of an algorithm.
In the following, we give the definitions of \emph{sublinear} and \emph{linear} convergence that will be frequently referred to in this paper.

\begin{definition}
(Sublinear convergence)~ Supposed  we have a sequence $\{\x^k\}_{k=0}^K$ converge to  the limit point $\x^{*}$ according to 
\begin{align*}
	\lim_{k \rightarrow \infty}\frac{\Vert \x^{k+1} - \x^{*}\Vert}{\Vert \x^k - \x^{*}\Vert} = 1,
\end{align*}
we say that the sequence $\{\x^k\}_{k=0}^K$  converges  at  a  sublinear convergence rate. 
\end{definition}

\begin{definition}
(Linear convergence)~Suppose we have a  sequence $\{\x^k\}_{k=0}^K$ converge to the limit point $\x^{*}$ according to
\begin{align*}
	\lim_{k \rightarrow \infty}\frac{\Vert \x^{k+1} - \x^{*}\Vert}{\Vert \x^k - \x^{*} \Vert} = C,  
\end{align*}
\end{definition}
where $C \in (0, 1)$ is a constant, we say that the sequence $\{\x_k\}_{k=0}^K$ converges at  linear convergence rate.

ADMM and  its many variants promise \emph{sublinear} convergence in the forms of $\mathcal{O}(1/k)$ or  $\mathcal{O}(1/\epsilon)$,  $\mathcal{O}(1/k^2)$ or $\mathcal{O}(1/\sqrt{\epsilon})$, where $k$ is the iteration counter and $\epsilon$ is a solution accuracy.  Those convergence rates are often associated with the (worst-case) iteration complexity of a distributed method.  A (worst-case) iteration complexity $\mathcal{O}(1/k)$ or $\mathcal{O}(1/\epsilon)$ states that the solution accuracy of the generated sequence would be the order $\mathcal{O}(1/k)$ after $k$ iterations, or equivalently  it would require at most $\mathcal{O}(1/\epsilon)$ iterations to approach a solution of  accuracy $\epsilon$.  
Despite those convergence rate characterizations are in different forms and orders, they all correspond to  \emph{sublinear} convergence  by definitions. However, it is clear that a second-order convergence rate  $\mathcal{O}(1/k^2)$ or  $\mathcal{O}(1/\sqrt{\epsilon})$   normally implies a much faster convergence rate with a method than a first-order convergence rate $\mathcal{O}(1/k)$ or  $\mathcal{O}(1/\epsilon)$.

Under some special or stricter conditions,   such as strong convexity and $\Lips$ differentiable,  some  ADMM variants can ensure \emph{linear} convergence. Note that we often prefer a \emph{linear}   than \emph{sublinear} convergence as the former secures a stable and fixed decay of the sub-optimality gap along the iterations. In contrast,  we often observe a decaying rate of the decrease of performance gap  with  \emph{sublinear} convergence. This is often referred to a ``tail convergence" property. This property is actually common with distributed methods. This implies that distributed optimization is generally  suitable for  applications that only require sufficiently accurate solutions,  and for the context that  extremely high solution accuracy is required, a centralized method should be more reliable. 

\subsection{Ergodic and non-ergodic convergence}
When it turns to examine the convergence property of generated sequences by  an iterative algorithm, there are two widely-used viewpoints,  which are \emph{ergodic} and \emph{non-ergodic}.
Basically, the \emph{non-ergodic} studies the convergence property of  generated sequences directly and the \emph{ergodic}  studies the convergence of time-averaged generated sequences \cite{davis2016convergence}. 
Specifically, suppose we have a  sequence $\{\x^k\}_{k=0}^K$ yield by an iterative algorithm,  the \emph{non-ergodic} studies the convergence of  $\{\x^k\}_{k=0}^K$ or certain measures  defined  on the sequence. 
In contrast,  the \emph{ergodic}  concerns  the convergence  of the time-averaged sequence $\{\bar{\x}^k\}_{k=0}^K $ with $\bar{\x}^k:= 1/(k + 1) \smallsum_{j=0}^k \x^j$ or its measures.

Both the \emph{ergodic} and \emph{non-ergodic} viewpoints have been widely used for examining the convergence and  convergence rate of ADMM and its variants. 
One often prefers  \emph{non-ergodic} over \emph{ergodic}  perspective for the former is more direct and informative. However,  an \emph{ergodic} perspective has the advantage of averaging out some bounded oscillation or noise of  generated sequences, thus not disrupting the convergence property held by an algorithm. 

\subsection{Mathematical notations and fundamentals}
In this survey, a little mathematics  will be included.  We use $\R$ and $\R^n$ to denote the real and $n$-dimensional real space. We have the bold alphabets $\x$, $\y$, $\x_i$, $\boldlambda$, $\bb$  represent vectors,   $\X, \Y$  represent subsets of real space, $\A$,  $\A_i$, $\B$, $\Q$, $\PP$ denote matrices, and $\mathbf{I}$ represents an identity matrix of suitable size. The operator $:=$ is meant  to give   definitions. 
We denote the standard Euclidean norm and inner product by  $\Vert \cdot \Vert$  and $\langle \cdot, \cdot\rangle$. We define $\Vert \x\Vert^2_{\Q}: = \x^\top \Q \x$ for any symmetric matrix $\Q$. 
We use parentheses to augment a vector or matrix, e.g.,  $(\x_i)_{i=1}^n: = (\x_1^\top, \x_2^\top, \cdots, \x_n^\top)^\top$ with $\x_i \in \R^n, i = 1, 2, \cdots, n$.   
We write $\x_{<i}: = (\x_1, \x_2, \cdots, \x_{i-1})$, $\x_{>i}: =(\x_{i+1}, \x_{i+2}, \cdots, \x_n)$ and $\x_{-i}: = (\x_1, \cdots, \x_{i-1}, \x_{i + 1}, \cdots, \x_n)$. Besides, we define $\A_{<i}\x_{<i} := \smallsum_{j <i}\A_j \x_j $ and $\A_{>i}\x_{>i}: = \smallsum_{j>i} \A_j \x_j$. 
We use curly brace to represent a collection, e.g., $\{\x^k\}_{k=0}^K$  denotes a sequence.   We express the indicator function of  subset $\X \subseteq \R^n$ by  $I_{\X}(\x)$, where  we have $I_{\X}(\x) = 0$ for any  $\x \in \X$ and otherwise $I_{\X}(\x) = \infty$. 
We denote the  projection on a subset $\X$ by  $[\cdot]_{\X}$.  We use ${\rm diag}(\A_1, \A_2, \cdot, \A_n)$ to denote a diagonal matrix formed by the sub-matrices $\A_i, i = 1, 2, \cdots, n$.  We use $1, 2, \cdots, n$ to indicate integers and $N: = \{1, 2, \cdots, n\}$  indicates the set formed by successive integers $1$ to $n$,  and by analogy we have  $M:=\{1, 2, \cdots, m\}$.  We use ${\rm Im}(\A)$ to denote the image of matrix $\A$. We denote the cardinality of subset  $\X$ by $\vert \X \vert$. We use $\mathbb{E}_{\xi}[\cdot]$ to characterize the expectation of a mathematical expressions w.r.t.  uncertain parameter $\xi$.  For a given function $f: \R^n \rightarrow \R$, we denote its domain by ${\rm dom}~f$ which implies $f(\x) \in (-\infty, \infty)$ for any $\x \in {\rm dom}~f$. 

We claim a matrix $\PP \in \R^{n \times n}$ to be \emph{positive definite} if for any $\x\in\R^n$  and $\x \neq \bm{0}$ we have $\x^\top \PP \x > 0$,  and the matrix  \emph{positive semidefinite} if for any $\x \in \R^n$ we have $\x^\top \PP \x \geq 0$.  We say function $f: \R^n \rightarrow \R$ \emph{$\mu$-strongly convex}  if we have $f(\x) - \frac{\mu}{2}\Vert \x \Vert^2$  convex.  We claim function $f:\R^n \rightarrow \R$ \emph{$L_f$-$\Lips$ smooth} if we have $\vert f(\x) - f(\y)\vert \leq L_f \vert \x - \y\vert, \forall \x, \y \in \R^n$.  We have function $f: \R^n \rightarrow \R$  \emph{$L_g$-$\Lips$ differentiable}  or equivalently $f$ has \emph{$L_g$-$\Lips$ continuous gradients}  if we have  $\vert f(\y) - f(\x) - \langle \nabla f(\x), \y - \x \rangle \vert \leq \frac{L_g}{2}\Vert \x - \y \Vert^2$ for all  $\x, \y \in \R^n$,   or equivalently  $\Vert \nabla f(\x) - \nabla(\y) \Vert \leq L_f \Vert \x - \y\Vert$ for all $\x, \y \in \R^n$.  In this paper, we interchangeably use the term  \emph{$\Lips$ differentiable} and \emph{$\Lips$ continuous gradients}. We say that function $f:\R^n \rightarrow \R$ has \emph{easily computable proximal mapping}, if the solution  $\x = \arg \min_{\x \in \R^n} f(\x) + \frac{\gamma}{2}\Vert \x - \y\Vert^2$ is easy to obtain for any given $\y \in \R^n$ and proximal parameter $\gamma>0$.  

Consider a general optimization $\{\min_{\x \in \R^n} f(\x) : h(\x) = \bm{0}, g(\x) \leq \bm{0}\}$, where $f:\R^n \rightarrow \R$, $h: \R^n \rightarrow \R^l$ and $g: \R^n \rightarrow \R^d$. We usually have $h$ and $g$ continuously differentiable. We claim a solution $\x^{*} \in \R^n$ to be a \emph{first-order stationary point} (or \emph{stationary point} for short) of the problem, if there exist Lagrangian multipliers $\boldlambda^{*} \in \R^l$ and $\bm{\gamma}^{*} \in \R^d$ together with $\x^{*} \in \R^n$ satisfy the first-order optimality conditions of the problem, i.e.,
\begin{align*}
\left\{	
\begin{array}{l|l}
	\x^{*} \in \R^n &\bm{0} \in \partial f(\x^{*}) + (\boldlambda^{*})^\top \nabla g(\x^{*}) + (\bm{\gamma}^{*})^\top \nabla h(x^{*}). \\
	\boldlambda^{*} \in \R^l &h(\x^{*}) = \bm{0}, g(\x^{*}) \leq \bm{0}. \\
	\bm{\gamma}^{*} \in \R^d & (\boldlambda^{*})^\top h(\x^{*}) = \bm{0}, (\bm{\gamma}^{*})^\top g(\x^{*}) = \bm{0}. \\
\end{array}
\right\}
\end{align*}
Note that we have assumed general nonsmooth $f$, if $f$ is smooth and continuously differentiable, the subgradient $\partial f(\x^{*})$ can be replaced by the gradient $\nabla f(\x^{*})$. Correspondingly, we have $\nabla f(\x^{*}) + (\boldlambda^{*})^\top \nabla g(\x^{*}) + (\bm{\gamma}^{*})^\top \nabla h(x^{*}) = \bm{0}$. 

A multi-agent system is often defined over  a network or  graph which characterizes the interactions or communications among the agents. For a multi-agent system with  nodes $N: =\{1, 2, \cdots, n\}$ and given adjacent relationship, i.e., $j \in N_i$ where $N_i$ denotes the set of neighbors of agent $i$ (not including itself), it is easy to construct a network or  graph  in the form of $\mathcal{G}(N, E)$ where $N$ is the set of nodes and $E$ is the set of edges. Clearly, we have $(i, j) \in E$ if $j \in N_i$ and $i \in N_j$. In this paper, we only consider undirected network or graph. 

To be clarified, this paper refers to \emph{linearly constrained} or \emph{non-linearly constrained} as the coupled constraints of a concerned optimization.

\section{Classical ADMM}
ADMM has a long history and  was  independently developed by   Glowinski \& Marroco \cite{glowinskiapproximation}  and  Gabay \& Mercier \cite{gabay1976dual}  in the 1970s. 
However,  it was  until the very recent decade that the method  began to experience the surge of  interest. This is mainly caused by the massive large-scale and data-distributed computation demands arsing from both computer science \cite{boyd2011distributed} and engineering systems \cite{zhong2019admm}. 

ADMM was primarily developed for solving linearly constrained two-block convex optimization. This class of problems takes the canonical formulation of  
\begin{align*}
& \label{pp:p1}\min_{\x, \y} f(\x) + g(\y) \tag{$\mathbf{P1}$} \\
&{\rm s.t.}~ \A \x + \B \y= \bm{0}, 
\end{align*}
where $f: \R^n \rightarrow \R$ and $g: \R^{n} \in \R$ are given convex  objective functions related to the decision variables $\x \in \R^n$ and $\y \in \R^n$ respectively. 
Functions  $f$ and $g$ are possibly nonsmooth and a usual case is that some local bounded convex constraints $\X \subseteq \R^n$ and $\Y \subseteq \R^n$  exist and are included  in the form of indicator functions.  In such context, the objective functions $f$ and $g$ can be distinguished by smooth and nonsmooth parts, i.e.,  
$\tilde{f}(\x) + I_{\X}(\x) + \tilde{g}(\y) + I_{\Y}(\y)$ with $\tilde{f}: \R^n \rightarrow \R$ and $\tilde{g}:\R^n \rightarrow \R$ denoting the smooth components, $I_{\X}(\x)$ and $I_{\Y}(\y)$ representing the nonsmooth components caused by the local constraints.  Note that if  $f$ or $g$ is claimed to be smooth, we implicitly have $\X \in \R^n$ or $\Y \in \R^n$. The coefficient matrices $\A \in \R^{m \times n}$ and $\B \in \R^{m \times n}$  encode  the linear couplings between  the decision variables $\x$ and $\y$. We enforce  $\bm{0}$ on the right-hand side of the constraints for simplification, but any constant $\bb \in \R^l$ is admissible by the model.

We refer to the well-known ADMM for solving the two-block convex optimization \eqref{pp:p1} as \emph{classical ADMM} \cite{boyd2011distributed}.  The method takes the iterative scheme
\begin{align*}
& \textbf{Classic ADMM:} \\
& \textbf{Primal update:}~~
\begin{cases}
	\x^{k+1} = \arg\min\limits_{\x} \Lag_{\rho}(\x, \y^k, \boldlambda^k) \\
	\y^{k+1} = \arg\min\limits_{\y} \Lag_{\rho}(\x^{k+1}, \y, \boldlambda^k) \\
\end{cases} \\
& \textbf{Dual update:}~~~~~~~ \boldlambda^{k+1} = \boldlambda^k + \tau \rho (\A \x^{k+1} + \B \y^{k+1})
\end{align*} 
where $\tau > 0 $ denotes a dual stepsize (it actually should be $\tau \rho$ but we often refer to $\tau$ as  stepsize because  $\rho$ is given penalty parameter).  

As documented in \cite{glowinski2014alternating},  \emph{classical ADMM} is a split version of ALM where the joint  ALM problem is decomposed into two subproblems by $\gauss$ decomposition. 
At the very beginning,  this method was  termed  ALG2 until its equivalence to Douglas-Rachford  alternating direction method was discovered when we have $\tau = 1$. This gave rise to the term ADMM  that we are familiar today. 
\emph{Classical ADMM} can 
be derived  from Douglas-Rachford  splitting method (DRSM) via a number of ways as documented in  \cite{han2022survey, eckstein2015understanding, davis2016convergence}. One popular way is to apply DRSM to the dual of problem  \eqref{pp:p1} which corresponds to finding the minimal of the sum of two convex functions. Viewing \emph{classical ADMM} from the perspective of DRSM is often helpful  in both studying  and understanding its convergence (see the comprehensive survey \cite{eckstein2015understanding}). 
We often see the trivial dual stepsize $\tau = 1$ with \emph{classical ADMM} due to its equivalence to DRSM,  whereas  the method  can take any  nontrivial  stepsizes $\tau \in (0, \frac{\sqrt{5} + 1}{2})$ , which is known as Fortin and Glowinski constant \cite{fortin2000augmented, glowinski1989augmented, fortin1983chapter}. Particularly, a larger dual stepsize is often advised to achieve faster convergence \cite{chen2012matrix, he2011solving}. 

Despite \emph{classical ADMM} was primarily developed as an inexact implementation of  ALM, its convergence behavior is quite different from real approximate ALM, i.e., solving the joint ALM problems relatively  accurate via multiple rounds of $\gauss$ decomposition instead of one. Surprisingly, \emph{classical ADMM} was found much more computationally efficient than ALM and its approximations.  Because of that, \emph{classical ADMM} was argued not a real approximate ALM.
The superior computation efficiency of \emph{classical ADMM}  somehow underlies the popularity and prevalence of the method, even over the real  ALM approximations, such as the Diagonal Quadratic Approximation (DQA) method \cite{ruszczynski1995convergence}.

The theoretical convergence of \emph{classical ADMM}  for convex optimization has  been long-established
(Gabay, 1983 \cite{fortin2000augmented}; Glowinski \& Tallec, 1989 \cite{glowinski1989augmented}; Eckstein \& Bersekas, 1992 \cite{eckstein1992douglas}). However, it was until the very recent decade that  its convergence rate and iteration complexity were established.  Monteiro \emph{et.al.} \cite{monteiro2010iteration} first established the $\mathcal{O}(1/k)$ iteration complexity in an ergodic sense and followed by He \emph{et.al.} \cite{he20121}. Later, the non-ergodic $\mathcal{O}(1/k)$ iteration complexity  was established by He \emph{et.al.} \cite{he2015non}.  
Among the literature, \cite{he20121, he2015non} have been recognized as the most general results regarding  the convergence and convergence rate of \emph{classical ADMM}.   
In addition, global linear convergence rate was established for some special cases, such as  linear programming  \cite{eckstein1990alternating} or one  objective function strictly convex and $\Lips$ differentiable  \cite{deng2016global}. 
For  general convex optimization, a global linear convergence can also be achieved by employing a sufficiently small dual stepsize $\tau$ \cite{hong2017linear}.

\section{ADMM Variants}
ADMM was  originally developed for solving linearly constrained two-block convex  optimization. In the past decade,  the method  has  experienced extensive developments.     On one hand, it has been generalized to broad classes of problems (i.e., multi-block, coupled objective, and nonconvex etc.).   Specifically, it has been extended  to  deal with \emph{five} major classes of problems: \emph{i)} two-block with separable objective,  \emph{ii)} multi-block with separable objective,  \emph{iii)} multi-block with coupled  objective,  \emph{iv)} consensus optimization,  and \emph{iv)} non-linearly constrained optimization.   On the other hand, the method has been reinforced  in diverse directions, including faster convergence rate, easier implementation, higher computation efficiency, flexible communication,  enhanced robustness and compatibility etc.  These developments lead to a plentiful of ADMM variants for different problems and situations.

\begin{figure}[h]
	\centering
	\includegraphics[width = 3.5 in]{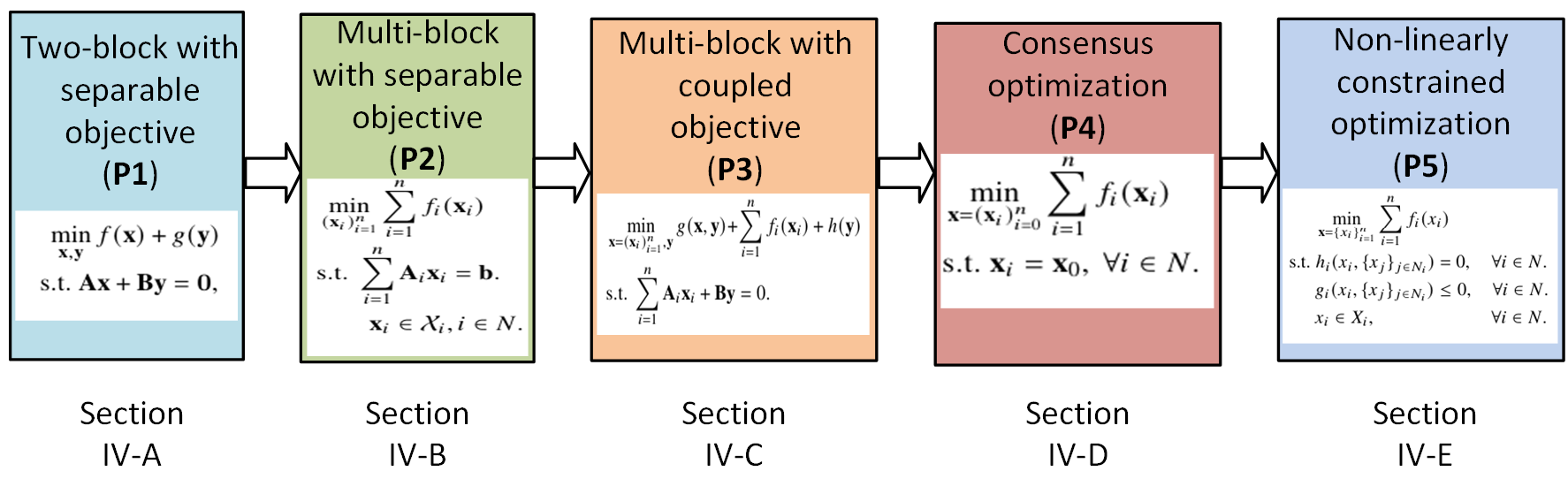}
	\caption{The roadmap of Section IV.  }
	\label{fig:section_VI}
\end{figure}

This section reviews  ADMM and its variants comprehensively and broadly for solving the \emph{five} major classes of problems. 
Specifically, each subsection is devoted to one class of problems followed by the related ADMM variants. We discuss the ADMM variants in terms of  main assumptions, decomposition scheme, convergence properties and main features.   A roadmap for this section can refer to Fig. \ref{fig:section_VI}. 

Throughout the section, we use $\Lag_{\rho}(\cdot)$ to denote an AL function with  penalty parameter $\rho > 0$ for a concerned problem. We assume that the AL functions can be easily derived from the context and thus do not discuss them in details. 
To be noted, in the algorithmic implementation of ADMM or its variants,  we often only indicate the related arguments in subproblems  for simplification. Without specifications, we use $\x, \y, \x_i$ to denote the primal  variables, $\boldlambda, \boldlambda_i$ to represent  Lagrangian multipliers and $r, s, \tau$ to indicate dual stepsizes.

\subsection{Two-block with separable objective}

\begin{figure}[h]
	\centering
	\includegraphics[width = 3.5 in]{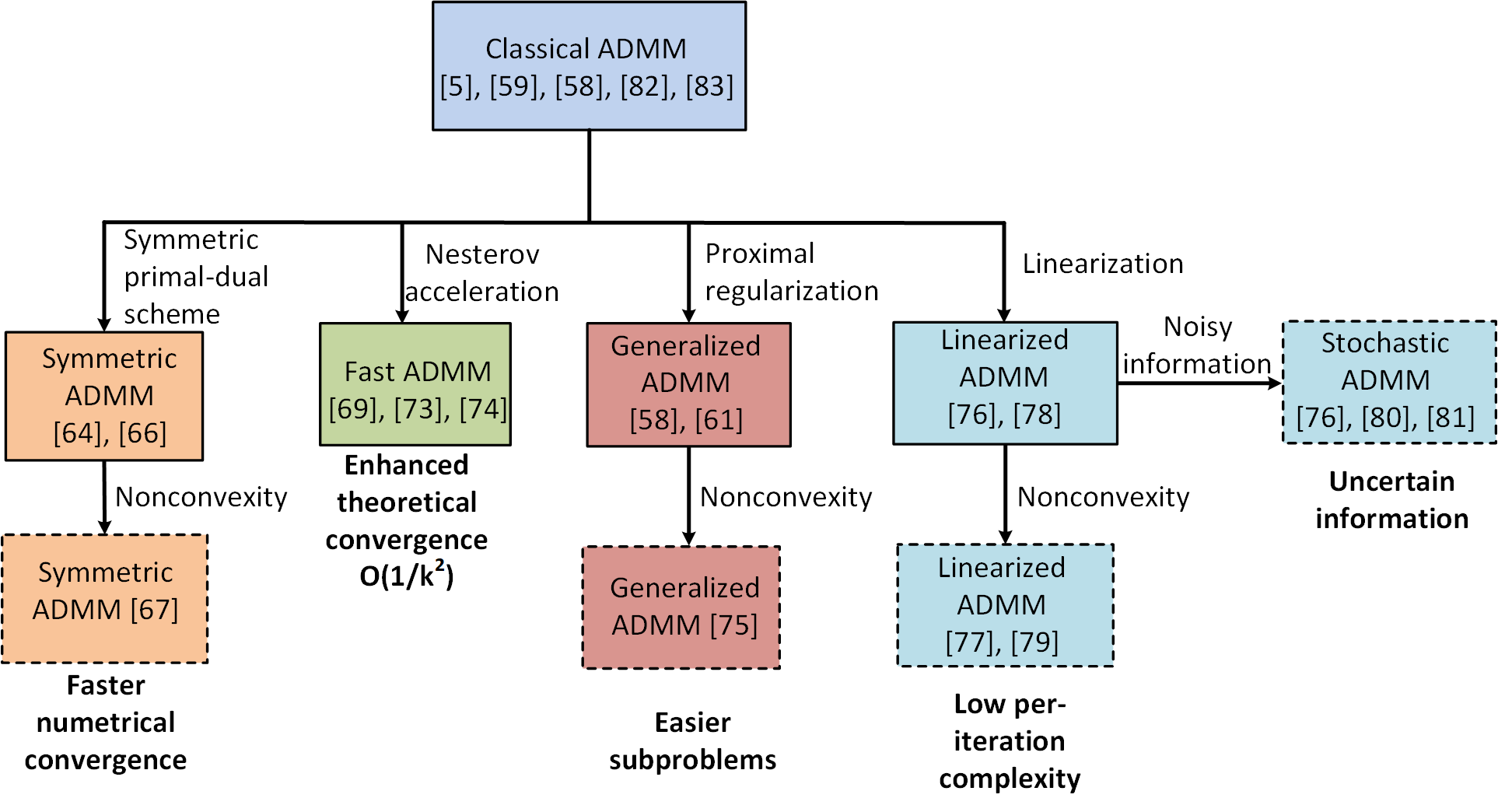}
	\caption{An overview of  ADMM variants for solving two-block problem \eqref{pp:p1} (solid and dashed boxes indicate convex and nonconvex methods respectively).}
	\label{fig:p1}
\end{figure}

In this part, we focus on the standard two-block  problem \eqref{pp:p1}.  In addition to classical ADMM,  a number of ADMM variants have been developed either for different situations or with different features. These ADMM variants range from \emph{symmetric ADMM}, \emph{fast ADMM}, \emph{generalized ADMM}, \emph{linearized ADMM} and \emph{stochastic ADMM}. They all can be viewed as the extension of classical ADMM with the integration of  certain techniques (i.e., symmetric primal-dual scheme, Nesterov acceleration, proximal regularization and linearization). Compared with classical ADMM, these ADMM variants are often celebrated by their distinguishing features, such as improved convergence rate, easier subproblems, low per-iteration complexity and compatible with uncertain information. An overview of the relationships and features of the ADMM variants for solving two-block problem \eqref{pp:p1}  is shown  in Fig. \ref{fig:p1}.  We distinguish the convex and nonconvex methods by solid and dashed boxes.    In the sequel, we introduce each of those methods.

\subsubsection{Symmetric ADMM} 
As a well-known ADMM variant, the main alternation of \emph{symmetric ADMM} over classical ADMM is that 
primal and dual variables are treated  in a symmetric manner. Specifically, a dual update follows each block of primal update.  The method takes the  iterative scheme 
\begin{align*}
&\textbf{Symmetric ADMM:} ~~\\
&\textbf{Primal update:}~  \x^{k+1} = \arg \min_{\x} \Lag_{\rho}(\x, \y^k, \boldlambda^k) \\
&\textbf{Dual update:}~~~ \boldlambda^{\kmid} = \lambda^k +  r \rho (\A \x^{k+1} + \B \y^k) \\
&\textbf{Primal update:}~   \y^{k+1} = \arg \min_{\y} \Lag_{\rho}(\x^{k + 1}, \y, \boldlambda^k) \\
&\textbf{Dual update:}~~~  \boldlambda^{k+1} = \boldlambda^{\kmid} +  s \rho(\A \x^{k+1} + \B \y^{k+1}) 
\end{align*}

\emph{Symmetric ADMM} was developed simultaneously as classical ADMM by 
Glowinski  in the 1970s \cite{glowinski1975approximation}. 
The method was primary termed ALG3 until  its equivalence to  Peaceman-Rachford splitting method (PRSM) was discovered  for trivial  dual stepsizes $r= 1$ and $s = 1$  \cite{fortin2000augmented, glowinski1989augmented, glowinski2003convergence}. 
Note that \textit{symmetric ADMM} degenerates into classical ADMM  when the dual stepsize is set as $r= 0$.  
The benefit of \emph{symmetric ADMM} over  classical ADMM is that it often yields  faster convergence  when convergent (see \cite{giselsson2014diagonal, he2014strictly} for some numerical examples).  

We often see the trivial stepsizes $r = 1$ and $s = 1$ with \emph{symmetric ADMM}, however they do not guarantee the convergence of the method  for  convex optimization as classical ADMM (see \cite{he2014strictly} for some divergent examples).  
This is because the generated sequence is  not strictly contractive \cite{he2014strictly}. 
To fix such issue,  a strictly contractive  \emph{symmetric ADMM} with damping  dual stepsizes  $r< 1$ and  $s <1$  was proposed in \cite{he2014strictly}. Besides, the ergodic convergence rate $\mathcal{O}(1/k)$ and non-ergodic convergence rate $\mathcal{O}(1/k)$  were established.  Further, it was argued  that the damping  stepsizes $r <1$ and $s < 1$  are not greeted and  one normally  prefers larger dual stepsizes to achieve faster convergence \cite{he2016convergence}. 
To deal with such a contradiction,   \cite{he2016convergence} comprehensively studied the dual stepsize  $(r, s)$  to ensure the convergence of \emph{symmetirc ADMM}  and  identified the admissible domain $\mathcal{D}=\{(s, r) \vert s \in (0, \frac{\sqrt{5} + 1}{2}), r \in (-1, 1),   r + s > 0, \vert r\vert < 1 + s -s^2 \}$. This implies that the dual stepsizes of \emph{symmertic ADMM} are actually not restricted to  $r<$ and $s < 1$. Moreover,  the admissible  domain $\mathcal{D}$ actually has  enlarged  the Fortin and Glowinski  constant $\tau \in (0, \frac{\sqrt{5} + 1}{2})$ with classical ADMM, which infers that  \emph{symmetric ADMM} enjoys larger flexibility in parameter settings over  classical ADMM.

The  above results  are for convex optimization. \emph{Symmetric ADMM} has already been generalized to nonconvex counterparts (i.e., $f$ and $g$ are  nonconvex).  Specifically, \cite{wu2017symmetric} established the convergence of the method in nonconvex setting under the conditions: \emph{i)} $g$ is $\Lips$ differentiable, and \emph{ii)} ${\rm Im}(\A) \subseteq {\rm Im}(\B)$, and the mapping $p(u) = \{\arg \min_{\y} g(\y): \B \y = u\}$ is $\Lips$ smooth. 
Actually, condition \emph{ii)} is a weaker assumption of full column rank $\B$ \cite{wang2019global}.  The faster convergence behaviors of \emph{symmetric ADMM} over classical ADMM have been  corroborated by many numerical studies \cite{wu2017symmetric}.

\subsubsection{Fast ADMM} As discussed,    classical ADMM  and symmetric ADMM only promise an   $\mathcal{O}(1/k)$ convergence rate.  
To enhance the convergence rate,   \cite{goldstein2014fast} proposed an accelerated ADMM variant  by 
combining classical ADMM with \emph{Nesterov acceleration}  technique.  This leads to  a \emph{fast ADMM} that  ensures  an  $\mathcal{O}(1/k^2)$ convergence rate for  a class of strongly convex problems (i.e., $f$ and $g$ are strongly convex, $g$ is convex quadratic). 
The \textit{Nesterov acceleration}  technique was originally developed  for  unconstrained smooth  convex optimization  \cite{nesterov1983method}. This technique is  attractive for it 
can improve the convergence rate of  first-order gradient methods by an order, i.e., from $\mathcal{O}(1/k)$ to $\mathcal{O}(1/k^2)$, which is argued to be the best attainable computation efficiency with  first-order information. This technique was later extended to a proximal gradient method for unconstrained nonsmooth and nonconvex  optimization, which has enjoyed wide success in the domain of machine learning  \cite{beck2009fast, beck2009fast, li2015accelerated}. Central to \textit{Nesterov acceleration}  is to introduce an interpolation step  in terms of the current and preceding iterates at each iteration. 
The combination of \textit{Nesterov acceleration} with classical ADMM is  reasonable  as  ADMM can  be seen as a first-order solver of  ALM. 
The implementation of  \emph{fast ADMM} is presented below. 
\begin{align*}
& \textbf{Fast ADMM:} ~~~~~\\
& \textbf{Primal update:}~  
\begin{cases}
	& \!\!\!\x^{k} = \arg \min_{\x} \Lag_{\rho}(\x, \hat{\y}^k, \hat{\boldlambda}^k)  \\
	& \!\!\!\y^{k} = \arg \min_{\y} \Lag_{\rho}(\x^k, \y, \hat{\boldlambda}^k)\\
\end{cases} \\
& \textbf{Dual update:}~~~~  \boldlambda^{k} = \hat{\boldlambda}^k +  \rho(\A \x^{k} + \B \y^{k})\\
& \textbf{Nesterov stepsize:}   ~a_{k + 1} = \frac{1 + \sqrt{1 + 4 a_k^2}}{2}\\ 
& \textbf{Interpolation step:}   \begin{cases}
	\hat{\x}^{k + 1} = \x^k + \frac{a_k - 1}{a_{k + 1}}(\x^k - \x^{k-1}) \\
	\hat{\y}^{k+1} = \y^k + \frac{a_k - 1}{a_{k + 1}}(\y^k - \y^{k-1}) \\
	\hat{\boldlambda}^{k+1} = \boldlambda^k + \frac{a_k - 1}{a_{k + 1}}(\boldlambda^k - \boldlambda^{k-1}) \\
\end{cases}
\end{align*}
where $\{a_k\}_{k=1}^K$ represents the interpolation  stepsize of \emph{Nesterov acceleration}.  
Note that the main alternation of  \emph{fast ADMM}  over classical  ADMM is that an interpolation  procedure is introduced to moderate the current and preceding primal-dual updates  $(\x^k, \y^k, \boldlambda^k)$ and $(\x^{k-1}, \y^{k-1}, \boldlambda^{k-1})$ generated by  classical ADMM at the end of each iteration. 
This leads to the modified primal-dual updates $(\hat{\x}^k, \hat{\y}^k, \hat{\boldlambda}^k)$ that  serve the  next update.  The work \cite{goldstein2014fast} established  the $\mathcal{O}(1/k^2)$ convergence rare of  \emph{fast ADMM} for the special case where   $f, g$ are  both strongly convex and $g$ is besides quadratic.  However, for more general  problems,  the theoretical results  are still open questions.

The difficulty to establish the convergence of \emph{fast ADMM} for  general  problems lies in  the fact that  classical ADMM is actually not a  first-order descent solver for ALM like gradient-based methods. In other words, we do not have the monotonically decreasing property of the objective value w.r.t. the iterations with ADMM.  This can be perceived from the perspective of  DRSM considering their equivalence  \cite{patrinos2014douglas}. However, it was argued that a descent solver may be constructed  by  adding  some monitoring and correction steps (see  \cite{goldfarb2013fast} for an example). This actually sheds some lights on the generalization of \emph{fast ADMM} to more general problems.

\subsubsection{Generalized ADMM} Note that the main computation burden with classical ADMM  lies in solving the subproblems iteratively. Therefore, it is significant to enable easier subproblems to improve computation efficiency.  
To achieve such a goal,  \emph{generalized ADMM} was proposed  as an advanced version of classical ADMM  \cite{deng2016global, he20121, li2015global}.  The main idea  is  to optimize some proximal surrogates of the subproblems which are often much easier than the original subproblems.  \emph{Generalized ADMM}  takes the iterative scheme
{	\setlength{\abovedisplayskip}{3pt}
	\setlength{\belowdisplayskip}{2pt}
\begin{align*}
&\textbf{Generalized ADMM:} \\
& \textbf{Primal update:} \\
& \!\!\!\!\begin{cases}
	&\!\!\!\x^{k+1} = \arg \min_{\x} \Lag_{\rho}(\x, \y^k, \boldlambda^k)  + \frac{1}{2}\Vert \x - \x^k\Vert^2_{\mathbf{P}}\\
	&\!\!\!\y^{k+1} = \arg \min_{\y} \Lag_{\rho}(\x^{k+1}, \y, \boldlambda^k)+ \frac{1}{2} \Vert \y- \y^k\Vert^2_{\mathbf{Q}}\\
\end{cases}\\
&\textbf{Dual update:}~~  \boldlambda^{k+1} = \boldlambda^k + \tau \rho(\A \x^{k+1} + \B \y^{k+1}) 
\end{align*} }
where $\mathbf{P} \in \R^{n \times n}$ and $\mathbf{Q} \in \R^{n \times n}$ are symmetric positive semidefinite matrices. 
Note that the main alternations  of \emph{generalized ADMM} over classical ADMM are the proximal terms $\frac{1}{2}\Vert \x - \x^k\Vert^2_\mathbf{P}$ and 
$\frac{1}{2} \Vert\y- \y^k\Vert^2_\mathbf{Q}$  added to the subproblems of primal update. The method is termed \textit{generalized ADMM} because it involves classical ADMM as a special case with \emph{zero} $\PP$ and $\Q$.  
The proximal terms are valuable for they bring benefits to the flexible implementation of the method. 
Specifically,  some potential structures of $f$ and $g$ can be exploited  to yield easier subproblems. For example,  if  $f$ is separable  across its coordinates, we can select   $\mathbf{P} = \tau \mathbf{I} - \rho \A^\top \A $ to yield an $\x$-subproblem
\begin{align}
\label{eq:prox-linear}\x^{k + 1} \!=\! \arg \min_{\x}  f(\x) \!+\! h(\x^k)^\top (\x - \x^k) \!+\! \frac{\tau}{2} \Vert \x - \x^k \Vert^2
\end{align}
where we have $h(\x^k) = \A^\top \boldlambda^k + \rho \A^\top (\A \x^k + \B \y^k )$ and the $\x$-subproblem  \eqref{eq:prox-linear} reduces to a number of one-dimensional subproblems.  Else if $f$ has easily computable proximal mapping, it is also beneficial because \eqref{eq:prox-linear} is exactly the proximal mapping of $f$, i.e.,  $\x^{k + 1} = \arg \min_{\x} f(\x) + \frac{\tau}{2}\Vert \x - (\x^k-\tau^{-1}h(\x^k)) \Vert^2$. 
Consider another case that  $f$ is quadratic with Hessian matrix $\mathbf{H}_f$ (this implies that $f$ can be expressed by  $f(\x) = f(\x^k) + \nabla f(\x^k)^\top (\x - \x^k) + \frac{1}{2}(\x - \x^k)^\top \mathbf{H}_f (\x - \x^k)$), we could  select $\mathbf{P} = \tau \mathbf{I} - \rho \A^\top\A - \mathbf{H}_f$ to yield an $\x$-subproblem
\begin{align}
\label{eq:proximal} \x^{k + 1} = \arg \min_{\x}   l(\x^k)^\top (\x - \x^k) + \frac{\tau}{2} \Vert \x - \x^k \Vert^2
\end{align}
where we have $l(\x^k) = \nabla f(\x^k) + \A^\top \boldlambda^k + \rho \A^\top (\A \x^k + \B \y^k)$.  Note that  \eqref{eq:proximal} admits a gradient-like closed-form solution $\x^{k+1} = \x^k - \tau^{-1} l(\x^k)$ and enjoys low per-iteration complexity. These are examples how \emph{generalized ADMM} can make use of the proximal terms to yield easier subproblems.   To be noted,  the proximal terms with positive semidefinite $\PP$ and $\Q$ will not disrupt the convergence property of classical ADMM. In other words, we do not require extra assumptions besides convexity to ensure the convergence of \emph{generalized ADMM}. 
This can be understood  that the proximal terms actually play the role of slowing down moving and enhancing convergence  since they  penalize the deviations from preceding updates.  The  $\mathcal{O}(1/k)$ iteration complexity of \emph{generalized ADMM} was   established  for general convex optimization in  \cite{he20121}.  For the special case where $f$ and $g$ are strongly convex and $\Lips$ differentiable, the matrices $\A, \B$ satisfy certain full row or column rank conditions, a global linear convergence rate  was established in \cite{deng2016global}.

The above results are for convex optimization. For the nonconvex counterpart (i.e., $f$ and $g$ are nonconvex),  the convergence of \emph{generalized ADMM}   towards stationary points was established under the conditions: \emph{i)} $g$ is $\Lips$ differentiable, and \emph{ii)} $\B$ has full row rank  \cite{li2015global}. Similar to the  convex counterpart, the proximal terms play the role of yielding easier subproblems and will not disrupt convergence. 

\subsubsection{Linearized ADMM} 
Note that the implementations of  above ADMM variants  assume that the subproblems of primal update are easy to be solved exactly.
There exist  cases that the objective functions $f$ and $g$ are complex  and  solving the subproblems exactly is expensive or not desirable due to the high computation complexity.  In such context, it is critical to figure out how to mitigate  per-iteration complexity. To address such an issue,   \textit{linearized ADMM} was proposed with the idea of optimizing local linear approximations  of subproblems, which often leads to some cheap  gradient iterates in place of solving the subproblems exactly  \cite{gao2018information, lu2021linearized}.  
The implementation of \emph{linearized ADMM}  takes the usual form of
\begin{align*}
&\textbf{Linearized ADMM:} ~~~~~\\
&\textbf{Primal update:}~  \x^{k+1} = \big[ \x^k - \alpha_k \nabla_{\x} \tilde{\Lag}_{\rho}(\x^k, \y^k, \boldlambda^k)\big]_{\X} \\
& \quad \quad \quad \quad  \quad \quad \quad \y^{k + 1} = \big[ \y^k -\beta_k \nabla_{\y} \tilde{\Lag}_{\rho}(\x^{k + 1}, \y^k, \boldlambda^k) \big]_{\Y} \\
& \textbf{Dual update:}~~~ \boldlambda^{k+1} = \boldlambda^k +  \rho(\A \x^{k+1} + \B \y^{k+1}) 
\end{align*}
where  $\tilde{\Lag}_{\rho}(\x, \y, \boldlambda) = \tilde{f}(\x) + \tilde{g}(\y) + \langle \boldlambda, \A \x + \B \y \rangle + \frac{\rho}{2}\Vert \A \x + \B \y \Vert^2$ aggregates the differentiable parts of  AL function; $\nabla_{\x} \tilde{\Lag}_{\rho}(\x, \y, \boldlambda)$ and $\nabla_{\y} \tilde{\Lag}_{\rho}(\x, \y, \boldlambda)$ denote the gradients of $\tilde{\Lag}_{\rho}(\x, \y, \boldlambda)$ w.r.t. $\x$ and $\y$; the subsets $\X$ and $\Y$ indicate the local constraints related to decision variable $\x$ and $\y$. As clarified in problem \eqref{pp:p1}, we have   $f(\x) = \tilde{f}(\x) + I_{\X}(\x)$ and $g(\y) = \tilde{g}(\y) + I_{\Y}(\y)$. 
Note that the primal update of  \emph{linearized ADMM} reduces to two projected gradient iterates. They  are actually derived from the proximal linearized subproblems
\begin{align*}
& \x^{k+1} \!=\! \arg \min_{\x \in \X} \langle \nabla_{\x} \tilde{\Lag}_{\rho}(\x^k, \y^k, \boldlambda^k), \x - \x^k  \rangle \!+\! \frac{1}{2 \alpha_k} \Vert \x - \x^k  \Vert^2 \\
& \y^{k+1} \!=\! \arg \min_{\y \in \Y} \langle \nabla_{\y} \tilde{\Lag}_{\rho}(\x^{k + 1}, \y^k, \boldlambda^k) , \y - \y^k\rangle \!+\! \frac{1}{2 \beta_k} \Vert \y - \y^k  \Vert^2
\end{align*}
To be noted,  the differentiable  AL function   $\tilde{\Lag}_{\rho}(\x, \y, \boldlambda)$ is linearized at $(\x^k, \y^k, \boldlambda^k)$ and $(\x^{k+1}, \y^k, \boldlambda^k)$ w.r.t. $\x$ and $\y$,  and besides some proximal terms $\frac{1}{2 \alpha_k} \Vert \x - \x^k  \Vert^2$ and $\frac{1}{2 \beta_k} \Vert \y - \y^k  \Vert^2$ are added in the subproblems to control the accuracy of local linear approximation.

\emph{Linearized ADMM} applies to both convex \cite{gao2018information, lin2017extragradient} and nonconvex optimization \cite{lu2021linearized, liu2019linearized}  but rests on  different conditions to ensure convergence. One common condition  is that  the differentiable objective components $\tilde{f}$ and $\tilde{g}$  are $\Lips$ differentiable.  This can  be understood that  the $\Lips$ differentiable property  makes it possible to  bound the linear approximation discrepancy by $\vert \tilde{f}(\x) - \tilde{f}(\x^k) - \langle \nabla \tilde{f}(\x^k), \x - \x^k\rangle \vert \leq \frac{L_{\tilde{f}}}{2} \Vert \x - \x^k \Vert^2$ where we assume $\tilde{f}$  $L_{\tilde{f}}$-$\Lips$ differentiable. 
For convex optimization, the convergence and     $\mathcal{O}(1/k)$ ergodic  convergence  rate   of  \emph{linearized ADMM}  were established in  \cite{gao2018information, lin2017extragradient}.    For  nonconvex counterpart,  the convergence of \emph{linearized ADMM} was established under slightly different conditions in \cite{lu2021linearized} and \cite{liu2019linearized}. Specifically, \cite{lu2021linearized} assumed  that $f$ and $g$ are $\Lips$ differentiable (i.e., $\X, \Y \subseteq \R^n$), and  \cite{liu2019linearized} made the assumptions that   $g$ is $\Lips$ differentiable (i.e., $\Y \in \R^n$), ${\rm Im}(\A)\subseteq {\rm Im}(\B)$ and $\B$ has  full column rank.
Actually,   these two works rely on the same key step to draw  convergence, i.e., identifying a sufficiently decreasing and lower bounded Lyapunov function.  To this end, they both require to  bound the Lagrangian multipliers updates $\Vert \boldlambda^{k + 1} - \boldlambda^k \Vert^2$ by the primal updates  $\Vert \x^{k + 1} - \x^k \Vert^2$ and $\Vert \y^{k + 1} - \y^k \Vert^2$. Though the  assumptions of there two works are different, they  are actually  used to  achieve such same objective. 

To be clarified,  we only require that  the objective functions to be linearized are $\Lips$ differentiable both in convex and nonconvex optimization. The method can be adapted to the case where  only  one objective function is $\Lips$ differentiable. In such case, we can only  linearize the subproblem with $\Lips$ differentiable objective  and solve the other one exactly. The established theoretical results still hold.

\subsubsection{Stochastic  ADMM} Another branch of  extension of ADMM  is to account for the incomplete and  inaccurate   information  in practical implementation.  A typical scenario is that  explicit formulas of objective functions are not available for a complex  engineering system  and instead only  noisy gradients regarding the system performance (i.e., the gradients of objective functions)  are accessible by means of sampling. 
In such situation,  it is impossible to solve the subproblems with classical ADMM or its variants exactly.   Therefore, 
\cite{gao2018information, ouyang2013stochastic, ouyang2012stochastic} studied a stochastic version of ADMM.  The basic idea is to perform some gradient-like iterates with the available noisy gradients at each iteration in place of solving the subproblems comprehensively.  The idea is natural since only gradient information is accessible in such situation.    \emph{Stochastic ADMM} takes the iterative scheme
\begin{align*}
&\textbf{Stochastic ADMM:} ~~~~~\\
&\textbf{Primal update:}~  \x_{k+1} = \big[ \x^k - \alpha_k \nabla_{\x} \tilde{\Lag}_{\rho}(\x^k, \y^k, \boldlambda^k, \xi^{k+1})\big]_{\X} \\
& \quad \quad \quad \quad  \quad \quad \quad \y^{k + 1} =\!\big[ \y^k -\beta_k \nabla_{\y} \tilde{\Lag}_{\rho}(\x^{k + 1}, \y^k, \boldlambda^k, \zeta^{k+1}) \big]_{\Y} \\
& \textbf{Dual update:}~~~ \boldlambda^{k+1} = \boldlambda^k +  \rho(\A \x^{k+1} + \B \y^{k+1}) 
\end{align*}

One may note that \emph{stochastic ADMM} resembles linearized ADMM. The only difference lies in the gradients used to perform  the primal update. Specifically, \emph{stochastic ADMM} uses some noisy gradients of AL function to perform the primal update whereas linearized ADMM uses deterministic and accurate ones. We have  $\xi^{k + 1}$ and $\zeta^{k + 1}$ characterize the estimation errors of  gradients for  the objective functions $\tilde{f}$ and $\tilde{g}$ at iteration $k$, which can be expressed by  $\nabla \tilde{f}(\x^k, \xi^{k+1})$  and $\nabla \tilde{g}(\y^k, \!\zeta^{k+1})$ and the corresponding noisy gradients of AL function are $\nabla_{\x} \tilde{\Lag}_{\rho}(\x^k, \y^k, \boldlambda^k, \xi^{k+1})$ and $\nabla_{\y} \tilde{\Lag}_{\rho}(\x^{k + 1}, \!\y^k, \!\boldlambda^k, \zeta^{k\!+1})$.

 Clearly, the convergence of \emph{stochastic ADMM} depends on the accuracy of the estimations of  gradients. For the usual case that the estimations are  unbiased and variance-bounded, \cite{gao2018information, ouyang2013stochastic, ouyang2012stochastic} established the convergence of the method for convex optimization with $\Lips$ differentiable objective functions $\tilde{f}$ and $\tilde{g}$.   
 Despite the similarity of \emph{stochastic ADMM} to  linearized ADMM, they actually show quite different  convergence behaviors due to the inaccurate and accurate information used in the iterative process.  Specifically, the convergence of linearized ADMM can be directly examined in a deterministic sense, whereas that can only be  evaluated in a stochastic space by studying the expectation of performance metrics with \emph{stochastic ADMM}.  Moreover, due to the inaccurate information, \emph{stochastic ADMM} only promises   $\mathcal{O}(1/\sqrt{k})$ convergence rate in contrast to the $\mathcal{O}(1/k)$ convergence rate of linearized ADMM with accurate information  \cite{gao2018information, ouyang2013stochastic, ouyang2012stochastic}. 
For the special case where $\tilde{f}$ and $\tilde{g}$ are strongly convex and $\Lips$ differentiable,  \emph{stochastic ADMM} ensures an $\mathcal{O}(\frac{\log k}{k})$ ergodic convergence rate  \cite{ouyang2013stochastic, ouyang2012stochastic}.  

 Similar to linearized ADMM,  \emph{stochastic ADMM} also 
 applies to the special case where only one objective function is $\Lips$ differentiable. In such context, the method can be applied if the other subproblem has explicitly  available objective function and can be solved exactly in the iterative process.   

\begin{table*}[h] 
\setlength\tabcolsep{2pt}
\renewcommand\arraystretch{2}
\centering
\caption{ADMM variants for solving two-block optimization \eqref{pp:p1}}
\label{tab:two-block_ADMM}
\begin{tabular}{lccccc}   
	\hline 
	\textbf{Methods}  & \makecell[l]{\textbf{Main assumptions}}  &  \makecell[l]{\textbf{Types}} & \makecell[l]{\textbf{Convergence}}  &  \makecell[l]{\textbf{Features}} &   \makecell[l]{\textbf{References}} \\
	\hline 
	\makecell[l]{~\\ ~\\ Classic ADMM}    & \makecell[l]{$f$ and $g$ convex. \\ Existence of saddle points. }  & Gauss-Seidel & \makecell[l]{Global convergence. \\Global optima. \\ Convergence rate $\mathcal{O}(1/k)$ (convex). \\ Linear convergence (strongly convex)} & \makecell[l]{Convex.} &  \makecell[l]{\cite{boyd2011distributed, he2015non} \\ \cite{he20121, monteiro2013iteration,nishihara2015general} } \\
	\hline
	\multirow{2}{*}{\makecell[l]{~\\ ~\\ Symmetric  ADMM}} & \makecell[l]{$f$ and $g$ convex.\\ $(r, s)$ properly selected.\\ Existence of saddle points.  }    & Gauss-Seidel  & \makecell[l]{ Global convergence. \\ Global optima.  \\  Convergence rate $\mathcal{O}(1/k)$.}  &  \makecell[l]{Faster  numerical convergence. \\ Convex. } & \makecell[l]{\cite{giselsson2014diagonal,  he2016convergence}} \\
	\cline{2-6}
	& \makecell[l]{$f$ and $g$ nonconvex. \\
		$g$ $\Lips$ differentiable. \\ ${\rm Im}(\A) \subseteq {\rm Im}(\B)$. \\ $\B$ full column rank.  }    & Gauss-Seidel  & \makecell[l]{Global convergence. \\ Stationary points. }  &  \makecell[l]{ Faster numerical convergence. \\ Nonconvex.} & \makecell[l]{\cite{wu2017symmetric}} \\
	\hline
	Fast ADMM  & \makecell[l]{$f$ and $g$  convex. \\ $g$ convex quadratic.\\ Existence of saddle points. }  & Gauss-Seidel  & \makecell[l]{Global convergence. \\ Convergence rate $\mathcal{O}(1/k^2)$.}  & \makecell[l]{Enhanced convergence rate. \\ Convex.} & \cite{goldstein2014fast, goldfarb2013fast, patrinos2014douglas} \\
	\hline
	\multirow{2}{*}{\makecell[l]{~\\~\\ Generalized ADMM}}  & \makecell[l]{$f$ and $g$ convex. \\Existence of saddle points. }  & Gauss-Seidel  & \makecell[l]{Global convergence. \\ Global optima. \\ Convergence rate $\mathcal{O}(1/k)$ (convex). \\ Linear convergence (strongly convex).}  & \makecell[l]{Easy subproblems. \\ Convex. } & \makecell[l]{\cite{deng2016global, he20121}} \\
	\cline{2-6}
	& \makecell[l]{ $\B$ has full row rank. \\ $g$ $\Lips$ differentiable.  }  & Gauss-Seidel  & \makecell[l]{Global convergence. \\ Global optima. \\ Linear convergence.}  & \makecell[l]{Easy subproblems \\ Nonconvex} & \makecell[l]{\cite{li2015global}} \\
	\hline
	
	\multirow{2}{*}{\makecell[l]{~\\ ~\\ Linearized  ADMM}} & \makecell[l]{$f$ and $g$ convex \\ $f$ and $g$ $\Lips$ differentiable. }    & Gauss-Seidel  & \makecell[l]{Global convergence. \\ Global optima. \\Convergence rate $\mathcal{O}(1/k)$. }  &  \makecell[l]{Low per-iteration complexity. \\ Convex. }  & \makecell[l]{\cite{gao2018information, lin2017extragradient}} \\
	\cline{2-6}
	& \makecell[l]{ $f$ and $g$  $\Lips$ differentiable. \\
		${\rm Im}(\A) \subseteq {\rm Im}(\B)$. \\ $\B$ full column rank. }    & Gauss-Seidel  & \makecell[l]{ Stationary. \\ Linear convergence. }  &  \makecell[l]{Low per-iteration complexity. \\ Nonconvex.}  & \makecell[l]{\cite{lu2021linearized, liu2019linearized}} \\
	\hline
	\makecell[l]{Stochastic ADMM}  & \makecell[l]{$f$ and $g$ convex.\\ $f$ and $g$ $\Lips$ differentiable. }  & Gauss-Seidel  & \makecell[l]{ Global convergence. \\ Global optima  under expectation. \\ Convergence rate $\mathcal{O}(1/\sqrt{k})$ (convex). \\ 
		Convergence rate $\mathcal{O}(\log k / k)$ \\ (strongly convex). } & \makecell[l]{Incomplete and \\ inaccurate information. \\ Convex.}&  \makecell[l]{\cite{gao2018information, ouyang2013stochastic, ouyang2012stochastic}} \\
	\hline		
\end{tabular}
\end{table*}

\emph{Summary:} In this part, we reviewed ADMM variants for solving the linearly constrained two-block  problem \eqref{pp:p1}.  In addition to classic ADMM that has been recognized as a benchmark, a number of variants are now available either with distinguishing features or suitable for different situations.  We report the ADMM variants (including classical ADMM) in terms of  main assumptions, decomposition schemes (i.e., type), convergence properties, main features and references in TABLE \ref{tab:two-block_ADMM}. We have the following main conclusions. 
ADMM and its variants provide \emph{sublinear} convergence (i.e., $\mathcal{O}(1/k)$, $\mathcal{O}(1/k^2)$) for general convex optimization. For the special case where the objective functions are strongly convex and $\Lips$ differentiable, linear convergence can be achieved (see \emph{classical  ADMM}, \emph{generalized ADMM} and \emph{linearized ADMM}).  Some of the ADMM variants have been generalized to nonconvex optimization but require certain $\Lips$ differentiable properties of the objective functions  and rank conditions on the coefficient matrices $\A, \B$ (see \emph{symmetric ADMM}, \emph{generalized ADMM} and \emph{linearized ADMM}). These ADMM variants can be  celebrated by their distinguishing features, such as faster numerical convergence (\emph{symmetric ADMM}), enhanced convergence rate (\emph{fast ADMM}), easier subproblems (\emph{generalized ADMM}), low per-iteration complexity (\emph{linearized ADMM}) and compatible with inaccurate information (\emph{stochastic ADMM}). 

\subsection{Multi-block with separable objective}
Previously, we have focused on two-block optimization and assumed two computing agents to undertake the computation. It is more than often that we have a multi-agent system and the computation is expected to be distributed across  multiple agents.  This usually corresponds to a
 constrained multi-block optimization where the objective is the sum of objectives of individual agents. This class of problems takes the  general formulation of 
\begin{align*}
\label{pp:p2}	& \min_{(\x_i)_{i=1}^{n}}\sum_{i = 1}^n  f_i(\x_i) \tag{$\PP2$}\\
&{\rm s.t.}~  \sum_{i = 1}^n \A_i \x_i = \bb.  \\
& \quad \quad  \x_i \in \X_i, i \in  N. 
\end{align*}
where  $f_i: \R^{n_i} \rightarrow \R$   are given objective functions of the agents defined on their decision variables $\x_i  \in \R^{n_i}, i \in N$;   $\X_i  \in \R^{n_i}$ are local bounded convex constraints  imposed on  decision variables $\x_i, i \in N$;  $\A_i \in \R^{l \times n_i}$ and $\bb \in \R^{l}$ are coefficient matrices that encode the linear couplings  across the agents.   Problem \eqref{pp:p2} can be viewed as an extension of  \eqref{pp:p1}, allowing arbitrary number of decision blocks instead of only two. By defining $\x:=(\x_i)_{i=1}^n$ and $\A: = (\A_i)_{i=1}^n$, we have the  linear couplings  take the compact format  $\A \x = \bb$. In \eqref{pp:p2}, we explicitly indicate the local bounded convex constraints $\X_i$  to show that  the problem is  entirely  nonsmooth  w.r.t. each decision block. This is an important problem characteristic to be considered while designing an ADMM variant for distributed optimization.

\begin{figure}[h]
	\centering
	\includegraphics[width = 3.5 in]{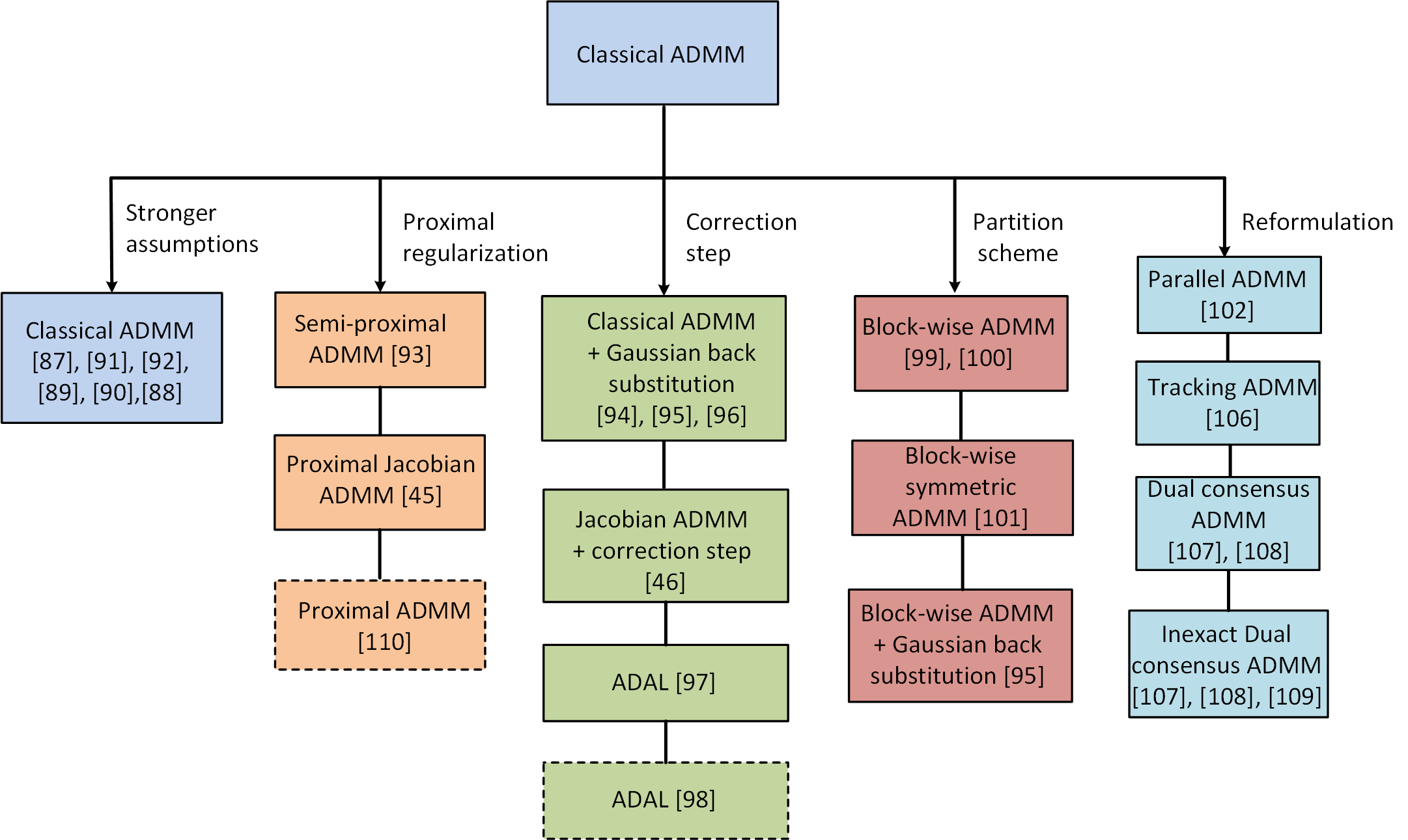}
	\caption{An overview of ADMM variants  resulting from the different modifications of classical ADMM for solving multi-block problem \eqref{pp:p2} (solid and dashed boxes  indicate convex and nonconvex methods respectively).}
	\label{fig:p2}
\end{figure}

In this part, we survey ADMM and its variants for solving problem \eqref{pp:p2} both in convex and nonconvex settings.  Since \eqref{pp:p1} is a special case of  \eqref{pp:p2} that involves only two decision blocks, the  methods of  this part  are readily  applicable to \eqref{pp:p1} provided that  corresponding conditions are satisfied.

Basically, the direct extension of classical ADMM to  multi-block problem \eqref{pp:p2} is not necessarily convergent   and some modifications are  required to ensure convergence \cite{chen2016direct}. In the literature, the modifications  are diverse and range from imposing stronger assumptions, adding proximal regularization,  adding some correction steps, leveraging partition schemes and  properly reformulating problems.  These  modifications lead to a plentiful of ADMM variants for solving multi-block problem \eqref{pp:p2}. An overview of the ADMM variants  resulting from the different modifications of classical ADMM  for solving multi-block problem \eqref{pp:p2}  is  shown in Fig. \ref{fig:p2}.  We distinguish the convex and nonconvex methods by solid and dashed boxes respectively.  
These ADMM variants are often preferred for different features, such as fast convergence, parallel implementatio, flexible communication (i.e., networked communication) and low per-iteration complexity etc. 
In what follows, we introduce  each of those methods.

\subsubsection{Classic ADMM} The direct application of \emph{classical ADMM} to multi-block problem \eqref{pp:p2} is  natural and takes the following iterative scheme. 
\begin{align*}
&	\textbf{Classic ADMM:} ~~~~~\\
& 	\textbf{Primal update:}~~ \x_i^{k+1} = \arg \min_{\x_i \in \X_i}    \Lag_{\rho}(\x_{<i}^{k + 1}, \x_i, \x_{>i}^k, \boldlambda^k), i \in N.  \\
& \textbf{Dual update:}~~~  \boldlambda^{k+1} = \boldlambda^k +  \rho(\A \x^{k+1}-\bb). 
\end{align*}

Despite the method has found many successful applications (see \cite{peng2012rasl, tao2011recovering} for examples),  the  convergence  is not secured for general  multi-block problem \eqref{pp:p2} (i.e., $n\geq 3$) in convex setting \cite{chen2016direct}.   Many efforts have been made in studying the  convergence conditions for the multi-block extensions \cite{lin2015sublinear, han2012note, chen2013convergence, cai2014direct}.  The results are diverse due to the different scenarios (i.e., different numbers of decision blocks) concerned and the different ways  used to draw  convergence. 
As an earlier work,  \cite{chen2013convergence}  focused on  the special case with $n=3$ blocks and established the convergence of the method under  the conditions:  \emph{i)}  $f_2, f_3$ are strongly convex,   and \emph{ii)} $f_1$ strongly convex or $\A_1$ has full column rank.  The results are  specialized to $n=3$ blocks and can not be directly generalized to arbitrary $n$  blocks.  Similarly for the 3-block case, \cite{cai2014direct} established the convergence  of the method under slightly different conditions: \emph{i)} $f_1,  f_2$ are convex and  $f_3$ is strongly convex, and \emph{ii)} $\A_2$ and $\A_3$ have full column rank. 
For general $n$-block case, \cite{han2012note} argued that  the convergence can be guaranteed provided that  all of the objective functions $f_i$ are  strongly convex.  Later,  the conditions were relaxed to $n-1$  strongly convex objective functions in   \cite{lin2014convergence, lin2015sublinear,lin2015global}.    Overall,  these  conditions  are  sufficient  instead of necessary conditions to guarantee the convergence of the method.  Presently, there is no consensus on the necessary  convergence conditions of \emph{classical ADMM} for the multi-block extension.

\subsubsection{Semi-proximal ADMM} While classic ADMM has focused on imposing stricter assumptions to ensure  convergence in multi-block setting, another line of works has  turned to modify the update scheme of classical ADMM. One example is   \textit{semi-proximal ADMM} that  proposes to   regularize  the subproblems by some proximal terms \cite{li2015convergent}.  The  method takes the following iterative procedures. 
\begin{align*}
&\textbf{Semi-proximal ADMM:}~~~~~\\
&\textbf{Primal update:}~  \\
&~~\x_i^{k+1} = \arg \min_{\x_i \in \X_i} \Lag_{\rho}(\x_{<i}^{k + 1}, \x_i, \x_{>i}^k, \boldlambda^k)
+ \frac{1}{2}\Vert \x_i - \x_i^k \Vert^2_{\PP_i}, i \in N. \\
& \textbf{Dual update:}~ \boldlambda^{k+1} = \boldlambda_k + \tau  \rho(\A \x^{k+1}-\bb) 
\end{align*}
where $\mathbf{P}_i \in \R^{n_i \times n_i}$ are positive semidefinite matrices.  Note that the major modifications  of \emph{semi-proximal ADMM} over   classical ADMM are the proximal terms $\frac{1}{2} \Vert \x_i - \x_i^k \Vert^2_{\PP_i}$ added to the subproblems.  The proximal terms were found to be able to enhance the convergence property of method and yield weaker convergence conditions than classical ADMM.   
Specifically for the $3$-block case,  it only requires  $f_2$ strongly convex  with sufficiently large proximal coefficient matrices $\PP_i$ \cite{li2015convergent}. 
In addition, it was proved that the method admits any nontrivial dual stepsizes $\tau \in (0, \frac{\sqrt{5} + 1}{2})$. 
However, the  results are limited to $n=3$ blocks and for general $n$-block case, the convergence conditions and theoretical convergence  remain to be addressed.

\subsubsection{Proximal Jacobian ADMM} Note that semi-proximal ADMM results from the combination of $\gauss$ decomposition with proximal regularization.  A natural idea is to consider the combination of  $\Jacobian$ decomposition and  proximal regularization. This has led to the \emph{proximal Jacobian ADMM} variant that takes the following iterative scheme \cite{deng2017parallel}. 
\begin{align*}
&\textbf{Proximal Jacobian ADMM:} ~~~~~\\
&\textbf{Primal update:}~  \\
&\quad \x_i^{k+1} = \arg \min_{\x_i \in \X_i} \Lag_{\rho}( \x_i, \x_{-i}^k, \boldlambda^k)
+ \frac{1}{2}\Vert \x_i - \x_i^k \Vert^2_{\PP_i}, i \in N. \\
& \textbf{Dual update:}~  \boldlambda^{k+1} = \boldlambda^k + \tau  \rho(\A \x^{k+1} - \bb) 
\end{align*}

The benefit of the $\Jacobian$ version is the parallelizable  implementation. 
Since $\Jacobian$ decomposition also only provides an approximation to the joint primal update, some proximal terms $\frac{1}{2}\Vert \x_i - \x_i^k \Vert^2_{\PP_i}$  are also required to control the $\Jacobian$ approximation accuracy. Clearly, the proximal regularization matrices $\PP_i$ should be selected sufficiently large. It was proved that we require $\PP_i > \rho(n-1)\A_i^\top \A_i$ (for the dual stepsize $\tau=1$)  to ensure  convergence of the method for general convex optimization \cite{deng2017parallel}. Under such condition, the $o(1/k)$ convergence rate of the method was further established. 
From the results, we note that the proximal coefficient matrices  are generally  required to be linearly increased  with the problem scale $n$.  Since the proximal terms play the role of  slowing down moving, slower convergence of the method is likely to be observed with larger-scale problems.

\subsubsection{Classic ADMM + Gaussian back substitution} Though the direct extension of classical ADMM to multi-block problem \eqref{pp:p2} is not necessarily convergent, it was found that a convergent sequence can be constructed by properly twisting the generated sequences \cite{he2012alternating}.   This leads to the  idea of using classical  ADMM to generate a  sequence as a prediction and then using some correction steps to twist a convergent sequence.  Following such idea,  a number of prediction-correction ADMM variants have been developed.  One of such methods is the \emph{classical ADMM + Gaussian back substitution} proposed in \cite{he2012alternating, fu2019block, he2012convergence}. The method takes the following iterative scheme.
\begin{align*}
&\textbf{Classic ADMM + Gaussian back substitution:} ~~~~~\\
&\textbf{Primal update:}~\tilde{\x}_i^{k} = \arg \min_{\x_i \in \X_i} \Lag_{\rho}(\tilde{\x}_{<i}^{k},  \x_i, \x_{>i}^k, \boldlambda^k), \forall i \in N. \\
& \textbf{Dual update:}~  \tilde{\boldlambda}^{k} = \boldlambda^k +   \rho(\A \tilde{\x}^k - \bb). \\
& \textbf{Gaussian back substitution}: \PP(\vbold^{k+1}- \vbold^k) = \alpha(\tilde{\vbold}^k - \vbold^k)\\
\end{align*}
where we have $\PP = \mathbf{H}^{-1}\mathbf{M}^\top$ with
\begin{align*}
& \mathbf{M} = \begin{pmatrix}
	\rho \A_2^\top \A_2 & \bm{0} & \cdots & \cdots & \bm{0} \\
	\rho \A_3^\top \A_2 & \rho \A_3^\top \A_3 & \cdots & \cdots & \bm{0}\\
	\vdots  & \vdots & \vdots & \vdots & \vdots \\
	\rho \A_n^\top \A_2 & \rho \A_n^\top \A_3 & \cdots & \rho \A_n^\top \A_n & \bm{0}\\
	\bm{0} & \bm{0} & \cdots & \bm{0} & \frac{1}{\rho} \mathbf{I}_l \\
\end{pmatrix} \\
& \mathbf{H} ={\rm diag}\big( \rho \A_2^\top \A_2, \rho \A_3^\top \A_3, \cdots, \rho \A_n^\top \A_n, \frac{1}{\rho} \mathbf{I}_l \big).
\end{align*}
where  $\vbold = (\x_2^\top, \x_3^\top, \cdots, \x_N^\top, \boldlambda^\top)^\top$ stacks the primal and dual variables excluding $\x_1$; the scalar $\alpha$ is a correction stepsize. The first block $\x_1$ is excluded from the \emph{Gaussian back substitution} (i.e., correction step) because $\x_1$ is an intermediate variable and  does not join next iterates.  
\emph{Classic ADMM + Gaussian back substitution} consists of two main steps: the first step uses classical ADMM to generate a prediction $\tilde{\vbold}^k$ and the second step uses \emph{Gaussian back substitution} to correct the generated sequence and obtain a modified sequence $\vbold^{k+1}$ for serving the next update. Specifically, the predictions step are performed in a forward manner, i.e., $\tilde{\x}_1^k \rightarrow \tilde{\x}_2^k \rightarrow \cdots \rightarrow \tilde{\x}_n^k \rightarrow \tilde{\boldlambda}^k$ and the correction steps are carried out  in a backward fashion, i.e., $\boldlambda^{k+1} \rightarrow \x_n^{k+1} \rightarrow \x_{n-1}^{k+1} \rightarrow \cdots \rightarrow \x_2^{k+1}$.  
The former results from the $\gauss$ scheme and the latter is induced from  the  upper triangle property of  matrix $\PP$ which is easy to infer from  $\mathbf{H}$ and $\mathbf{M}$. The convergence and  $\mathcal{O}(1/k)$ convergence rate in both ergodic and noner-godic sense together with the admissible correction stepsizes $\alpha \in (0, 1)$ of the method were established  for convex optimization in \cite{he2012alternating, fu2019block, he2012convergence}. One may note that  the \emph{Gaussian back substitution} can be converted to $\vbold^{k+1} = \vbold^k - \alpha\mathbf{M}^{-\top} \mathbf{H}(\vbold^k-\tilde{\vbold}^k)$, this is not necessary considering the upper triangle  property of $\PP$, which can be directly exploited to enable easy computation and avoid calculating the inverse and transpose  matrix $\mathbf{M}^{-\top}$.

\subsubsection{Jacobian ADMM + correction step} \emph{Jacobian ADMM + correction step} is another typical prediction-correction  ADMM variant for solving multi-block  problem \eqref{pp:p2} \cite{he2015full}. The basic idea is to obtain a convergent sequence by twisting the sequence generated by a Jacobian ADMM. The method takes the following iterative procedures. 
\begin{align*}
&\textbf{Jacobian ADMM + correction step:} ~~~~~\\
& \textbf{Primal update:}~~  \tilde{\x}_i^k = \arg \min_{\x_i \in \mathcal{\X}_i} \Lag_{\rho}( \x_i, \x_{-i}^k, \boldlambda^k), \forall i \in N.\\
& \textbf{Dual update:}~~  \quad   \tilde{\boldlambda}^k = \boldlambda^k + \rho (\A \tilde{\x}^k - \bb)\\
& \textbf{Correction step:}~~  \w^{ k+ 1} = \w^k - \alpha (\w^k - \tilde{\w}^k) 
\end{align*}
where  $\w = (\x_1^\top, \x_2^\top, \cdots, \x_N^\top, \boldlambda^\top)^\top$ stacks the primal and dual variables; the scalar $\alpha$ is a correction stepsize.  Very similar to classical ADMM + Gaussian back substitution, this method is also composed of  the prediction and correction steps.  The major difference is that this method  uses a $\Jacobian$ ADMM to generate a prediction $\tilde{\w}^k$ instead of the $\gauss$ counterpart. 
One may note that this brings difference to the correction steps where the first block $\x_1$ is involved in contrast to classical ADMM + Gaussian back substitution.  This is because all primal and dual updates are required to proceed next iterates by a $\Jacobian$ ADMM. 
It was shown  that any  correction stepsizes  $\alpha = \gamma (1 - \sqrt{{n}/{(n + 1)} }$ with $\gamma \in (0, 2)$ are admissible by the method \cite{he2015full}. In addition, the worst-case iteration complexity $\mathcal{O}(1/k)$ or $\mathcal{O}(1/\epsilon)$ was established both in an ergodic and non-ergodic sense \cite{he2015full}.

\subsubsection{ADAL} Another ADMM variant that results from the  combination of  $\Jacobian$ ADMM with correction scheme is the accelerated distributed augmented method (ADAL) proposed in \cite{chatzipanagiotis2015augmented}. 
This method originated from Diagonal Quadratic Approximation (DQA) method which relies on a loop of  $\Jacobian$ decomposition to solve the  joint primal update accurately  at each iteration \cite{ruszczynski1995convergence}. To reduce the iteration complexity, \cite{chatzipanagiotis2015augmented} proposed to eliminate the loop and perform a single iterate instead.  This reduces to the $\Jacobian$ ADMM that we are familiar.  However, $\Jacobian$ ADMM can not ensure convergence even for two-block optimization due to the insufficient approximation accuracy as discussed. To ensure convergence in general multi-block setting, ADAL also relies on a correction step to  twist the generated sequence.  The major difference from the other prediction-correction ADMM variants is that the correction is only imposed on the primal updates. Specifically, \emph{ADAL} takes the following iterative scheme.  
{	\setlength{\abovedisplayskip}{3pt}
	\setlength{\belowdisplayskip}{3pt}
\begin{align*}
&\textbf{ADAL:} ~~~~~\\
&\textbf{Primal update:}~~~  \hat{\x}_i^{k} =  \arg \min_{\x_i \in \mathcal{\X}_i} \Lag_{\rho}( \x_i, \x_{-i}^k, \boldlambda^k), \forall i \in N.\\
& \textbf{Correction step:}~~    \x^{k+1} = \x^k + \tau(\hat{\x}^k - \x^k) \\
&\textbf{Dual update:}~\quad ~ \boldlambda^{k+1} = \boldlambda^k + \tau \rho (\A \x^{k+1} - \bb) 
\end{align*} }
where  $\tau$ denotes both the correction and dual stepsize. Different from the other prediction-correction  ADMM variants where the correction is imposed  on both  primal and dual sequences, ADAL only performs correction  on the primal sequence. 
For general convex problems (i.e., $f_i$ are all convex) with a correction stepsize  $\tau \in (0, q^{-1})$ ($q$ denotes the maximum degree of the network characterizing the couplings across the agents), \cite{chatzipanagiotis2015augmented} established the convergence of the method towards global optima. Further, this method was  extended to nonconvex counterpart (i.e., $f_i$ are nonconvex but continuously differentiable) in  \cite{chatzipanagiotis2017convergence}. By assuming the existence of second-order stationary points, local convergence of the method towards stationary points was proved. The notion of local convergence is that if the method starts with some points sufficiently close to some local optima, the convergence towards the local optima can be guaranteed.

\subsubsection{Block-wise ADMM}  For multi-block problem \eqref{pp:p2}, another natural solution  is to artificially split the  decision variables into two groups and then apply two-block  ADMM variants. This idea is natural and reasonable since the convergence behaviors of ADMM variants for two-block optimization  have been well studied.  Following this idea, a number of ADMM variants have been proposed. One of them is  \emph{block-wise ADMM} resulting from the combination of the splitting scheme and classical ADMM \cite{he2015block, he2015splitting}.  To present the method, we first define some notations. Suppose the decision variables  of \eqref{pp:p2} are split into two groups with $p$ and $q$ blocks respectively (i.e., $p+q=n$).  We indicate the decision variables in the first group by  $\x:=(\x_i)_{i=1}^p$ and the second group by  $\y:=(\y_j)_{j=1}^q$. Correspondingly, we denote the objective functions by $\{f_i\}_{i=1}^p$ and $\{f_j\}_{j=1}^q$, the coefficient matrices of linear constraints by $\A:=(\A_i)_{i=1}^p$ and $\B:=(\B_j)_{j=1}^q$.  We besides distinguish the local bounded convex constraints  by  $\{\X_i\}_{i=1}^p$ and $\{\Y_j\}_{j=1}^q$.  By adopting the above notations,  \emph{block-wise ADMM} takes the following iterative scheme. 

{\small 
\vspace{-7mm}
{	\setlength{\abovedisplayskip}{3pt}
	\setlength{\belowdisplayskip}{3pt}
\begin{align*}
&\textbf{Block-wise ADMM:} ~~~~~\\
&\textbf{Primal update:}~\\
&  \x_i^{k+1} \!=\! \arg \min_{\x_i \in \X_i} \Lag_{\rho}(\x_i, \x_{-i}^k, \y^k, \boldlambda^k)  \!+\!  \frac{\tau_{\x}}{2} \Vert \A_i(\x_i \!-\! \x_i^{k}) \Vert^2, i \in P. \\
&  \y_j^{k+1} \!=\! \arg \min_{\y_j \in \Y_i} \Lag_{\rho}(\x^{k+1}, \y_j, \y_{-j}^k, \boldlambda^k)  \!+\!  \frac{\tau_{\y}}{2} \Vert \B_j(\y_j \!-\! \y_j^{k}) \Vert^2, j \!\in \! Q. \\
&\textbf{Dual update:}~  \boldlambda^{k+1} = \boldlambda^k +\tau \rho  (\A \x^{k+1} + \B \y^{k+1}) 
\end{align*}} }
where $P:=\{1, 2, \cdots, p\}$ and $Q: = \{1, 2, \cdots, q\}$ indicate the two-group partition of the decision variables with problem \eqref{pp:p2}. 

\emph{Block-wise ADMM} can be understood as the result of combining $\gauss$ and $\Jacobian$ decomposition to approximate the multi-block  joint primal update of ALM.  Specifically, the primal update of the method shows a two-level modular structure:   
 the upper level uses $\gauss$ scheme to enable a serial update   of the two groups $\x:=(\x_i)_{i=1}^p$ and $\y:=(\y_j)_{j=1}^q$ and the lower level employs $\Jacobian$ decomposition  to enable   parallel  updates within each group. 
Since directly applying $\Jacobian$ decomposition to approximate a joint primal  update tends to disrupt the convergence due to the insufficient approximation accuracy, some proximal terms $\frac{\tau_{\x}}{2}\Vert \A_i(x_i - \x_i^k) \Vert^2$ and $\frac{\tau_{\y}}{2}\Vert \B_j(\y_j - \y_j^{k})\Vert^2$ are required in the lower level   to control the approximation accuracy. It has been proved that the proximal coefficients should be sufficiently large, i.e.,  $\tau_{\x} \geq (p - 1)\rho$ and $\tau_{\y} \geq (q - 1) \rho$ \cite{he2015block}. The convergence and  $\mathcal{O}(1/k)$ convergence rate  of the method in both ergodic and non-ergodic sense were  established  \cite{he2015block, he2015splitting}. Note that if the upper level also employs a $\Jacobian$ decomposition, the method reduces to the proximal Jacobian ADMM. Actually, we may view proximal Jacobian ADMM as a special case of \emph{block-wise ADMM} with a partition of $p = n$ and $q = 0$.  
The benefit of  \emph{block-wise ADMM} over proximal $\Jacobian$ ADMM  is the smaller proximal coefficients are required, which is expected to yield a faster convergence rate. Note that the proximal coefficient depends on the group size of partition. As a result,  proximal Jacobian ADMM  requires the proximal coefficients to be  larger than $(n-1)\rho$ (i.e., $\PP_i \geq (n-1)\rho\A_i^\top\A_i$),  whereas \emph{block-wise ADMM} only requires them to be larger than  $(p - 1)\rho$ and $(q - 1) \rho$ ($p + q = n$). 
Since the proximal terms will slow down moving,  \emph{block-wise ADMM} with smaller proximal coefficients is expected to yield a faster convergence.


\subsubsection{Block-wise symmetric ADMM} \emph{Block-wise symmetric ADMM} is another example of applying two-block ADMM to solving multi-block problem \eqref{pp:p2} by  partition  \cite{bai2018generalized}. The method resembles block-wise ADMM with the only modification of using symmetric ADMM in place of classical ADMM. 
Using the same notations with block-wise ADMM, the implementation of \emph{block-wise symmetric ADMM} is presented below. 

{\small
{	\setlength{\abovedisplayskip}{3pt}
	\setlength{\belowdisplayskip}{3pt}
\begin{align*}
&\textbf{Block-wise symmetric ADMM:} ~~~~~\\
&\textbf{Primal update:}~ \\
&~\x_i^{k+1} \!=\!\arg \min_{\x_i \in \X_i} \Lag_{\rho}(\x_i, \x_{-i}^k, \y^k, \boldlambda^k)  +  \frac{\tau_{\x}}{2} \Vert \A_i(\x_i - \x_i^{k}) \Vert^2, i \in P. \\
& \textbf{Dual update:}~ \boldlambda^{\kmid} = \boldlambda^k +  r\rho (\A \x^{k+1} + \B \y^{k}) \\
&~\y_j^{k+1} \!\!=\! \arg \min_{\y_j \in \Y_i} \Lag_{\rho}(\x^{k+1}, \y_j, \y_{-j}^k, \boldlambda^{\kmid}) \!+\! \frac{\tau_{\y}}{2} \Vert \B_j(\y_j \!-\! \y_j^k)\Vert^2, j \in Q. \\
&\textbf{Dual update:}~  \lambda^{k+1} = \boldlambda^k +  s\rho (\A \x^{k+1} + \B \y^{k+1})  
\end{align*}} }

Note that the major difference of \emph{block-wise symmetric ADMM} over block-wise ADMM is that  the Lagrangian multipliers are updated  twice at each iteration due to  symmetric primal-dual scheme.  
Like symmetric ADMM, \emph{block-wise symmetric ADMM} allows to impose larger dual stepsizes  to achieve faster convergence.   It has been proved that any dual stepsizes $(r, s) \in \mathcal{D}=\{(r, s) \vert r + s >0,  r\leq 1, -r^2 - s^2 - r s + r  + s + 1 > 0\}$ are admissible by the method  \cite{bai2018generalized}. The  $\mathcal{O}(1/k)$ convergence rate  of the method was further established for general multi-block convex optimization \eqref{pp:p2} provided with sufficiently large  proximal coefficients, i.e., $\tau_{\x} > (p-1) \rho$ and $\tau_{\y} > (q-1) \rho$. Note that the proximal coefficients  agree with that of block-wise ADMM for any given partition $(p, q)$.

\subsubsection{Block-wise ADMM + Gaussian back substitution} As discussed, the proximal coefficients of block-wise ADMM variants depend on the group sizes of partition.
Considering the convergence rate, we normally prefer smaller group sizes to yield a faster convergence rate. To this end, a natural solution  is to combine  multi-block ADMM variants with a multi-group partition scheme.
One method following such idea is the \emph{block-wise ADMM + Gaussian back substitution} proposed in  \cite{fu2019block}. 
To present this method, we first define some notations.  Suppose the decision variables are split into $M$ groups and  each group $m$ involves $n_m$ blocks (clearly we have $\sum_{m=1}^M n_m = n$).  We indicate the decision variables in  group $m$ by $\x_m: = (\x_{mj})_{j \in n_m}$ where $\x_{mj}$ denotes the $j$-th decision variable of group $m$.  
Correspondingly, the objective functions are denoted by $\{f_{mj}\}_{j=1}^{n_m}$ and the coefficient matrices are indicated by $\A_m: = (\A_{mj})_{j=1}^{n_m}$, the local bounded convex constraints are represented by $\{\X_{mj}\}_{j=1}^{n_m}$.  By adopting the above notations, the implementation of \emph{block-wise ADMM + Gaussian back substitution} can be written as below. 
{	\setlength{\abovedisplayskip}{3pt}
	\setlength{\belowdisplayskip}{3pt}
\begin{align*}
&\textbf{Block-wise ADMM + Gaussian back substitution:} ~~~~~\\
&\textbf{Primal update:}~  \\
&\quad \tilde{\x}_{mj}^{k} = \arg \min_{\x_{mj} \in \X_{mj}} \Lag_{\rho}(\tilde{\x}_{<m}^{k},  \underbrace{\tilde{\x}_{<mj}^k, \x_{mj}, \x_{>mj}^k}_{\x_m}	,  \x_{>m}^k, \boldlambda^k),  \\
& \quad \quad \quad \quad \quad \quad  + \frac{\tau_{m}}{2}\Vert \A_{mj}( \x_{mj} - \x_{mj}^k) \Vert^2, \forall  j \in N_m, m \in M. \\
& \textbf{Dual update:}~  \tilde{\boldlambda}^{k} = \boldlambda^k +   \rho(\A \tilde{\x}^{k} - \bb). \\
& \textbf{Gaussian back substitution}: \PP(\vbold^{k+1}- \vbold^k) = \alpha(\tilde{\vbold}^k - \vbold^k)\\
\end{align*}}

where we have $N_m:= \{1, 2, \cdots, n_m\}$;  $\vbold =(\x_2^\top, \x_3^\top, \cdots, \x_M^\top, \boldlambda^\top)^\top$  stack the  primal and dual variables excluding $\x_1$,  and $\mathbf{P} = \mathbf{H}^{-1}\mathbf{Q}^\top$ defined by
\begin{align*}
& \mathbf{Q} = \begin{pmatrix}
	\rho \A_2^\top \A_2 & \bm{0} & \cdots & \cdots & \bm{0} \\
	\rho \A_3^\top \A_2 & \rho \A_3^\top \A_3 & \cdots & \cdots & \bm{0}\\
	\vdots  & \vdots & \vdots & \vdots & \vdots \\
	\rho \A_M^\top \A_2 & \rho \A_M^\top \A_3 & \cdots & \rho \A_M^\top \A_M & \bm{0}\\
	\bm{0} & \bm{0} & \cdots & \bm{0} & \frac{1}{\rho} \mathbf{I}_l \\
\end{pmatrix} \\
& \mathbf{H} ={\rm diag}\big( \rho \mathcal{D}_2,  \rho\mathcal{D}_3, \cdots,  \rho\mathcal{D}_M, \frac{1}{\rho} \mathbf{I}_l \big). \\
& \mathcal{D}_m = (\tau_m/\rho + 1) {\rm diag}(\A_m^\top\A_m),  \tau_m \geq (n_m -1)\rho.
\end{align*}
We have ${\rm diag}(\A_m^\top\A_m)$ denote a diagonal matrix formed by the diagonal elements of matrix $\A_m^\top\A_m$.  Similar to block-wise ADMM,  \emph{block-wise ADMM + Gaussian back substitution} also relies on the $\gauss$ and  $\Jacobian$  decomposition to approximate the multi-block joint primal update via a two-level scheme. Specifically, the upper level updates the $M$ groups of decision variables sequentially by a $\gauss$ pass and the lower level enables  parallel and distributed updates  within each group by  $\Jacobian$ decomposition. The convergence  and the $\mathcal{O}(1/k)$  worst-case iteration complexity  in both  ergodic and non-ergodic sense of the method were established for general multi-block convex optimization \eqref{pp:p2} provided with sufficient large proximal coefficients, i.e., $\tau_m > (n_m-1)\rho, \forall m \in M$  \cite{fu2019block}. 

One benefit of the method over the other block-wise ADMM variants is that smaller proximal coefficients are required  due to the multi-group partition. As a result, the method is expected to yield a faster convergence than the two-group counterparts.




\subsubsection{Parallel ADMM} \emph{Parallel ADMM} has been long-established for  multi-block convex problem \eqref{pp:p2}  (see Ch3, pp. 250, \cite{bertsekas2015parallel}) and has found many successful applications  \cite{baroche2019exogenous, rivera2016distributed}. However,  the method was not much covered in recent reviews as expected. The key idea of \emph{parallel ADMM} is to split the local and global constraints to two decision copies and then apply classic ADMM to solve the resulting two-block optimization. Specifically, for the concerned problem \eqref{pp:p2}, we have the  equivalent two-block formulation
{	\setlength{\abovedisplayskip}{3pt}
	\setlength{\belowdisplayskip}{3pt}
\begin{align}
& \label{pp:p2-v1}\min_{\x:=(\x_i)_{i=1}^{n}, \y:=(\y_i)_{i=1}^{n}}\sum_{i = 1}^n f_i(\x_i) \tag{$\PP2-1$} \\
&{\rm s.t.}~  \A_i \x_i - \y_i = \bb_i, \forall i \in N. \notag\\
& \quad \sum_{i=1}^n \y_i = \bm{0}. \notag\\
& \quad ~\x_i \in \X_i, i \in N. \notag
\end{align}}
where $\y: =(\y_i)_{i \in N}$  are slack variables introduced as the mappings of $\x:=(\x_i)_{i = 1}^n$;  $\{\bb_i\}_{i = 1}^n$ is a partition of  constant $\bb$ with $\sum_{i=1}^n\bb_i = \bb$.    
By treating the collections $\x:=(\x_i)_{i=1}^{n}$ and $\y:=(\y_i)_{i=1}^{n}$ as two blocks, problem \eqref{pp:p2-v1} can be readily handled by classical ADMM and  some problem structures can be exploited  to enable easier implementation. Specifically, we have the  primal update of $\x:=(\x_i)_{i=1}^n$ naturally decomposable across the decision components $\x_i, i \in N$. 
Besides,  we have the joint primal updates of slack variables $\y:=(\y_i)_{i=1}^n$  admit closed-form solutions, i.e., $\y_i^{k + 1} = \A_i \x_i^{k+1} - \bb_i - \mathbf{d}^{k+1}, \forall i \in N$ with $\mathbf{d}^{k+1} = \frac{1}{n} (\sum_{i=1}^n \A_i \x_i^{k+1} - \bb)$ denoting the average violation of linear couplings yield by the updates of iteration $k$. 
By substituting the closed-form solutions of $\y$ to $\x$ subproblems, we have the succinct  implementation of  \emph{parallel ADMM} presented below. 
{	\setlength{\abovedisplayskip}{3pt}
	\setlength{\belowdisplayskip}{3pt}
\begin{align*}
&\textbf{Parallel ADMM}: ~~~~~\\
& \textbf{Primal update:}~  \\
& \x_i^{k+1} \!=\!\arg \!\min_{\x_i \in \X_i} f_i(\x_i) \!+\! \langle \boldlambda^k, \A_i \x_i \rangle \!+\! \frac{\rho}{2} \Vert \A_i \x_i \!-\! \A_i \x_i^k \!+\! \mathbf{d}^k \Vert^2, \forall i \in N. \\
& \textbf{Dual update:}~   \mathbf{d}^{k+1} = \frac{1}{n} (\sum_{i=1}^n \A_i \x_i^{k+1} -\bb)\\
&\quad \quad \quad \quad \quad \quad  \boldlambda^{k+1} = \boldlambda^k +  \rho \mathbf{d}^{k+1}\\
\end{align*} }

One distinguishing feature of  \emph{Parallel ADMM} is that it does not rely on any extra conditions or corrections to ensure convergence. It is actually a direct application of classical ADMM by a proper problem reformulation.  This is the major difference from the other multi-block ADMM variants discussed before. Considering that both the proximal regularization and correction steps are likely to slow down convergence,  \emph{parallel ADMM} is expected to  yield a faster convergence for applications. 
This has been empirically  verified in \cite{he2016proximal} which shows that \emph{parallel ADMM} can be viewed as  a special case of proximal Jacobian ADMM  with minimal proximal regularization, i.e., $\PP_i = (n-1)\ \rho \A_i^\top \A_i$. 
Another salient feature of \emph{parallel ADMM} is the full parallel primal update which allows the computing agents to behave simultaneously to achieve high computation efficiency.  In addition, the method favors privacy since the agents  only require to communicate with a central  coordinator regarding some composite  information, i.e., the average violation of linear couplings $\mathbf{d}^k$ and Lagrangian multipliers $\boldlambda^k$. Clearly,  the convergence and convergence rate of \emph{parallel ADMM} directly follows classical ADMM, which has been thoroughly studied.

\subsubsection{Tracking ADMM}  \emph{Tracking ADMM} is an advanced version of parallel ADMM with the capability of accounting for networked communication \cite{falsone2020tracking}. Networked communication means that there exist a network or a graph characterizing the interactions of  agents in distributed computation. 
Parallel ADMM admits a \emph{master-workers}  communication scheme in which  all computing agents communicate  with a central coordinator to achieve coordination. 
There exist cases that the system is fully decentralized and a central coordinator does not exist. In such context, the agents are constrained to communicate with their interconnected agents (often called \emph{neighbors}) and parallel ADMM is not applicable.  
To address such an issue,  \emph{tracking ADMM} was proposed by  combining parallel ADMM with  an averaging consensus mechanism.  The key idea is that each agent holds a local estimate of the composite system-wide information (i.e., $\mathbf{d}^k$ and $\boldlambda^k$ with parallel ADMM) and uses the local estimate to perform primal updates. To achieve coordination, the agents iteratively communicate with their neighbors to exchange their local estimates   via an averaging consensus mechanism.  The implementation of \emph{tracking ADMM} is presented below. 
\begin{align*}
&\textbf{Tracking ADMM:} ~~~~~\\
&\textbf{Averaging consensus mechanism:}~  
\begin{cases}
	& \!\!\!\!\bm{\delta}_i^k =  \sum_{j \in N_i} w_{ij} \mathbf{d}_j^k\\ 
	&\!\!\!\! \bm{\ell}_i^k = \sum_{j \in N_i} w_{ij} \boldlambda_{j}^k\\
\end{cases} \\
&\textbf{Primal update:}~ \\
& \x_i^{k+1} = \arg \min_{\x_i \in \X_i} f_i(\x_i) + \langle \bm{\ell}_i^k, \A_i \x_i \rangle + \frac{\rho}{2} \Vert \A_i \x_i - \A_i \x_i^k + \bm{\delta}_i^k \Vert^2 \\
&\textbf{Dual update:}~~   \mathbf{d}_i^{k+1} = \bm{\delta}_i^k + \A_i(\x_i^{k+1} - \x_i^k)\\
& \quad \quad \quad  \quad \quad   \quad \boldlambda_i^{k+1} = \bm{\ell}_i^k +  \rho \mathbf{d}_i^{k+1}, \quad \quad \forall i \in N.\\
\end{align*}
where $\mathbf{d}_i^k$ and $\boldlambda_i^k$ are the local estimates of $\mathbf{d}^k$ and $\boldlambda^k$  held by agent $i$; $N_i$ represents the set of neighbors of agent $i$ (including itself);   $\mathbf{W} = (w_{ij})_{i, j = 1}^n$ is a consensus matrix  defining  the averaging consensus mechanism for the agents (Row $i$ corresponds to agent $i$,  and $w_{ij}>0$ if agents $i, j$ are connected, otherwise $w_{ij}=0$). We have  $\mathbf{W}$ a symmetric and doubly stochastic matrix \footnote{The sum of each row and each column equals to 1}.  
At each iteration, the local estimates $\mathbf{d}_i^k$ and $\boldlambda_i^k$ held by agent $i$  are  updated based on the exchanged estimates  from  its neighbors. Specifically,  $w_{ij}$ is the weight characterizing  how agent $i$ values the information from agent $j$. Afterwards, the updated estimates  $\bm{\delta}_i^k$ and $\bm{\ell}_i^k$  are used for the  primal update of agent $i$. 
For convex multi-block \eqref{pp:p2}, the convergence of \emph{tracking ADMM} was  established  provided with positive semidefinite consensus matrix $\mathbf{W}$ and connected communication network \footnote{We claim a network is connected if there exists at least one path from one node to another} \cite{falsone2020tracking}.


\subsubsection{Dual consensus ADMM} \emph{Dual consensus ADMM} is another ADMM variant that explores the direct application of classical ADMM to the multi-block problem \eqref{pp:p2} by problem reformulation. 
The key idea  is to apply  classical ADMM to  solve the dual   of multi-block problem \eqref{pp:p2}  \cite{chang2014multi, chang2016proximal}. 
Specifically, by defining  Lagrangian multipliers $\boldlambda \in \R^l$ for the linear couplings, we have the Lagrangian function $\Lag(\x, \boldlambda) = \sum_{i=1}^n f_i(\x_i) - \boldlambda^\top (\sum_{i=1}^n \A_i \x_i - \bb)$ and the resulting dual problem 
\begin{align}
\label{eq:dual_problem}\min_{\boldlambda \in \R^l} \sum_{i=1}^n  f_i^{*}(\A_i^\top \boldlambda) - \mathbf{b}^\top \boldlambda
\end{align}
where $f_i^{*}(\y) = \sup_{\x_i} \{\y^\top \x - f_i(\x_i): \x_i \in \X_i\}$ is the Fenchel conjugate of $f_i$, which is convex for convex $f_i$ and bounded convex subset $\X_i$.   Note that the dual problem \eqref{eq:dual_problem} corresponds to  minimizing  the  sum of $n$ convex functions that are correlated through the commonly owned dual variables $\boldlambda \in \R^{l}$. This class of problems is  well known as consensus optimization and can be handled by classical ADMM after proper reformulation, i.e.,
\begin{align}
\label{eq:dual_problem2}	& \min_{\boldlambda:= (\boldlambda_i)_{i=1}^n, \mathbf{t}: = (t_{ij})_{i, j = 1}^n} \sum_{i=1}^n  \big( f_i^{*}(\A_i^\top \boldlambda_i) - \frac{\mathbf{b}^\top}{n} \boldlambda_i)  \tag{$\PP2-2$}\\
& \label{eq:6a} {\rm s.t.~} \boldlambda_i = t_{ij}, ~\forall  j \in N_i, i \in N. \tag{6a} \\
& \label{eq:6b} \quad~~ \boldlambda_j = t_{ij}, ~\forall j\in N_i, j \in N.  \tag{6b}
\end{align}
where  $\boldlambda_i \in \R^l$ denotes a local  copy of  Lagrangian multipliers held by agent $i$;  $N_i$ denotes the set of neighbors of agent $i$ (not including itself) and $\mathbf{t}:=(t_{ij})_{i, j = 1}^n$ are linking variables used to enforce the consistency of Lagrangian copies among the agents. 

By viewing decision variables $\boldlambda:=(\boldlambda_i)_{i = 1}^n$ and $\mathbf{t}: = (t_{ij})_{i, j = 1}^n$ as two blocks, problem \eqref{eq:dual_problem2} is a two-block convex optimization  and can be handled by classical ADMM. Particularly, a closed-form solution for the linking variables $\mathbf{t}: = (t_{ij})_{i,j = 1}^n$ can be derived. By substituting such closed-form solutions back to classical ADMM and switching the optimization from  dual space to the primal space based on  strong duality theorem, we have  the  implementation of \emph{dual consensus ADMM} below. 
\begin{align*}
&\textbf{Dual consensus ADMM:} ~~~~~\\
&\textbf{Primal update:}~  \\
& \x_i^{k+1} \!=\!\arg \min_{\x_i \in \X_i} f_i(\x_i) \!+\!\frac{1}{4   \rho \vert N_i\vert } \Big\Vert \A_i \x_i \!-\! \frac{\bb}{n} - \mathbf{p}_i^k + \rho \!\!\sum_{j \in N_i} (\boldlambda_i^k + \boldlambda_j^k ) \Big\Vert^2\\ 
&\boldlambda_i^{k+1} = \frac{1}{2 \rho \vert N_i \vert }( 
\A_i \x_i^k -\frac{\bb}{n}- \mathbf{p}_i^{k}) + \frac{1}{2 \vert N_i \vert} \sum_{j \in N_i}( \boldlambda_i^k +\boldlambda_j^k) \\
& \textbf{Dual update:}~   \mathbf{p}_i^{k+1} = \mathbf{p}_i^k + \rho \!\!\sum_{j \in N_i }(\boldlambda_i^k - \boldlambda_j^k), \forall i \in N.
\end{align*}
where  $\mathbf{p}_i^{k}$  can be interpreted as the Lagrangian multipliers associated with the linking constraints of agent $i$ (i.e., $\boldlambda_i^k = \boldlambda_j^k, \forall j \in N_i$).  
One  salient feature of \textit{dual consensus ADMM}  is the parallel implementation. In addition, the method can accommodate networked communication schemes.  The convergence of the method for convex optimization directly follows classical ADMM  since it is a direct application \cite{chang2014multi, chang2016proximal}.

\subsubsection{Inexact dual consensus ADMM} \emph{Inexact dual consensus ADMM} is an advanced version of dual consensus ADMM with
the advantage of low per-iteration complexity. 
Note  that  \textit{dual consensus ADMM} requires to solve the subproblems exactly at each iteration. This could be computationally expensive or not efficient when the objective functions $f_i$ are complex. To reduce per-iteration complexity,  \cite{chang2014multi, chang2016proximal} proposed to linearize the subproblems and perform some inexact updates at each iteration to reduce per-iteration complexity.  The implementation of the method is presented below. 
\begin{align*}
&\textbf{Inexact dual consensus ADMM:} ~~~~~\\
&\textbf{Primal update:}~   \x_i^{k+1} \!=\! \arg \min_{\x_i \in \X_i} \langle g_i(\x_i^k), \x_i - \x_i^k  \rangle + \frac{\tau_i}{2}\Vert \x_i - \x_i^k \Vert^2\\ 
&\quad \quad \quad \boldlambda_i^{k+1} = \frac{1}{2\rho n_i }(
\A_i \x_i^k -\frac{\bb}{n} - \mathbf{p}_i^{k}) + \frac{1}{2  n_i} \sum_{j \in N_i}( \boldlambda_i^k +\boldlambda_j^k)\\
&\textbf{Dual update:}~   \mathbf{p}_i^{k+1} = \mathbf{p}_i^k + \rho \!\!\! \sum_{j \in N_i }(\boldlambda_i^k - \boldlambda_j^k), \forall i \in N.
\end{align*}
where $g_i(\x_i^k) = \nabla f_i(\x_i^k) + \frac{1}{2\rho n_i} \A_i^\top (\A_i \x_i^k - \frac{\bb}{n} - \mathbf{p}_i^k + \rho \smallsum_{j \in N_i}(\boldlambda_i^k + \boldlambda_j^k))$ denotes the gradients of  the subproblems of dual consensus ADMM. 
It is easy to note that the primal update  reduces to some projected gradient descent iterates, i.e., $\x_i^{k+1} = \big[ \x_i^k - \tau^{-1} g(\x_i^k)\big]_{\X_i}, i \in N$,  which are easy and simple if the subsets $\X_i$ have easily computable projections. The method is guaranteed to converge when  $f_i$ are convex and  $\Lips$ differentiable  \cite{chang2014multi}. For the special case where \emph{i)} $f_i$ are strongly convex and $\Lips$ differentiable, and \emph{ii)} $\A_i$ have full column rank,  a  linear convergence rate can be  achieved \cite{chang2014multi}. 

As  certain extensions, the method was further generalized to handle local polyhedral constraints (i.e., $\mathbf{c}_i^\top \x_i \leq \mathbf{d}_i$) in an efficient manner in \cite{chang2016proximal},  and account for  random ON/OFF behaviors of agents in \cite{chang2015randomized}.  Particularly, the convergence and the $\mathcal{O}(1/k)$  convergence rate of \emph{inexact dual consensus ADMM} in an ergodic sense were established  when the computing agents are randomly activated with certain positive probability in the iterative process  \cite{chang2015randomized}.

\subsubsection{Proximal ADMM} The above ADMM variants are all for convex optimization except for  ADAL which has been generalized to a nonconvex counterpart  with local convergence guarantee. When problem \eqref{pp:p2} is nonconvex (i.e., $f_i$ are nonconvex), the other multi-block ADMM variants  are  not applicable in general.   To address a nonconvex multi-block \eqref{pp:p2},  \cite{yang2022proximal} proposed a \emph{proximal ADMM}  with global convergence guarantee. \emph{Proximal ADMM} takes the iterative scheme
\begin{align*}
&\textbf{Proximal ADMM:} ~~~~~\\
&\textbf{Primal update:}~  \\
& \quad \quad \x_i^{k+1} \!=\! \arg \min_{\x_i \in \X_i} \Lag_{\rho}(\x_i, \x_{-i}^k, \boldlambda^k) \!+\! \frac{1}{2}\Vert \x_i - \x_i^k \Vert^2_{\PP_i}, ~\forall i \in N. \\
&\textbf{Dual update:}~  \boldlambda^{k+1} = (1 - \tau) \boldlambda^k + \rho(\A \x^{k+1} - \bb)
\end{align*}

Note that \emph{proximal ADMM} resembles  proximal Jacobian ADMM in primal update where the $\Jacobian$ decomposition and proximal regularization are employed. 
The major difference lies in the  dual update where a discounted factor  $(1-\tau)$ is imposed. 
As argued in   \cite{yang2022proximal}, the discounted dual update scheme is critical to ensure  the  convergence of the method for solving nonconvex \eqref{pp:p2}.  Specifically, it ensures the boundness of Lagrangian multipliers and makes it possible to identify a proper sufficiently decreasing and lower bounded Lyapunov function, which is a general key step to establish  convergence of  ADMM and its variants in nonconvex setting. The  convergence of the method towards approximate stationary points was  established for  $\Lips$ differentiable objective functions $f_i$ \cite{yang2022proximal}. 


\emph{Summary:} This section reviews ADMM and its variants for solving the linearly constrained multi-block problem \eqref{pp:p2}.  We report the ADMM variants in terms of  main assumptions, decomposition schemes (i.e., Types), convergence properties, main features and references in TABLE \ref{tab:multi-block_ADMM}. 
We have the following main conclusions. For multi-block convex \eqref{pp:p2},  diverse ADMM variants are now available  and most of them  promise an $\mathcal{O}(1/k)$ convergence rare. However, it is not clear which one behaves the best. It may depends on  applications and requires further examinations.  For convex \eqref{pp:p2}, many ADMM variants are favorable to enable parallel computation (e.g., \emph{proximal Jacobian ADMM}, \emph{Jacobian ADMM + relaxation step},  \emph{ADAL}, \emph{parallel ADMM}, \emph{tracking ADMM}, \emph{dual-consensus ADMM} and \emph{inexact consensus ADMM}). Particularly, \emph{inexact dual consensus ADMM} has made an effort to reduce per-iteration complexity. \emph{Tracking ADMM}, \emph{dual-consensus ADMM}, and \emph{inexact dual consensus} are able to accommodate networked communication  schemes.  For nonconvex \eqref{pp:p2}, the existing results are quite limited.  \emph{ADAL} can work as a solution but only provides local convergence guarantee.   \emph{Proximal ADMM} is able to provide a global convergence guarantee but only ensures approximate stationary points.  It is nontrivial to develop an ADMM variant 
for solving  \eqref{pp:p2}   in nonconvex setting as discussed in  \cite{yang2022proximal}. 
This is attributed to the entirely nonsmooth structure of problem \eqref{pp:p2} caused by the local bounded convex constraints imposed on each decision block. Whereas for nonconvex optimization, we generally require a   smooth \emph{well-behaved}  last bock  to ensure  convergence of ADMM or its variants. This will be discussed in the subsequent section.

\begin{table*}[h]
\setlength\tabcolsep{2pt}
\renewcommand\arraystretch{1}
\centering
\caption{ADMM variants for solving multi-block problem \eqref{pp:p2}}
\label{tab:multi-block_ADMM}
\begin{tabular}{llllll}   
	\hline 
	\textbf{Methods}  & \textbf{Main assumptions}  & \textbf{Types} & \textbf{Convergence}  & \textbf{Features} &  \textbf{References} \\
	\hline
	\multirow{4}{*}{\makecell[l]{~\\~\\~\\~\\~\\Classic ADMM}}   & \makecell[l]{$n-1$ $f_i$ strongly convex. \\ Existence of saddle points.} & Gauss-Seidel &  \makecell[l]{Global convergence. \\ Global optima. \\ Convergence rate $\mathcal{O}(1/k)$.}  & \makecell[l]{Arbitrary $n$ blocks. \\ Convex.} & \cite{lin2014convergence, lin2015sublinear,lin2015global}\\
	\cline{2-6}
	& \makecell[l]{$n = 3$. \\ $f_1$ strongly convex \\ or $\A_1$ full column rank. \\
		$f_2$ and $f_3$ strongly convex. \\ Existence of saddle points.} & Gauss-Seidel &  \makecell[l]{Global  convergence. \\ Global optima.}  & \makecell[l]{ $n=3$ blocks. \\ Convex.}  & \cite{chen2013convergence}\\
	\cline{2-6}
	& \makecell[l]{$n = 3$. \\ $f_3$ strongly convex. \\
		$f_1$ and $f_2$ convex. \\ $\A_2$ and $\A_3$ full column rank. \\ Existence of saddle points. } & Gauss-Seidel &  \makecell[l]{Global  convergence. \\ Global optima. \\ Convergence rate $\mathcal{O}(1/k)$.}  &  \makecell[l]{ $n=3$ blocks. \\ Convex.} & \cite{cai2014direct}\\
	\cline{2-6}
	& \makecell[l]{~\\ $f_i$ all strongly convex. \\ Existence of saddle points. } & Gauss-Seidel &  \makecell[l]{Global  convergence. \\ Global optima. }  & \makecell[l]{Arbitrary $n$ blocks. \\ Convex} & \cite{han2012note}\\
	\hline
	Semi-proximal ADMM   & \makecell[l]{$n = 3$. \\ $f_1$ and $f_3$ convex. \\
		$f_2$ strongly convex. \\ Existence of saddle points.} & Gauss-Seidel &  \makecell[l]{Global  convergence. \\ Global optima. }  &   \makecell[l]{ $n=3$ blocks. \\ Convex.}   & \cite{li2015convergent}\\					
	\hline 
	\makecell[l]{Proximal Jacobian ADMM} &  \makecell[l]{$f_i$ convex. \\ Existence of saddle points.}  & Jacobian &  \makecell[l]{Global convergence. \\ Global optima. \\ Convergence rate $\mathcal{O}(1/k)$.} & \makecell[l]{ Parallel computation. \\ Convex.}  & \cite{deng2017parallel}\\
	\hline 
	\makecell[l]{Classic ADMM \\ + Gaussian back substitution} &  \makecell[l]{$f_i$ convex. \\ Existence of saddle points.}  & $\gauss$  &  \makecell[l]{Global convergence. \\ Global optima. \\ Convergence rate $\mathcal{O}(1/k)$.} & \makecell[l]{ Convex.}  & \cite{he2012alternating, fu2019block, he2012convergence}\\
	\hline
	\makecell[l]{Jacobian ADMM \\ + correction step} &  \makecell[l]{$f_i$ convex. \\ Existence of saddle points.}  & Jacobian &  \makecell[l]{Global convergence. \\Global optima. \\ Convergence rate $\mathcal{O}(1/k)$.} & \makecell[l]{ Parallel computation. \\ Convex.}  & \cite{he2015full}\\
	\hline
	\makecell[l]{Block-wise ADMM} &  \makecell[l]{$f_i$ convex. \\ Existence of saddle points.}  & \makecell[l]{Gauss-Seidel \\+ Jacobian} &  \makecell[l]{Global convergence. \\ Global optima. } & \makecell[l]{Convex. \\ Two-group partition.}   & \cite{he2015block, he2015splitting}\\
	\hline
	\makecell[l]{  Block-wise \\ Symmetric ADMM} &  \makecell[l]{$f_i$ convex. \\ Existence of saddle points.}  & \makecell[l]{Gauss-Seidel \\+ Jacobian} &  \makecell[l]{Global convergence. \\ Global optima.} & \makecell[l]{Convex. \\ Two-group partition.}  & \cite{bai2018generalized}\\
	\hline 
	\makecell[l]{Block-wise ADMM \\ + Gaussian back substitution} &  \makecell[l]{$f_i$ convex. \\ Existence of saddle points.}  & \makecell[l]{Gauss-Seidel \\+ Jacobian}  &  \makecell[l]{Global convergence. \\Global optima. \\ Convergence rate $\mathcal{O}(1/k)$.} & \makecell[l]{ Convex. \\ Multi-group partition.}  & \cite{fu2019block}\\
	\hline 
	\multirow{2}{*}{\makecell[l]{~\\  ADAL}}  &  \makecell[l]{$f_i$ convex. \\ Existence of saddle points.}  & Jacobian &  \makecell[l]{Global convergence. \\Global optima.} & \makecell[l]{Parallel computation. \\Convex.}   & \cite{chatzipanagiotis2015augmented}\\
	\cline{2-6} 
	&  \makecell[l]{$f_i$ nonconvex. \\
		$f_i$ continuously differentiable. \\
		Existence of saddle points.}  & Jacobian &  \makecell[l]{Local convergence. \\ Local optima.} &    \makecell[l]{Parallel computation. \\ Nonconvex.} &  \cite{chatzipanagiotis2017convergence}\\
	\hline
	\makecell[l]{Parallel 	ADMM}   & \makecell[l]{$f_i$ convex. \\ Existence of saddle points.} &  Jacobian &  \makecell[l]{Global convergence. \\ Global optima. \\Convergence rate $\mathcal{O}(1/k)$.}   & \makecell[l]{ Parallel computation. \\ Convex. } & \cite{bertsekas2015parallel}\\
	\hline
	\makecell[l]{Tracking ADMM}   & \makecell[l]{$f_i$ convex. \\ Existence of saddle points. \\ Connected  network.} &  Jacobian &  \makecell[l]{Global convergence. \\Global optima.}  & \makecell[l]{ Networked communication. \\ Parallel computation. \\ Convex.} &  \cite{falsone2020tracking}\\
	\hline
	\makecell[l]{Dual-consensus \\ ADMM}   & \makecell[l]{$f_i$ convex. \\ Existence of saddle points. \\ Connected  network. } &  Jacobian &  \makecell[l]{Global convergence. \\Global optima. \\Convergence rate $\mathcal{O}(1/k)$.}  & \makecell[l]{Networked communication. \\ Parallel computation. \\Convex.} & \cite{chang2014multi, chang2016proximal}\\
	\hline
	\makecell[l]{Inexact dual \\ consensus  ADMM}   & \makecell[l]{$f_i$ strongly convex 
		\\ and $\Lips$ differentiable.  \\ Existence of saddle points. \\ Connected  network.} &  Jacobian &  \makecell[l]{Global convergence. \\ Global optima.}  & \makecell[l]{ Networked communication. \\ Parallel computation. \\ Low per-iteration complexity. \\ Convex.} & \cite{chang2014multi, chang2016proximal, chang2015randomized}\\
	\hline
	\makecell[l]{Proximal ADMM}   & \makecell[l]{$f_i$ nonconvex. \\ $f_i$ $\Lips$ differentiable.} &  Jacobian &  \makecell[l]{Global convergence. \\ Approximate stationary points.}  & \makecell[l]{ Parallel computation. \\Nonconvex.}& \cite{yang2022proximal}\\
	\hline
\end{tabular}
\end{table*}

\subsection{Multi-block with coupled objective}
So far, we have focused on constrained optimization with an  objective that is separable across  the decision variables of agents. There exist cases that some coupled objective components exist.   This usually corresponds to a class of linearly constrained multi-block optimization with a coupled objective, which takes the general formulation of 
\begin{align}
\label{pp:p3}~& \min_{\x =(\x_i)_{i=1}^n, \y} \phi(\x, \y) \!=\! g(\x, \y) \!+\! \sum_{i=1}^n f_i(\x_i) + h(\y) \tag{$\PP3$}\\
&\quad \quad  {\rm s.t.}~ \sum_{i = 1}^n \A_i \x_i + \B \y = \bm{0}. \notag
\end{align}
where $f_i: \R^{n_i} \rightarrow \R$ and $h: \R^{d} \rightarrow \R$ are separable objective functions  related to decision variables $\x_i \in \R^{n_i}, i \in N$ and $\y \in \R^d$ respectively;  $g: \R^{\sum_i n_i + d}$ represents a coupled objective  component  that correlates multiple decision variables, such as $\x_i, i \in N$ and $\y$. Note that $f_i$ may be nonsmooth and a usual case is that some local constraints are involved in the form of indicator functions. 
In the formulation, we differentiate decision variable $\y$ from $\x_i, i \in N$ for $\y$ often plays a special role in the ADMM variants to be discussed (i.e., working as last block).  

By defining $\x: = (\x_i)_{i=1}^n$ and $\A:= (\A_i)_{i=1}^n$, the coupled linear constraints take the compact format  $\A \x + \B \y = \bm{0}$. For the special case that 
$\x$ degenerates into one block, problem \eqref{pp:p3} reduces to $\{\min_{\x, \y}  g(\x, \y) + f(\x) + g(\y), {\rm s.t.}~ \A \x + \B \y = \bm{0}. \}$, which is a two-block optimization with a coupled objective.  

\begin{figure}[h] 
	\centering
	\includegraphics[width = 3.5 in]{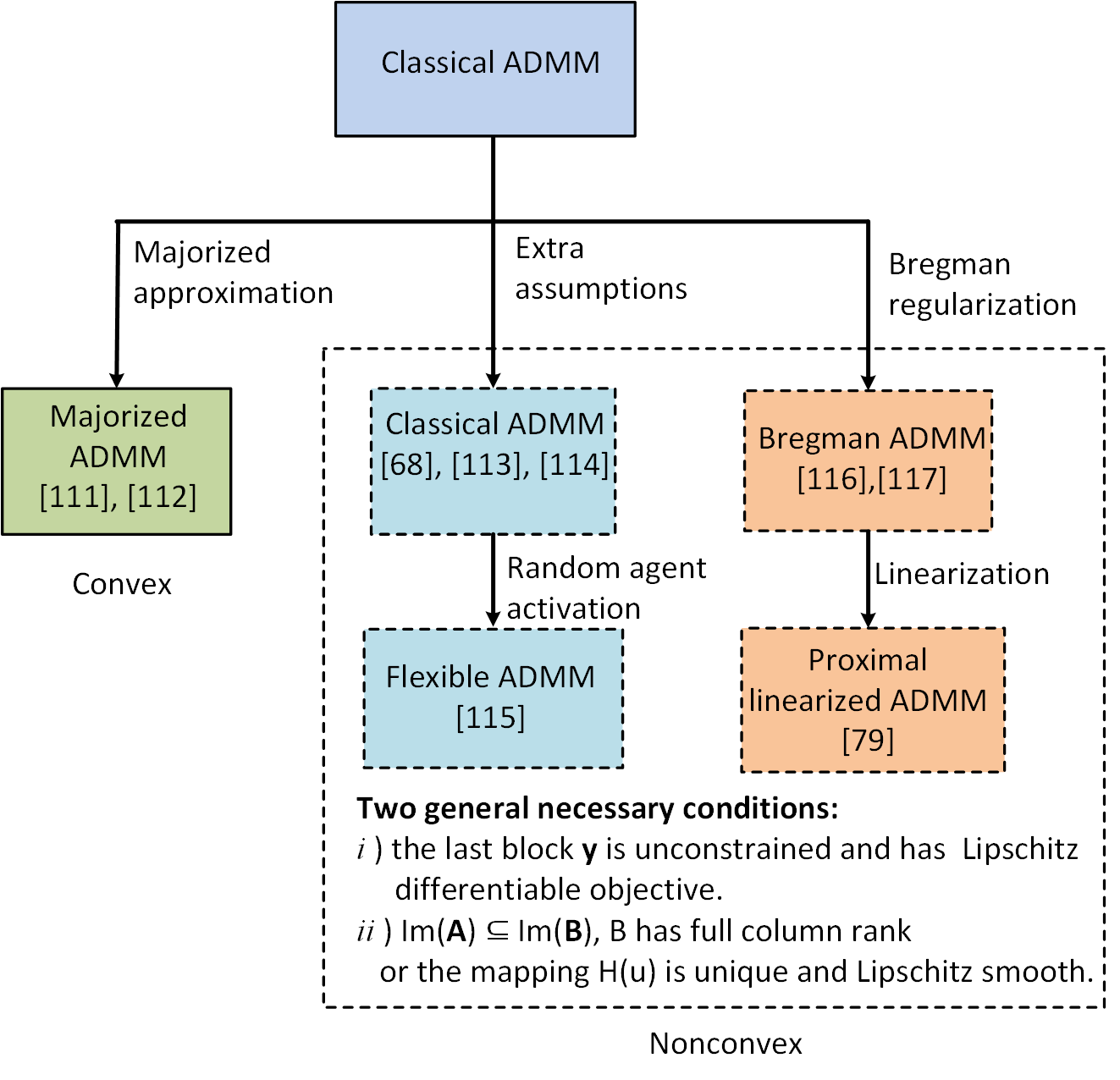}
	\caption{An overview of the relationships of ADMM variants for solving multi-block  coupled problem \eqref{pp:p3} (solid and dashed boxes indicate convex and nonconvex methods respectively).}
	\label{fig:p3}
\end{figure}

For problem \eqref{pp:p3}, a number of ADMM variants have been developed either for convex or nonconvex setting. These ADMM variants are generally developed from classical ADMM with the integration of  majorized approximation, extra assumptions or Bregman regularization.  
An overview of the relationships of the ADMM variants for solving the multi-block  coupled problem \eqref{pp:p3} is shown in Fig. \ref{fig:p3}. We distinguish the convex and nonconvex methods by solid and dashed boxes, respectively. The convex and nonconvex methods generally use different techniques to handle the coupled objective component for distributed computation. Specifically, convex methods generally rely on a proper majorized approximation,  which is separable across the decision  variables of agents. 
The  nonconvex counterparts usually use block coordinate methods or linearization to handle the coupled objective component. In terms of convergence, convex methods  generally require certain strong conveixty of subproblems.  In contrast, nonconvex methods generally   require a \emph{well-behaved} last block $\y$ satisfying the  two necessary conditions: \emph{i)} the last block $\y$ is unconstrained and has $\Lips$ differentiable objective, and \emph{ii)} ${\rm Im}(\A)\subseteq {\rm Im}(\B)$, $\B$ has full column rank or the mapping $H(\uu) = \{\arg \min_{\y} \phi(\x, \y), {\rm s.t.~} \B \y = \uu\}$ is unique and $\Lips$ smooth.  Notably, the ADMM variants both for convex and nonconvex optimization require  the coupled objective component $g$ to be $\Lips$ differentiable. 
In the following, we introduce each of those methods. 

\subsubsection{Majorized ADMM} For convex multi-block coupled problem \eqref{pp:p3} (i.e., $f$, $g$ and $h$ are convex),  \emph{majorized ADMM} has been the main solution  \cite{lu2017unified,cui2015convergence}. 
The key idea  is to optimize a majorized  surrogate of AL problem in a distributed manner. A majorized surrogate represents some approximation of a function from above.  For problem \eqref{pp:p3}, \emph{majorized ADMM} takes the following general iterative framework. 
{	\setlength{\abovedisplayskip}{3pt}
	\setlength{\belowdisplayskip}{3pt}
\begin{align*}
	&\textbf{Majorized ADMM:} ~~\\
	&\textbf{Primal update:}~~  \x_i^{k+1} = \arg \min_{\x_i} 
	\hat{\Lag}_{\rho}( \x_i, \x_{-i}^k, \y^k, \boldlambda^k), i \in N. \\ 
	& \quad \quad \quad \quad \quad \quad \quad  \y^{k+1} = \arg \min_{\y} 
	\hat{\Lag}_{\rho}(\x^{k}, \y, \boldlambda^k) \\ 
	&\textbf{Dual update:}~ \quad\boldlambda^{k+1} = \boldlambda^k + \rho (\A \x^{k+1} + \B \y^{k+1})
\end{align*} }
where $\hat{\phi}(\x, \y)$  represents a majorized surrogate or  majorized approximation of objective function $\phi(\x, \y)$.  A majorized surrogate means $\hat{\phi}(\x, \y) \geq \phi(\x, \y), \forall \x, \y$.  Correspondingly,  $\hat{\Lag}_{\rho}(\x, \y, \boldlambda) =\hat{\phi}(\x, \y) + \langle\boldlambda, \A \x + \B \y \rangle + \frac{\rho}{2}\Vert \A \x + \B \y \Vert^2$ can be viewed as a majorized surrogate of the AL function.    

 \emph{Majorized ADMM} is a conceptual algorithmic framework and its practical implementation requires to determine a proper majorized surrogate $\hat{\phi}(\x, \y)$. In terms of such issue,  some general instructions  have been identified \cite{lu2017unified}.   Specifically,   let   $\z: = (\x^\top, \y^\top)^\top$ be the stacked primal variables, we have $g(\z):= g(\x, \y)$ and $\phi(\z):=\phi(\x, \y)$.  A majorized surrogate $\hat{\phi}(\x, \y)$ for \emph{majorized ADMM} requires that
\begin{itemize}
\item[\emph{i)}] $\hat{\phi}(\z) \geq \phi(\z), \forall \z \in {\rm dom}~\phi$. 
\item[\emph{ii)}] there exists $\uu \in {\rm dom}~\phi$ and positive semidefinite matrix $\mathbf{L}$ such that $\vert \hat{\phi}(\z) - \phi(\z) \vert \leq \frac{1}{2}\Vert \z- \uu \Vert^2_{\mathbf{L}}, \forall \z \in {\rm dom}~\phi$.
\item[\emph{iii)}]  $\hat{\phi}(\z)$ is $\mathbf{P}$-strongly convex ($\PP$ is positive semidefinite),  which states that for any given  $\z \in {\rm dom}~\phi$,  we have $\phi(\z) - \frac{1}{2}\Vert \z - \vbold \Vert^2_{\mathbf{P}}$  convex. 
\end{itemize}
Some typical  majorized  surrogates $\hat{\phi(\z)}$ for \emph{majorized ADMM} include \cite{lu2017unified}
\begin{itemize}
\item \textbf{Proximal surrogate}:  $\hat{\phi}(\z) = \phi(\z) + \frac{1}{2}\Vert \z - \z^k\Vert_{\mathbf{L}}$ for any positive semidefinite matrix $\mathbf{L}$. 
\item \textbf{$\Lips$ surrogate}: $\hat{\phi}(\z) = \phi(\z^k) + \langle \nabla \phi(\z^k), \z - \z^k \rangle  + \frac{1}{2}L_{\phi}\Vert \z - \z^k \Vert^2$, where $\phi(\z)$ is  $L_{\phi}$-$\Lips$ differentiable.  
\item \textbf{Proximal gradient surrogate}: 
$\hat{\phi}(\z) = \sum_{i=1}^n f_i(\x_i) + h(\y) + \langle \nabla g(\z^k), \z - \z^k \rangle  + \frac{1}{2}L_{g}\Vert \z - \z^k \Vert^2$,  where $g(\z)$ is $L_g$-$\Lips$ differentiable.  
\end{itemize}
where $\z^k:=\big((\x^k)^\top, (\y^k)^\top \big)^\top$ is the generated primal updates at iteration $k$. These  majorized surrogates can be viewed as some local approximations of the coupled objective $\phi(\z)$ from above at each iteration. 

Note that  \emph{majorized ADMM} relies on a $\Jacobian$ decomposition to approximate the joint primal update ($\y$ and $\x_i$ play the same role in this method).  We usually prefer a majorized surrogate $\hat{\phi}(\z)$ that is separable across the decision blocks $\x_i, i \in N$ and $\y$,  which favors distributed computation. Besides, it was found that a tight majorized approximation indicated by smaller $\vert \hat{\phi}(\z) - \phi(\z) \vert$  favors convergence speed of the method \cite{lu2017unified}. For general convex multi-block  coupled problem \eqref{pp:p3}, the convergence and $\mathcal{O}(1/k)$ ergodic convergence rate of the method were established for $\Lips$ differentiable  objective $g$   \cite{lu2017unified}.  For the special two-block convex  case (i.e.,  $\x$ degenerates into one  block), \cite{cui2015convergence} established the similar theoretical convergence for the method.

\subsubsection{Classic ADMM} As a celebrated work, \cite{wang2019global} studied the direct extension of classical ADMM to multi-block coupled problem  \eqref{pp:p3}  in general nonconvex setting. The  application of \emph{classical ADMM} to problem \eqref{pp:p3} is standard and takes the following iterative scheme. 
\begin{align*}
&\textbf{Classic ADMM:} ~~\\
&\textbf{Primal update:}~   \x_i^{k+1} = \arg \min_{\x_i} 
\Lag_{\rho}(\x_{<i}^{k+1}, \x_i, \x_{>i}^k, \y^k, \boldlambda^k), \forall i \in N.\\ 
&\quad \quad  \quad \quad  \quad \quad\quad  \y^{k+1} = \arg \min_{\y} 
\Lag_{\rho}(\x^{k+1}, \y, \boldlambda^k)\\ 
&\textbf{Dual update:}~ \quad\boldlambda^{k+1} = \boldlambda^k +  \rho (\A \x^{k+1} + \B \y^{k+1})
\end{align*}

\emph{Classical ADMM} relies on a  $\gauss$ decomposition to split the joint primal update. For the  coupled objective component $g$, a $\gauss$ decomposition is equivalent to a block coordinate method.  The global convergence of \emph{classical ADMM} towards stationary points was established  under the conditions \cite{wang2019global}: 
\begin{itemize}
\item[\textbf{a)}] $f_i$ are restricted prox-regular.
\item[\textbf{b)}] $g$ and $h$ are $\Lips$ differentiable.
\item[\textbf{c)}]  $\A_i$ have full column rank or  the mappings $F_i(\uu) = \{\arg \min_{\x_i} \phi_i(\x_{<i}, \x_i, \x_{>i}, \y): \A_i \x_i = \uu\}$  are unique and $\Lips$ smooth. 
\item[\textbf{d)}] ${\rm \A}\subseteq {\rm \B}$, $\B$ has full column rank or the mapping $H(\uu) = \{\arg \min_{\y} \phi(\x, \y): \B \y = \uu\}$ is unique and $\Lips$ smooth.  
\end{itemize}

Restricted prox-regular functions were argued to be broad including  but  not limited to   $\Lips$ differentiable functions, weakly convex functions and  $\ell_q$ quasi-norms ($0 < q < 1$) \cite{wang2019global}. The assumptions related to  $F_i(\uu)$ and $H(\uu)$ were claimed as some weaker assumptions of  full column rank $\A_i$ and $\B$.   
To our best knowledge, \textbf{a)}-\textbf{d)} have been the  most general conditions  regarding the convergence of  \emph{classical ADMM} in  nonconvex setting. 

Note that no convexity is assumed for the problem. This implies that \emph{classical ADMM} is actually applicable to broad problems provided with conditions \textbf{a)}-\textbf{d)} regardless of convexity.  
This seems a contraction to our previous results where we concluded that the direct extension of \emph{classical ADMM} to multi-block optimization is not necessary convergent without any extra strong convexity assumptions (see Section IV-B). Actually, this is caused by the different  structures  of problem \eqref{pp:p3} from \eqref{pp:p2}. Specifically,  \eqref{pp:p3} has a special last block $\y$ that satisfies the two conditions: \emph{i)} the last block $\y$ is unconstrained and has $\Lips$ differentiable objective (i.e., $g$ and $h$ are $\Lips$ differentiable w.r.t. $\y$), and \emph{ii)} the coefficient matrices $\A, \B$ satisfy  ${\rm Im}(\A) \subseteq {\rm Im}(\B)$ and $\B$ has full column rank (or the mapping $H(\uu)$ is unique and $\Lips$ smooth).  We refer to such a last block as a \emph{well-behaved}  last block. This is in the sense that such a last block makes it possible to bound the dual updates $\Vert \boldlambda^{k+1} -\boldlambda^k\Vert^2$ by the primal updates $\Vert \x^{k+1} - \x^{k+1}\Vert^2$ and $\Vert \y^{k+1} - \y^k\Vert^2$ in the iterative process. This is critical to provide the sufficiently decreasing property of the AL function (viewed as a Lyapunov function) over the iterations, implying the convergent property of generated primal and dual sequences.

While  \cite{wang2019global} studied a general multi-block coupled problem \eqref{pp:p3} in nonconvex setting, some other  works considered some special cases. For example,  \cite{guo2017convergence} studied the convergence of \emph{classical ADMM} for a special two-block nonconvex optimization without any coupled objective, i.e., $n=1$, $g=0$, and $\B=\mathbf{I}$.
Besides, \cite{yang2017alternating} studied a special $3$-block nonconvex optimization  arising from image processing with $n=3$,  $g=0$ and $h$ strongly convex and quadratic. These works can be regarded as the special cases of \cite{wang2019global}  in terms of both problem structures and convergence conditions. 





\subsubsection{Flexible ADMM} \emph{Flexible ADMM}  can be viewed as a flexible version of classical ADMM  \cite{hong2016convergence}. The key idea is to activate the computing agents by a random mechanism. 
 This is often preferred when the computing agents have limited energy or suffer  certain random ON/OFF behaviors. The implementation of 
 \emph{flexible ADMM} is presented below.  
 
{	\setlength{\abovedisplayskip}{-10pt}
	\setlength{\belowdisplayskip}{3pt}
\begin{align*}
	&\textbf{Flexible ADMM:} ~~\\
	& \textbf{Primal update:}~ \\
	& \x_i^{k+1} \!\!=\!\! 
	\begin{cases}
		\arg \min_{\x_i}\!\Lag_{\rho}(\x_{<i}^{k+1}, \! \x_i, \! \x_{>i}^k, \!\y^k, &\!\!\!\!\!\boldlambda^k), \text{ if agent  $\x_i$ is activated.} \\
		\x_i^k,  &~~~\text{ if agent $\x_i$ not activated.}
	\end{cases}\\
	& \y^{k+1} \!\!=\!\! \begin{cases}
		\arg \min_{\y} \Lag_{\rho}(\x^{k+1}, \y, \boldlambda^k),  &\quad \quad  \text{ if agent $\y$ is activated.} \\
		\y^k, & \quad \quad  \text{ if agent  $\y$  not activated.} 
	\end{cases}\\
	&\textbf{Dual update:}~   \boldlambda^{k+1} = \boldlambda^k + \rho (\A \x^{k+1} + \B \y^{k+1})
\end{align*}}

 \emph{Flexible ADMM} employs a random agent activation mechanism in the primal update. As a result, a decision block will be updated only if the related computing agent is activated, otherwise the old update will be held and communicated to the others at each iteration.  Clearly, the convergence of \emph{flexible ADMM} depends on the random agent activation mechanism. For the two usual cases that the agents are  all  activated  with  positive probability   or  in an essentially cyclic manner  (i.e., activated at least once within $K$ iterations),   
 the convergence of \emph{flexible ADMM} towards stationary points has been  established for a class of problems with the following structures \cite{hong2016convergence}. 
\begin{itemize}
	\item[\textbf{a)}]  $f_i$ are convex or $\Lips$ differentiable.
	\item[\textbf{b)}]  $h$ is $\Lips$ differentiable and $g=0$.
	\item[\textbf{c)}] $\A_i$ have full column rank.
	\item[\textbf{d)}] $\B = \mathbf{I}$.
\end{itemize}
Note that the above conditions \textbf{a)}-\textbf{d)} are a special case of conditions \textbf{a)}-\textbf{d)} with classical ADMM.  To our understanding, the more general conditions  \textbf{a)}-\textbf{d)} with classical ADMM are readily extensible to the \emph{flexible ADMM} since the two methods share the same procedures  to establish convergence and the related conditions  play the same roles.

\subsubsection{Bregman ADMM} For multi-block coupled problem \eqref{pp:p3},  \emph{Bregman ADMM} is another  ADMM variant for nonconvex optimization  \cite{wang2018convergence}. Central to the method is to regularize the subproblems with classical ADMM by some Bregman divergence. Bregman divergence or Bregman distance is one metric for measuring the discrepancy of two points, which includes Eculidean distance as a special case.  Compared with classical ADMM,  \emph{Bregman ADMM} does not impose any restrictions on the  objective functions $f_i$ (i.e., restricted prox-regular) and the coefficient matrices $\A_i$, making the method  applicable to more general problems. \emph{Bregman ADMM} takes the following general framework. 
{	\setlength{\abovedisplayskip}{3pt}
	\setlength{\belowdisplayskip}{3pt}
\begin{align*}
&\textbf{Bregman ADMM:} ~~\\
&\textbf{Primal update:}~  \\
&~\x_i^{k+1} = \arg \min\limits_{\x_i}\Lag_{\rho}(\x_{<i}^{k+1}, \x_i, \x_{>i}^k, \y^k, \boldlambda^k) + \Delta_{\phi_i}(\x_i, \x_i^k), \forall i \in N.\\ 
&~\y^{k+1} = \arg \min\limits_{\y} \Lag_{\rho}(\x^{k+1}, \y, \boldlambda^k) + \Delta_{\psi}(\y, \y^k)\\
& \textbf{Dual update:}~   \boldlambda^{k+1} = \boldlambda^k + \rho (\A \x^{k+1} + \B \y^{k+1})
\end{align*}}
where $\Delta_{\phi_i}(\cdot, \cdot)$ and  $\Delta_{\psi}(\cdot, \cdot)$ are  Bregman divergence   defined on the  convex functions $\phi_i$ and $\psi$, respectively.  For a general  convex differentiable function $\phi:\R^n \rightarrow \R$,  the related  Bregman divergence is defined by 
{	\setlength{\abovedisplayskip}{3pt}
		\setlength{\belowdisplayskip}{3pt}
\begin{align*}
\Delta_{\phi}(\x, \y) = \phi(\x) - \phi(\y) - \langle \nabla \phi(\y), \x - \y\rangle, ~\forall \x, \y \in \R^n.  
\end{align*} }
Note that  Bregman distance reduces to Eculidean distance  $\Delta_{\phi}(\x, \y) = \Vert \x - \y\Vert^2$ with $\phi(\x) = \Vert \x \Vert^2$,  which has been widely used in proximal ADMM variants as discussed before. For  $\phi(\x) = \Vert \x \Vert^2_{\Q}$, we have the resulting Bregman distance $\Delta_{\phi}(\x, \y) = \Vert \x - \y \Vert^2_{\Q}$  which has been used in many generalized ADMM  variants.  From this perspective, we see that the \emph{Bregman ADMM} defines a  general  algorithmic  framework for solving problem \eqref{pp:p3}. The practical implementation requires to determine the specific Bregman divergence  $\Delta_{\phi_i}(\cdot, \cdot)$ and  $\Delta_{\psi}(\cdot, \cdot)$. 
The convergence of  \emph{Bregman ADMM} towards first-order stationary points has been established for a class of problems with the following structures \cite{wang2018convergence}.  
\begin{itemize}
\item[\textbf{a)}]  $g=0$.
\item[\textbf{b)}]  $h$, $\phi_i$, $\varphi$ are $\Lips$ differentiable.
\item[\textbf{c)}]  $f_i$ or $\phi_i$,  $h$ or $\varphi$ are strongly convex.
\item[\textbf{d)}]  $\B$ has full row rank .
\end{itemize}
Consider a usual  case that Eculidean distance is used as  Bregman divergence (i.e., $\phi_i(\x) = \Vert \x \Vert^2$ and $\varphi(\x) = \Vert \x \Vert^2$), we clearly have $\phi_i$ and $\varphi$ strongly convex and $\Lips$ differentiable. In such context, conditions  \textbf{a)}-\textbf{d)} reduce to $g=0$, $h$ $\Lips$ differentiable and $\B$ has full row rank.  Actually, this represents a  special case of conditions \textbf{b)}-\textbf{d)} with classical ADMM.  To our understanding,  the general conditions  \textbf{b)}-\textbf{d)} with classical ADMM can be directly extended to \emph{Bregman ADMM} if some strongly convex and $\Lips$ differentiable Bregman divergence functions  $\phi_i$ and $\varphi$ are selected. This is because the two methods follow the same procedures  to establish convergence and the related conditions play similar roles.   
One benefit resulting from Bregman divergence  is that the restricted prox-regular conditions on $f_i$ with classical ADMM and the rank assumptions on the coefficient matrices $\A_i$ can be relaxed, making the methods applicable to more general problems.

 The above results are for the general  \emph{Bregman ADMM}. Some works have studied  
 some special instances of \emph{Bregman ADMM}.  For example,  \cite{jiang2019structured} proposed two proximal ADMM variants where  the proximal terms $\frac{\tau}{2}\Vert \x_i - \x_i^k \Vert^2_{\PP_i}$ and $\frac{L_{\y}}{2}\Vert \y - \y^k\Vert^2$ are used as Bregman divergence. The convergence of the methods was proved under the conditions: \emph{i)} $g$ and $h$ are $\Lips$ differentiable, and \emph{ii)} $\B$ has full row rank, which are in line with \textbf{a)} - \textbf{d)}.

\subsubsection{Proximal linearized ADMM} Note that the above ADMM variants for  problem \eqref{pp:p3} require to solve the subproblems exactly at each iteration. This could be expensive or not efficient when the objective functions are complex and coupled.  To  reduce per-iteration complexity,  \cite{liu2019linearized} proposed a \textit{proximal linearized ADMM} as an inexact ADMM variant. The key idea is to linearize some complex (i.e., often hard to compute proximal mappings) but differentiable parts of the objective components and then solving the resulting easier proximal linearized subproblems.  \emph{Proximal linearized ADMM} takes the following iterative scheme. 
{	\setlength{\abovedisplayskip}{3pt}
	\setlength{\belowdisplayskip}{3pt}
\begin{align*}
	&\textbf{Proximal linearized ADMM:} ~~\\
	&\textbf{Primal update:}~ \\
	&  \x_i^{k+1} \!=\! \arg \min_{\x_i} 
	f_i(\x_i) \!+\! l_i(\x^k, \y^k, \boldlambda^k)^\top(\x_i \!-\!\x_i^k) \!+\! \frac{\tau_{\x}}{2}\Vert \x_i \!-\!\x_i^k \Vert^2, i \in N.\\ 
	&  \y^{k+1} \!=\! \arg \min_{\y} u(\x^{k + 1}, \y^k, \boldlambda^k)^\top(\y- \y^k) + \frac{\tau_{\y}}{2} \Vert \y - \y^k \Vert^2 \\
	& \textbf{Dual update:}~   \boldlambda^{k+1} = \boldlambda^k + \rho ( \A \x^{k+1} + \B \y^{k+1})
\end{align*}}
where  $l_i(\x^k, \y^k, \boldlambda^k)$ and $ u(\x^{k+1}, \y^k, \boldlambda^k)$
denote the gradients of the differentiable parts of AL function w.r.t. $\x_i$ and $\y$ respectively. Specifically,  $l_i(\x^k, \y^k, \boldlambda^k)$ is obtained by differentiating $g(\x, \y) + (\boldlambda^k)^\top \A \x + \frac{\rho}{2}\Vert\A \x + \B \y \Vert^2$ w.r.t. $\x_i$ at $(\x^k, \y^k)$,  and $u(\x^{k+1}, \y^k, \boldlambda^k)$ is obtained by differentiating $h(\y) + g(\x, \y) + (\boldlambda^k)^\top \B \y + \frac{\rho}{2}\Vert \A \x + \B \y \Vert^2$ w.r.t. $\y$ at $(\x^{k+1}, \y^k)$. Here we consider a general  $f_i$ that is probably nonsmooth. If $f_i$ is smooth and differentiable, it can be linearized  and involved in $l_i(\x^k, \y^k, \boldlambda^k)$ similarly.
Since a first-order linearization only provides an inaccurate approximation to the AL function, some proximal terms $\frac{\tau_{\x}}{2}\Vert \x_i - \x_i^k \Vert^2$ and $\frac{\tau_y}{2}\Vert \y - \y^k\Vert^2$ are required to control the approximation accuracy.

Slightly different from classical ADMM where \eqref{pp:p3} is considered as a multi-block optimization and $\gauss$ decomposition is used to enable a multi-block serial update, \emph{proximal linearized ADMM} was derived by  treating \eqref{pp:p3} as a two-block optimization and applying $\gauss$ decomposition to the two decision blocks  $\x:=(\x_i)_{i=1}^n$ and $\y$. Due to the linearization, we have subproblems  $\x$  naturally decomposable across the decision components $\x_i$. This brings the benefit of parallel implementation.   Note that  subproblems $\x_i$ in the primal update reduce to  the proximal mappings of $f_i$, i.e., $\x_i^{k+1} = \arg \min_{\x_i} f_i(\x_i) + \frac{\tau_x}{2}\Vert \x_i - (\x_i^k - \tau_{\x}^{-1} l_i(\x^k, \y^k, \boldlambda^k)) \Vert^2$, and subproblems $\y$ reduce to a gradient-like iterate, i.e., $\y^{k+1} = \y^k - \tau_{\y}^{-1} u(\x^{k+1}, \y^k, \boldlambda^k)$. If $f_i$ has easily computable proximal mappings, we would have  low per-iteration complexity with the method.  The convergence of \emph{proximal linearized ADMM} towards stationary points was proved under the conditions \cite{liu2019linearized}: 
\begin{itemize}
	\item[\textbf{a)}]$g$ and $h$ are $\Lips$ differentiable.  
	\item[\textbf{b)}] ${\rm Im}(\A) \subseteq {\rm Im}(\B)$, and $\B$ has full column rank.
\end{itemize}
Similar to other linearized ADMM variants discussed before, the \emph{linearization  technique} with the method also requires  the objective functions to be $\Lips$ differentiable.

One salient feature of \emph{proximal linearized ADMM} is that no restrictions on the objective functions $f_i$ and the coefficient matrices $\A_i$ are required  as with classical ADMM. 
This is caused by the proximal regularization  $\frac{\tau_{\x}}{2}\Vert \x - \x^k\Vert^2$ that enhances the convexity of subproblems.  As discussed with classical ADMM,
the full column rank  $\B$ can be replaced by the weaker assumption, i.e., the mapping $H(\uu) = \{\arg \min_{\y} \phi(\x, \y): \B \y = \uu\}$ is unique and $\Lips$ smooth. This holds for \emph{proximal linearized ADMM} since the two methods follow  the same procedures to establish convergence and the related assumptions play the same roles. 
This implies that \emph{proximal linearized ADMM} depends on weaker conditions than classical ADMM.  As a result, the method is applicable to more general applications, such as sparsity regression \cite{chen2014convergence} and integer programming \cite{liu2019linearized}. Due to the low per-iteration complexity and parallel implementation, \emph{proximal linearized ADMM} was shown  to be advantageous than  classical ADMM  in computation efficiency \cite{liu2019linearized}.

\begin{table*}[h] 
\setlength\tabcolsep{3pt}
\renewcommand\arraystretch{2}
\centering
\caption{ADMM variants for solving multi-block coupled problem \eqref{pp:p3}}
\label{tab:multi-block_coupled_ADMM}
\begin{tabular}{llllll}   
	\hline 
	\textbf{Methods}  & \textbf{Main assumptions}  & \textbf{Types} & \textbf{Convergence} & \textbf{Features} &   \textbf{References} \\
	\hline 
	\makecell[l]{Majorized ADMM}  &  \makecell[l]{ $f, g$ and $h$ convex. \\ $g$ $\Lips$ differentiable.}  &  \makecell[l]{$\Jacobian$ \\ or Gauss-Seidel}&  \makecell[l]{Global convergence. \\Global optima. \\Convergence rate $\mathcal{O}(1/k)$. } &  \makecell[l]{Convex.}& \cite{cui2015convergence, lu2017unified} \\
	\hline				
	\multirow{3}{*}{\makecell[l]{~\\ ~\\ ~\\ ~\\ Classic ADMM}}  &  \makecell[l]{$f_i$ restricted prox-regular. \\ $g$ and $h$  $\Lips$ differentiable.  \\$\A_i$ full column rank \\ or $F_i(\uu)$ unique and $\Lips$ smooth. \\ $\text{Im}(\A) \subseteq \text{Im}(\B)$. \\ $\B$ full column rank \\ or $H(\uu)$ unique and $\Lips$ smooth. }  &  Gauss-Seidel&  \makecell[l]{Global convergence. \\ Stationary points} & \makecell[l]{ Weak assumptions. \\ Nonconvex. }& \cite{wang2019global} \\
	\cline{2-6} 
	& \makecell[l]{$g =0, n =1$.\\ h $\Lips$ differentiable. \\$\A_1$ full column rank. \\ $\B = \mathbf{I}$ }  & Gauss-Seidel & \makecell[l]{Global convergence. \\ Stationary points} & \makecell[l]{2-block. \\ No restrictions on $f$. \\ Nonconvex. }  &  \cite{ guo2017convergence} \\
	\cline{2-6} 
	& \makecell[l]{$g =0, n =3$.\\ $h$ quadratic . \\ $f_1$ convex. \\ $f_2$ nonconvex and nonsmooth. \\ $\A_1$ and $\A_3$ full row rank.}  & Gauss-Seidel & \makecell[l]{Global convergence. \\ Stationary ponts} & \makecell[l]{3-block. \\ Nonconvex. }  &  \cite{yang2017alternating}  \\
		\hline 		
	Flexible ADMM  &  \makecell[l]{g = 0. $\B = \mathbf{I}$. \\ $f_i$ convex or $\Lips$ differentiable. \\$\A_i$ has full column rank.}  &  Gauss-Seidel&  \makecell[l]{Global convergence. \\Stationary points.} &   \makecell[l]{
		Random agent activation.\\Nonconvex. }& \cite{hong2016convergence} \\
	\hline
	\multirow{2}{*}{\makecell[l]{Bregman ADMM}}  &  \makecell[l]{g = 0. \\ $f_i$ or $\phi_i$, $h$ or $\psi$ strongly convex.\\
		$h$, $\phi_i$, $\psi$ $\Lips$ differentiable. \\
		$\B$ has full row rank.}  &  Gauss-Seidel&  \makecell[l]{Global convergence. \\ Stationary points.} & \makecell[l]{No restrictions on $f_i$ and $\A_i$. \\ Nonconvex. }& \cite{wang2018convergence} \\
	\cline{2-6}
	&  \makecell[l]{ $g$ and $h$  $\Lips$ differentiable. \\ 
		$\B = \mathbf{I}$}  &  Gauss-Seidel&  \makecell[l]{Global convergence. \\ Stationary points.} & \makecell[l]{No restrictions on $f_i$ and $\A_i$. \\ Nonconvex. }& \cite{jiang2019structured} \\
		\hline 
	\makecell[l]{Proximal \\linearized ADMM}  &  \makecell[l]{ $g$ and $h$  $\Lips$ differentiable. \\ $\text{Im}(\A) \subseteq \text{Im}(\B)$. \\ $\B$  full column rank. }  &  \makecell[l]{Gauss-Seidel \\+ Jacobian}&  \makecell[l]{Global convergence. \\ Stationary points. } &  \makecell[l]{No restrictions on $f_i$ and $\A_i$. \\ Low per-iteration complexity. \\ Nonconvex.}& \cite{liu2019linearized} \\
		\hline 			
\end{tabular}
\end{table*}

\emph{Summary:} This section surveys ADMM variants  for solving multi-block coupled problem \eqref{pp:p3}. Several ADMM variants have been developed either for   convex or nonconvex  optimization. We report those ADMM variants in terms of  main assumptions, decomposition schemes (i.e., type), convergence properties, main features and references in TABLE \ref{tab:multi-block_coupled_ADMM}. From the results, we draw the following overview. 
For convex optimization, \emph{majorized ADMM} is the main solution and guarantees the convergence towards global optima. For nonconvex optimization, \emph{classical ADMM}, \emph{flexible ADMM}, \emph{Bregman ADMM} and \emph{proximal linearized ADMM} are applicable and ensure the convergence towards stationary points. 
 The nonconvex ADMM variants can be distinguished by their different features. Specifically, \emph{Bregman ADMM} and \emph{proximal linearized ADMM} do not require any restrictions on the objective functions $f_i$ and the coefficient matrices $\A_i$ and thus are applicable to broader applications. \emph{Proximal linearized ADMM} is favorable for its low per-iteration complexity. \emph{Flexible ADMM} enables random agent activation mechanism which is beneficial when the computing agents have limited energy or suffer certain random ON/OFF behaviors. 
  Particularly, the ADMM variants for nonconvex optimization  generally depend on the two necessary conditions to ensure convergence: \emph{i)} the last block $\y$ is unconstrained and $\Lips$ differentiable, and \emph{ii)} ${\rm Im}(\A)\subseteq {\rm Im}(\B)$, $\B$ has full column rank or the mapping $H(\uu) = \{\arg \min_{\y} \phi(\x, \y): \B \y = \uu\}$ is unique and $\Lips$ smooth.

\vspace{-5mm}
\subsection{Consensus optimization}
Previously, we have focused on multi-block optimization where the  agents hold  disjoint decision variables. In practice, it is also common to see multi-agent systems where the agents share commonly-owned decision variables but hold private objectives.  The leads to the well-known consensus optimization that takes the general formulation of
{	\setlength{\abovedisplayskip}{3pt}
	\setlength{\belowdisplayskip}{3pt}
\begin{align}
\label{pp:p4} ~& \min_{\x}  F(\x) = \sum_{i = 1}^n f_i(\x) \tag{$\PP4$}
\end{align}}
where $f_i: \R^n \rightarrow \R$ denote the private objectives of the agents;  $\x \in \R^n$ is the commonly-owned decision variables;  $f_i$ may be nonsmooth and a usual case is that some local constraints related to the agents are included in the form of indicator functions. 

\begin{figure}[h]   
	\centering
	\includegraphics[width = 3.5 in]{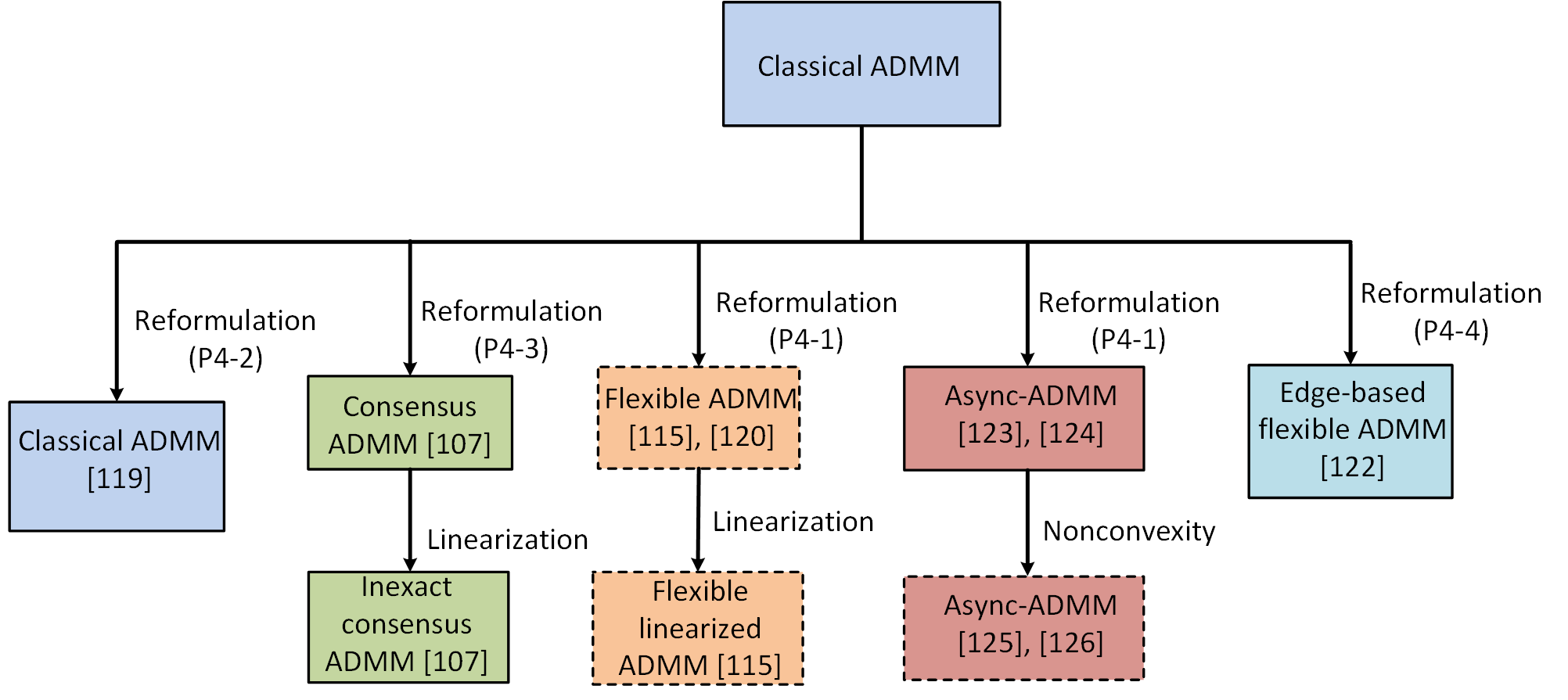}
	\caption{An overview of the relationships of the ADMM variants for solving consensus problem \eqref{pp:p4} (the solid and dashed boxes to indicate convex and nonconvex methods respectively).}
	\label{fig:p4}
\end{figure}

Note that consensus problem \eqref{pp:p4} involves only one block of decision variables that is shared by the agents. This does not correspond to the practice of ADMM or its variants which essentially explore the problem decomposition across decision blocks for distributed optimization.  To address such an issue,  ADMM variants for consensus optimization \eqref{pp:p4} are generally developed based on certain equivalent multi-block reformulations. 
There are four widely-used reformulations indicated by \eqref{pp:p4_v1} - \eqref{pp:p4_v3}.  
A number of ADMM variants have been developed for both convex and nonconvex optimization by applying classical ADMM to those reformulations. 
An overview of the relationships of the ADMM variants for solving consensus problem \eqref{pp:p4} is shown in Fig. \ref{fig:p4}. 
We distinguish the convex and nonconvex methods by solid and dashed boxes.  These methods are often preferred for different features, such as parallel implementation, networked communication schemes, low per-iteration complexity, random agent activation, and asynchronous computing (robust to communication delays or losses). 

In the sequel, we introduce each of those methods. Particularly, we first present the four commonly-used reformulations which are helpful to understand the main ideas of related ADMM variants.   The reformulations all depend on 
introducing local decision copies for individual agents and then enforcing  the consistency
 of the decision copies. Their major differences lie in the ways of enforcing  consistency constraints. 
Specifically for problem \eqref{pp:p4}, one natural and commonly-used  reformulation  is 
{	\setlength{\abovedisplayskip}{3pt}
	\setlength{\belowdisplayskip}{3pt}
\begin{align}
\label{pp:p4_v1} ~& \min_{\x:=(\x_i)_{i=0}^n}  F(\x) = \sum_{i = 1}^n f_i(\x_i)  \tag{$\PP4$-1}\\
&{\rm s.t.}~  \x_i = \x_0, ~ \forall i \in N. \notag
\end{align}}
where $\x_i \in \R^{n}$ represents the local decision copy held by agent $i$. To guarantee the consistency of  decision copies, a global decision copy  $\x_0$  is introduced, which is  managed by a central coordinator. By enforcing  the consistency  of all local  copies to the global one, i.e.,  $\x_i = \x_0, \forall i \in N$, we directly have the  equivalence of   \eqref{pp:p4_v1} and  \eqref{pp:p4}.

In reformulation \eqref{pp:p4_v1},  the agents are assumed to communicate via a central coordinator. There exist cases that a system is fully decentralized and a central coordinator  does not exist. In such context,  the agents are restricted to communicate over a network or a graph. We refer to that as \emph{networked communication}.  One direct reformulation that accounts for the networked communication is
{	\setlength{\abovedisplayskip}{3pt}
		\setlength{\belowdisplayskip}{3pt}
\begin{align}
	\label{pp:p4_v4} ~  &  \min_{\x:=(\x_i)_{i=1}^n} F(\x) =\sum_{i = 1}^n f_i(\x_i)   \tag{$\PP4$-2}\\
	& {\rm s.t.~} \x_i = \x_j, \forall (i, j) \in E, i< j.  \notag
\end{align} }
where $\x_i \in \R^n$ denotes the local decision copy held by agent $i$; $E$ indicates the edges of  communication network. To ensure the consistency of decision copies, a hard equality constraint $\x_i = \x_j$ is imposed for each pair of connected agents, i.e., $(i, j) \in E$. Since we only consider an undirected communication network,  one consistency constraint is sufficient for each pair of connected agents  and we therefore have $i < j$ in the reformulation.

When  the agents are restricted to communicate over a network or a graph,  another  reformulation is 
{	\setlength{\abovedisplayskip}{3pt}
	\setlength{\belowdisplayskip}{3pt}
\begin{align}
\label{pp:p4_v2} ~ & \min_{\x:=(\x_i)_{i=1}^n, \tbold:=(t_{ij})_{i, j  = 1}^n} ~ F(\x) = \sum_{i=1}^n f_i(\x_i)  \tag{$\PP4$-3}\\
& {\rm s.t.}~~ \x_i = t_{ij}, \quad \forall j \in N_i, i \in N. \notag\\
& \quad \quad \x_j = t_{ij}, \quad \forall j \in N_i, i \in N. \notag
\end{align} }
where $\x_i \in \R^n$ denotes the local decision copy held by agent $i$ and  $N_i$ denotes the set of its neighbors (not including itself). To ensure the consistency of decision copies across the agents, a linking variable  $t_{ij} \in \R^n$  (also known as slack variable) is introduced for each pair of connected agents (i.e., $j \in N_i$ and $i \in N_j$). It is easy to show  that \eqref{pp:p4_v2} is equivalent to \eqref{pp:p4} for any connected communication network. 

Another similar reformulation  accounting  for networked communication scheme is
{	\setlength{\abovedisplayskip}{3pt}
	\setlength{\belowdisplayskip}{3pt}
\begin{align}
\label{pp:p4_v3} ~  & \min_{\x:=(\x_i)_{i=1}^n, \z: =(\z_{ei}, \z_{ej})_{e \in E}} F(\x) = \sum_{i=1}^n f_i(\x_i)  \tag{$\PP4$-4}\\
&{\rm s.t.~} \x_i = \z_{ei}, ~\x_j = - \z_{ej},  \z_{ei} + \z_{ej}= \bm{0}, \forall e \in E.  \notag 
\end{align}}
where $\x_i \in \R^n$ denotes the local decision copy held by agent $i$; $E$ denotes the edges of the communication network, i.e., if  agent $i, j$ are neighbors, we have $(i, j) \in E$.  To ensure the consistency of decision copies held by the agents, two slack variable $\z_{ei} \in \R^n$ and $\z_{ej} \in \R^n$ are introduced for each edge $e \in E$ that connects agent $i, j$. Clearly, formulation \eqref{pp:p4_v3} is equivalent to \eqref{pp:p4} for any connected communication network.

Note that reformulations \eqref{pp:p4_v1}-\eqref{pp:p4_v3} all correspond to  multi-block linearly constrained optimization where the objective is the sum of private objectives held by the agents. The ADMM variants to be discussed  are exactly developed based on the above reformulations. Most of them are the results of  the direct application of classical ADMM to the reformulations. However, different problem structures  were fruitfully exploited to favor easier implementations.

\subsubsection{Classical ADMM} To account for networked communication scheme, \cite{wei2012distributed} studied the direction application of \emph{classical ADMM}  to  reformulation \eqref{pp:p4_v4}. Note that problem  \eqref{pp:p4_v4} can be treated as a multi-block optimization with decision variables $\x_i, i \in N$.  The application of \emph{classical ADMM} to solve multi-block \eqref{pp:p4_v4} is standard and  takes the following iterative scheme. 
{	\setlength{\abovedisplayskip}{3pt}
		\setlength{\belowdisplayskip}{3pt}
\begin{align*}
	&\textbf{Classical ADMM:} ~~\\
	&\textbf{Primal update:}~ \\
	&~~\x_i^{k + 1} = \arg \min_{\x_i} \Lag_{\rho}\big(\x_i, (\x_j^{k+1})_{j <i}, (\x_j^k)_{j >i}, (\boldlambda_{ji}^k)_{j < i}, (\boldlambda_{ij})^k_{i<j}\big), \\
	& \quad \quad \quad \quad \quad \quad \quad \quad \quad \forall i \in N. \\
	&  \textbf{Dual update:} \\
	&~~\lambda_{ij}^{k + 1} = \lambda_{ij}^k + \rho(\x_i^{k+1}  - \x_j^{k+1}), \forall (i, j) \in E, i < j. \\ 
\end{align*} }
 As discussed, the direct extension of \emph{classical ADMM} to multi-block optimization generally requires some  extra conditions (e.g., strongly convex)  to ensure convergence.  In line with the previous results,  \cite{wei2012distributed} established the convergence and   $\mathcal{O}(1/k)$  convergence rate of the method for a class of problems with  strongly convex objective functions $f_i$. Note that the method applies to other consensus problems \eqref{pp:p4} that satisfy the necessary conditions of  multi-block extension of \emph{classical ADMM} discussed in Section VI-B.

\subsubsection{Consensus ADMM} \emph{Consensus ADMM} results from the direct application of classical ADMM to reformulation \eqref{pp:p4_v2} \cite{chang2014multi}.  Note that problem \eqref{pp:p4_v2} corresponds to a two-block optimization by treating the collections $\x:=(\x_i)_{i=1}^n$ and $\tbold: = (t_{ij})_{i, j =1}^n$ as two decision  blocks. When the problem is convex, classical ADMM is directly applicable. For the first block $\x:=(\x_i)_{i=1}^n$, we have the subproblem naturally decomposable across the decision components $\x_i, i \in N$, and for the second block $\tbold: = (t_{ij})_{i, j  = 1}^n$, we have  quadratic subproblems that admit  closed-form solutions. 
By leveraging  the closed-form solutions of the slack variables $\tbold$, we have the succinct  implementation of \emph{consensus ADMM} below.
{	\setlength{\abovedisplayskip}{3pt}
	\setlength{\belowdisplayskip}{3pt} 
\begin{align*}
&\textbf{Consensus ADMM:} ~~\\
&\textbf{Primal update:}~ \\
&  \x_i^{k+1} = \min_{\x_i} f_i(\x_i) + (\boldlambda_i^{k})^\top \x_i + \rho \sum_{j \in N_i} \Big\Vert \x_i \!-\!\frac{\x_i^k - \x_j^k}{2}\Big\Vert^2, \forall i \in N. \\
& \textbf{Dual update:}~   \boldlambda_i^{k+1} = \boldlambda_i^k + \rho \sum_{j \in N_i} (\x_i^{k+1} - \x_j^{k+1}), \forall i \in N. 
\end{align*}}

One salient feature of \textit{consensus ADMM} is the parallelizable implementation. Specifically,  the agents are allowed to perform primal updates in a parallel  manner.  In addition, the method can accommodate a networked communication scheme. Since \emph{Consensus ADMM} is a direct application of classical ADMM to a two-block convex optimization, its convergence and convergence rate directly follow  classical ADMM.

\subsubsection{Inexact consensus ADMM} Consensus ADMM requires to solve the subproblems exactly at each iteration.  The could be expensive or not efficient when the objective functions $f_i$ are complex. To reduce per-iteration complexity,  \emph{inexact consensus ADMM} was proposed as an inexact version of consensus ADMM  \cite{chang2014multi}. The key idea to linearize the subproblems with consensus ADMM and perform an inexact update at each iteration. The method takes the following iterative scheme.
{	\setlength{\abovedisplayskip}{3pt}
	\setlength{\belowdisplayskip}{3pt} 
\begin{align*}
&\textbf{Inexact consensus ADMM:} ~~\\
&\textbf{Primal update:}~   \\
&\quad \x_i^{k+1} = \min_{\x_i} \langle \nabla f_i(\x_i^k),  \x_i - \x_i^k \rangle + (\boldlambda^{k}_i)^\top \x_i\\
&\quad \quad \quad \quad + \rho \sum_{j \in N_i} \Big\Vert \x_i - \frac{\x_i^k - \x_j^k}{2}\Big\Vert^2 + \frac{\tau_i}{2}\Vert \x_i - \x_i^k \Vert^2, \forall i \in N. \\
& \textbf{Dual update:}~  \boldlambda_i^{k+1} = \boldlambda_i^k + \rho \sum_{j \in N_i} (\x_i^{k+1} - \x_j^{k+1}), \forall i \in N. 
\end{align*}}

Note that objective function $f_i$ in the primal update is replaced by  $\langle \nabla f_i(\x_i^k),  \x_i - \x_i^k \rangle + \frac{\tau_i}{2}\Vert \x_i - \x_i^k\Vert^2$ which is a local first-order approximation of $f_i$ plus a proximal term (the constant $f_i(\x_i^k)$ is discarded). It is easy to note that the resulting subproblems of the primal update admit closed-form solutions and yield low per-iteration complexity. 
The convergence of \emph{inexact consensus ADMM} was proved  for a class of problems:  $f_i(\x_i) = \phi_i(\A_i \x_i)$ where $\phi_i(\x)$  are strongly convex and $\Lips$ differentiable ($\A_i$ can be rank deficient)  \cite{chang2014multi}. Particularly, some convex (possibly nonsmooth) objective functions $g_i(\x_i)$ are admissible and  the subproblems should be adapted to $\x_i^{k+1} = \min_{\x_i} g_i(\x_i) + \langle \nabla f_i(\x_i^k),  \x_i - \x_i^k \rangle + (\boldlambda^{k}_i)^\top \x_i + \rho \sum_{j \in N_i} \big\Vert \x_i - \frac{\x_i^k - \x_j^k}{2}\big\Vert^2 + \frac{\tau_i}{2}\Vert \x_i - \x_i^k \Vert^2$,  which are proximal linearized subproblems.



\subsubsection{Flexible ADMM} \emph{Flexible ADMM} results from the direction application of classical ADMM to reformulation \eqref{pp:p4_v1} together with the deployment of a random agent activation mechanism  \cite{hong2016convergence, wang2019distributed}. Note that \eqref{pp:p4_v1} is a two-block optimization if  $\x_0$ and  $\x: =(\x_i)_{i=1}^n$ are treated as two decision blocks. This corresponds to the problem structures of classical ADMM.  The random agent activation mechanism means that  the computing agents are activated in a random manner in the iterative process.  This is often preferred when the computing agents have limited energy or suffer certain random ON/OFF behaviors.  \emph{Flexible ADMM} takes the following iterative scheme. 
{	\setlength{\abovedisplayskip}{3pt}
	\setlength{\belowdisplayskip}{3pt}
\begin{align*}
&\textbf{Flexible ADMM:} ~~\\
& \textbf{Primal update:}~ \\
& \x_0^{k+1} \!\!=\! \begin{cases}
	\arg\min\limits_{\x_0} \Lag_{\rho}(\x_0, (\x_i^{k})_{i=1}^n, (\boldlambda_i^k)_{i=1}^n),  \!\!\!\!& \text{if agent $0$ activated} \\
	\x_0^k,  \!\!\!\!& \text{if agent $0$ not activated.} \\
\end{cases}\\
& \x_i^{k+1} \!=\! \begin{cases}
	\arg\min\limits_{\x_i} \Lag_{\rho}(\x_0^{k+1}, \x_i,\boldlambda_i^k), &\quad \quad~~ \text{if agent $i$ activated} \\
	\x_i^k,  & \quad \quad~~ \text{if agent $i$ not activated}
\end{cases}\\
&\textbf{Dual update:}~   \boldlambda_i^{k+1} = \boldlambda_i^k + \rho (\x_i^{k+1} - \x_0^{k+1}), i \in N. 
\end{align*}}

Note that the primal update of the second block $\x:=(\x_i)_{i=1}^n$ is performed in a distributed and parallel manner due to the naturally decomposable problem structure. 
Similar to other flexible ADMM variants, the convergence of the method depends on the random agent activation mechanism. For the two usual cases that  each agent is activated either with a positive probability at each iteration or in an essentially cyclic manner (i.e., each agent is activated at least once within $K$ iterations),  \cite{hong2016convergence, wang2019distributed} established the convergence of the method for  nonconvex optimization  with $\Lips$ differentiable objective functions $f_i$.   To be noted,  the method directly applies to the convex counterpart since it is a direct application of classical ADMM \cite{wang2016group}. Particularly, to improve communication efficiency,  \cite{wang2016group} proposed a group-based communication mechanism for the method where the computing nodes are grouped and the information of nodes are  aggregated within groups before communication. Moreover,  for convex optimization,  the two decision blocks $\x_0$  and  $\x: = (\x_i)_{i=1}^n$  are interchangeable  as the first and second block. This is different from the  nonconvex optimization where only $\x:=(\x_i)_{i=1}^n$ can work as the last block. This is because we required a \emph{well-behaved} last block  that satisfies certain $\Lips$ differentiable and rank properties to ensure convergence in nonconvex setting (see the discussions in Section IV-C). 

\subsubsection{Flexible linearized  ADMM}  As an advanced version, \textit{flexible linearized ADMM} was  proposed with the purpose of reducing per-iteration complexity with flexible ADMM  \cite{hong2016convergence}. The basic idea is to linearize smooth subproblems and perform an inexact update instead of solving the subproblems exactly at each iteration. \emph{Flexible linearized ADMM} takes the following iterative scheme. 
\begin{align*}
&\textbf{Linearized flexible ADMM:} ~~\\
&\textbf{Primal update:}~   \\
&\x_0^{k+1} \!=\! 
\begin{cases}
	\arg\min\limits_{\x_0} \Lag_{\rho}(\x_0, (\x_i^k)_{k=1}^n, \boldlambda^k), ~~\text{if agent $0$  activated.}& \\
	\x_0^k, \quad \quad \quad \quad \quad  \quad \quad \quad \quad \quad ~~~\text{if agent $0$  not activated.} &
\end{cases}\\
&\x_i^{k+1} = \begin{cases}
	\arg \min\limits_{\x_i}\langle \nabla f_i(\x_0^{k+1}), \x_i^k, \boldlambda_i^k), \x_i - \x_i^k \rangle  + \langle \boldlambda_i^k, \x_i\rangle  & \\
	\quad  + \frac{\tau_i}{2}\Vert \x_i - \x_i^k\Vert^2, \quad \quad \quad~~~~~ \text{if agent $i$ activated.} &  \\
	\x_i^k, \quad \quad \quad \quad \quad  \quad \quad \quad \quad \quad   ~\text{if agent $i$ not activated.} &
\end{cases}\\
&\textbf{Dual update:}~   \boldlambda_i^{k+1} = \boldlambda_i^k + \rho (\x_i^{k+1} - \x_0^{k+1}), \forall i \in N. 
\end{align*}

Note that the objective functions $f_i$ are linearized at $\x_i = \x_0^{k+1}$ and some proximal terms  $\frac{\tau_i}{2}\Vert \x_i - \x_i^k\Vert^2$ are added to the subproblems to control the  linear approximation accuracy. The convergence of \emph{flexible linearized ADMM} was established under the same conditions as with \emph{flexible ADMM}, i.e., the objective functions $f_i$ are $\Lips$ differentiable \cite{hong2016convergence}. Note that we can also choose to linearize $f_i$ at $\x_i = \x_i^k$ and the theoretical proof  can be adapted accordingly.

\subsubsection{Edge-based flexible ADMM} \emph{Edge-based flexible ADMM} is another ADMM variant that adopts a random agent activation mechanism  \cite{wei20131}. The major difference  from flexible ADMM and its linearized version  is that the agents are randomly activated by the edges of their communication network. Specifically at each iteration, a subset of the edges of the communication network are randomly selected  and  the  agents connected to those edges will be activated  to perform  new updates. This is often referred to \emph{random edge-based agent activation mechanism}. \emph{Edge-based flexible ADMM} can be regarded as the result of applying classical ADMM to the reformulation \eqref{pp:p4_v3} together with the deployment of the \emph{random edge-based agent activation mechanism}. Note that problem  \eqref{pp:p4_v3} corresponds to a two-block optimization by viewing the collections  $\x:=(\x_i)_{i=1}^n$ and $\z:=(\z_{ei}, \z_{ej})_{(i, j) \in e, e \in E}$ as two decision blocks.   \emph{Edge-based flexible ADMM} takes the general iterative framework 
{	\setlength{\abovedisplayskip}{3pt}
	\setlength{\belowdisplayskip}{-8pt}
\begin{align*}
& \textbf{Edge-based flexible ADMM:} ~~\\
& \text{if edge $e \in E$ is selected, the corresponding agents $(i, j)\in e$}\\ & \text{will perform the following primal and dual updates.} \\
& \textbf{Primal update:}~\\ 
& \begin{cases}
	\x_i^{k+1} = \arg \min_{\x_i} \Lag_{\rho}(\x_i, \z_{ei}^k, u_e^k) \\
	\x_j^{k+1} = \arg \min_{\x_j} \Lag_{\rho}(\x_j, \z_{ej}^k, v_e^k) \\
\end{cases} \\
& (\z_{ei}^{k+1}, \z_{ej}^{k+1})  = \begin{cases}
	\arg \min_{\z_{ei}, \z_{ej}} \Lag_{\rho}(\x_i^{k + 1}, \x_j^{k + 1}, \z_{ei}, \z_{ej},  u_e^k, v_e^k)\\ 
	{\rm s.t.~} \z_{ei} + \z_{ej} = 0. \\
\end{cases}\\
&\textbf{Dual update:}~  u_{e}^{k+1} = u_e^k + \rho(\x_i^{k+1} - \z_{ei}^{k+1}) \\
& \quad \quad \quad \quad \quad \quad v_{e}^{k+1} = v_e^k + \rho(\x_j^{k+1} + \z_{ej}^{k+1}) \\ 
\end{align*} }
where $u_{e} \in \R^n$ and $v_{e} \in \R^n$ are Lagrangian multipliers associated with the consistency constraints $\x_i = \z_{ei}, ~\x_j = - \z_{ej}$ in the reformulation \eqref{pp:p4_v3}.   

One may note that the subproblems related to  decision variables $\z_{ei}$ and $\z_{ej}$ are quadratic and admit close-form solutions: $\z_{ej}^{k+1} = -\frac{u_e^k + v_e^k}{2\rho} - \frac{1}{2}(\x_i^{k+1} + \x_j^{k+1})$ and $\z_{ei}^{k+1} = \frac{u_e^k + v_e^k}{2\rho} + \frac{1}{2}(\x_i^{k+1} + \x_j^{k+1})$, which can be easily derived based on KKT conditions. 
Similar to other flexible ADMM variants,  \emph{edge-based flexible ADMM}   enables flexible implementation and is often preferred  when the agents have limited energy. The convergence of the method also depends on the  \emph{random edge-based agent activation mechanism}. For the usual case where each edge is activated with a positive probability at each iteration, the convergence of the method towards global optima was established for convex optimization  \cite{wei20131}. Further, an $\mathcal{O}(1/k)$ convergence rate that characterizes the difference of objective gap and the feasibility violation under expectation was established.

\subsubsection{Async-ADMM}  Note that the above ADMM  variants are all implemented in a synchronous manner. The \emph{synchronization} is in two senses: \emph{i)} each agent  waits for the  information of its all interconnected agents to perform a new update \footnote{Flexible ADMM variants are slightly different due to the random agent activation mechanism.}, and  \emph{ii)}  each agent  uses the updated information of  its interconnected agents to perform a new update.   In practice, there often exist communication delays or losses caused by multiplied factors, such as limited network bandwidth, diverse processor configurations or inhomogeneous agent workloads, etc. In such context, the computing agents of a synchronous ADMM may have to wait a long time for the updated information of their interconnected agents over the network, thus disrupting or degrading the
the computation efficiency and scaling property of the method. 
To address such an issue,  a number of works have studied  the asynchronous implementation of  ADMM variants. In contrast, an \emph{asynchronous} implementation  means that \emph{i)} each agent only waits for the information of its partial interconnected agents to perform a new update, and \emph{ii)} each agent is allowed to use some outdated information of its interconnected agents to perform a new update.  As some typical examples, \cite{zhang2014asynchronous, li2020synchronous, chang2016asynchronous, chang2016asynchronous2} studied the asynchronous implementation of flexible ADMM. In flexible ADMM, the agents are distinguished  by \emph{one} \emph{master} and $n$ \emph{workers}.  The \emph{master} refers to the central coordinator  managing the global decision copy $\x_0$ and the \emph{workers} are  agents updating the local decision copies $\x_i, \forall i \in N$.  Note the communications of the method are restricted to the  \emph{master}  and  the  $n$ \emph{workers}.  An synchronous implementation defines that the  \emph{master} waits for  the  information of  all \emph{workers} to proceed a new update. Clearly, when the number of \emph{works}  is large, the \emph{master} may have to wait a long time due to the possible communication delays or losses. As a remedy, \cite{zhang2014asynchronous, li2020synchronous, chang2016asynchronous, chang2016asynchronous2} proposed to allow the \emph{master} to perform a new update only with  the information of  partial \emph{workers}. Moreover, the information is not necessarily updated due to the lack of global synchronized iteration counter.  This leads to the \emph{asynchronous ADMM} (\emph{async-ADMM}) that takes the following iterative scheme. 
\begin{align*}
	&\textbf{Async-ADMM:} ~~\\
	&\textbf{Primal update:}~ \\
	&\text{Master}:\begin{cases}
		& \x_i^k = \begin{cases}
			\hat{\x_i}, & \text{if} ~i \in \mathcal{A}_k  \\ 
			\x_i^k,  & \text{if} ~ i \in \mathcal{A}_k^c 
		\end{cases} \\
		&  \boldlambda_i^k  = \begin{cases} 
			\hat{\boldlambda}_i, & \text{if} ~i \in \mathcal{A}_k  \\
			\boldlambda_i^k,   & \text{if} ~ i \in \mathcal{A}_k^c\\
		\end{cases}\\
		& d_i = \begin{cases}
			0, & \text{if} ~i \in \mathcal{A}_k  \\ 
			d_i + 1,  & \text{if} ~ i \in \mathcal{A}_k^c 
		\end{cases} \\
		&\!\!\! \text{Wait until  $\vert \mathcal{A}_k\vert \geq S$ and $d_i \leq \tau -1$:} \\
		& \!\!\!\quad \x_0^{k+1} =\arg\min\limits_{\x_0} \Lag_{\rho}(\x_0, (\x_i^k)_{k=1}^N, \boldlambda^k),  \\
		& \!\!\!\text{Send $\x_0^{k+1}$ to the nodes $\mathcal{A}_k$.} \\ 
	\end{cases}\\
	&\text{Worker}~i: \begin{cases}
		&\!\!\!\text{Wait until $\hat{\x}_0 $ arrives}:\\
		&\!\!\! \x_i^{k_i+1} = \arg\min\limits_{\x_i \in \X_i} \Lag_{\rho}(\hat{\x}_0, \x_i,  \boldlambda_i^{k_i})\\
	\end{cases}, ~\forall i \in N. \\
	& \textbf{Dual update:}~  \boldlambda^{k_i+1}_i = \boldlambda_i^{k_1} + \rho (\x_i^{k_i+1} - \hat{\x}_0), ~\forall i \in N.
\end{align*}
where the decision variables with a hat (i.e., $\hat{\x}_i$, $\hat{\boldlambda}_i$ and $\hat{\x}_0$) denote the new information received  from related agents by the \emph{master}. Since the agents behave in an asynchronous manner,  each agent holds a private iteration counter.  We denote the iteration counter of \emph{master} and \emph{worker} by $k$ and $k_i, i \in N$, respectively. 
In \emph{async-ADMM},  the \emph{master} and \emph{workers} behave in a slightly different manner due to the communication scheme. 

For the \emph{master},   the \emph{workers} are distinguished by whether their new updates have arrived or not at each iteration $k$. Specifically,  $\mathcal{A}_k$ and $\mathcal{A}_k^c$  indicate  the sets of \emph{workers} whose  new updates have arrived or not, respectively. To ensure convergence, the \emph{master} will set a maximum tolerable delay  bound  $\tau$ for each individual \emph{worker}. This states that  at least one new update  must be served by each individual \emph{worker} within  $\tau$ successive iterations of the \emph{master}.  In other words, the information of the \emph{workers} used by the \emph{master} can be at most $\tau$ outdated. To enforce the maximum tolerable delay constraints, a delay counter $d_i$ is set for each individual \emph{worker}. 
At each iteration, if a new update  is  received from \emph{worker} $i$, the delay counter will be cleared (i.e., $d_i = 0$), otherwise it will be increased by $1$ (i.e., $d_i = d_i + 1$). 
In \emph{async-ADMM}, instead of waiting for  the new updates of all \emph{workers} to arrive, the \emph{master} will proceed a new update with the new information of only partial \emph{workers}. This is referred to  \emph{partial synchronization}. A  specific \emph{partial synchronization mechanism} $S$ states that the \emph{master} will proceed  a new iterate after receiving  the new updates from any  $S$ \emph{workers} and confirming that the remaining \emph{workers} are not to exceed  the maximum tolerable delay bound $\tau$. For the special case that certain remaining \emph{workers} are to exceed the bound $\tau$ ($d_i = \tau - 1$),   the \emph{master} will hold on and wait for  the   new updates of those \emph{workers}.  Note that when we have $S = n$ and $\tau = 1$, \emph{async-ADMM} reduces to synchronous ADMM.   In the algorithmic implementation, we use $\x_i^k$ and $\boldlambda_i^k$ to denote the latest updates of \emph{worker} $i$ hold by the \emph{master} at iteration $k$. To be noted, they may not be the latest update of the \emph{workers} due to the communication delays or losses. Note that if a new update $\hat{\x}_i$ is received from \emph{worker} $i$, $\x_i^k$ will be updated accordingly, i.e.,  $\x_i^k = \hat{\x}_i$ by the \emph{master}, otherwise the old updates will be hold to perform a new update.

For the \emph{worker}, the update scheme is slightly different.  Since each \emph{worker} only requires  information from the single \emph{master}, each \emph{worker} $i$ will wait  the new information of the \emph{mater} to perform a new update.

The convergence of \emph{async-ADMM} has been studied for both convex \cite{zhang2014asynchronous, li2020synchronous} and nonconvex  \cite{chang2016asynchronous, chang2016asynchronous2} optimization. 
One  common and important condition to ensure convergence  is that the communication delay is bounded which states that a new  update from each individual \emph{worker} must be served to  the \emph{master} with a limited number of iterations (i.e., $\tau$). For convex optimization, \cite{zhang2014asynchronous} established the worst-case iteration complexity  $\mathcal{O}(\frac{n\tau}{k S})$. We imply that the convergence rate of \emph{async-ADMM} closely relates to the problem scale $n$, the maximum tolerable  delay bound configuration $\tau$, and the partial synchronization mechanism $S$. There  usually exists  a trade-off between the iteration complexity (the number of iterations) and the waiting time of \emph{async-ADMM}. Specifically, a larger tolerable delay bound $\tau$ and smaller synchronization $S$ often lead to less waiting time but more iterations to converge and vice verse.  
For the special convex case where $f_i$ are strongly convex and $\Lips$ differentiable, a linear convergence of \emph{async-ADMM} was established in \cite{chang2016asynchronous}.

\subsubsection{Summary} This section reviewed ADMM variants for solving consensus problem  \eqref{pp:p4}. A number of ADMM variants have been developed either for convex or nonconvex settings.  We report those ADMM variants in terms of  main assumptions, decomposition schemes (i.e., type), convergence properties, main features and references in TABLE \ref{tab:multi-block_consensus}. We see that these ADMM variants can be distinguished by their features.  
 \emph{Consensus ADMM} and \emph{inexact consensus ADMM} enables full parallel  computation and can accounts for networked communication. \emph{Flexible ADMM}, \emph{flexible linearized ADMM} and \emph{edge-based flexible ADMM} allow a random agent activation mechanism to save the energy consumption of agents. \emph{Async-ADMM} enables asynchronous computing and is robust to communication delays or losses.   
 \emph{Inexact consensus ADMM} and \emph{flexible linearized ADMM}  are advantageous with low  per-iteration complexity.  Particularly,  we conclude that the ADMM variants for nonconvex consensus problem \eqref{pp:p4}  require the objective functions $f_i$ to be $\Lips$ differentiable. This is in line with our discussions on the necessary conditions of nonconvex optimization in Section IV-C.

\begin{table*}[h] 
\setlength\tabcolsep{3pt}
\renewcommand\arraystretch{2}
\centering
\caption{ADMM variants for solving consensus problem \eqref{pp:p4}}
\label{tab:multi-block_consensus}
\begin{tabular}{llllll}   
	\hline 
	\textbf{Methods}  & \textbf{Main assumptions}  & \textbf{Types} & \textbf{Convergence} & \textbf{Features} &   \textbf{References} \\
		\hline
	\makecell[l]{Classical ADMM}  &  \makecell[l]{$f_i$ strongly convex.}  &  Gauss-Seidel &  \makecell[l]{Global convergence. \\Global optima. \\ Convergence rate $\mathcal{O}(1/k)$.} &  \makecell[l]{Networked communication.  \\ Convex.} & \cite{wei2012distributed} \\	
	\hline				
	Consensus ADMM  &  \makecell[l]{$f_i$ convex. }  &  Gauss-Seidel&  \makecell[l]{Global convergence. \\ Global optima. \\} &   \makecell[l]{Parallel implementation. \\Networked communication. \\ Convex.} & \cite{chang2014multi} \\
	\hline					
	\makecell[l]{Inexact consensus \\ ADMM}  &  \makecell[l]{$f_i(\x) = \phi_i(\A_i \x_i) + g_i(\x_i)$\\
		$\phi_i$ strongly convex \\and $\Lips$ differentiable. \\
		$g_i$ convex (possibly nonsmooth). \\
		$\A_i$ can be rank deficient.}  &  Gauss-Seidel&  \makecell[l]{Global convergence. \\ Global optima. \\} &   \makecell[l]{Parallel implementation.\\Networked communication. \\Low per-iteration complexity. \\ Convex.}& \cite{chang2014multi} \\
	\hline
	Flexible ADMM  &  \makecell[l]{$f_i$ $\Lips$ differentiable. }  &  Gauss-Seidel&  \makecell[l]{Global convergence. \\ Stationary points.} &   \makecell[l]{
		Random agent activation. \\ Nonconvex }& \cite{hong2016convergence, wang2019distributed} \\
	\hline
	\makecell[l]{Flexible \\linearized ADMM}  &  \makecell[l]{$f_i$ $\Lips$ differentiable. }  &  Gauss-Seidel&  \makecell[l]{Global convergence. \\Stationary points.} &  \makecell[l]{Random agent activation. \\Low per-iteration complexity. \\ Nonconvex.}& \cite{hong2016convergence} \\	
		\hline
	\makecell[l]{Edge-based \\ flexible ADMM}  &  \makecell[l]{$f_i$ convex. }  &   Gauss-Seidel&  \makecell[l]{Probability convergence. \\Global optima. \\ Convergence rate $\mathcal{O}(1/k)$.} &   \makecell[l]{Random agent activation.\\ Networked communication. \\ Convex.}& \cite{wei20131} \\		
	\hline
	\multirow{2}{*}{\makecell[l]{~\\Async-ADMM}}  &  \makecell[l]{$f_i$ convex. \\Bounded delay $\tau$. }  &  Gauss-Seidel&  \makecell[l]{Global convergence. \\ Global optima. \\ Convergence rate $O(\tau/k)$.} &   \makecell[l]{Robust to communication delays \\(asynchronous computing). \\ Networked communication. \\ Convex. }& \cite{zhang2014asynchronous, li2020synchronous} \\
	\cline{2-6}
	&  \makecell[l]{$f_i$ $\Lips$ differentiable.   \\Bounded delay $\tau$. }  &   Gauss-Seidel&  \makecell[l]{Global convergence. \\Stationary points.} &   \makecell[l]{Robust to communication delays. \\ (asynchronous computing) \\Networked communication.\\ Nonconvex.}& \cite{chang2016asynchronous, chang2016asynchronous2} \\
	\hline
\end{tabular}
\end{table*}

\subsection{Non-linearly constrained optimization}
So far, we have focused on linearly constrained optimization. In other words, the couplings across the agents can be characterized by some  linear constraints. In practice, many other  systems  exist in which the interactions across the agents or subsystems are complex, leading to non-linear coupled constraints. Examples include but not limited to building thermal comfort management \cite{yang2020hvac, yang2021distributed} and power system control \cite{low2014convex, sun2021two}. 
 This class of problems takes the general formulation of 
 {	\setlength{\abovedisplayskip}{3pt}
 	\setlength{\belowdisplayskip}{3pt}
\begin{align}
\label{pp:p5}  & \min_{\x= \{x_i\}_{i =1}^n}  \sum_{i=1}^n  f_{i}(x_{i}) \tag{$\PP5$}\\
\text{s.t.}~ & h_{i}(x_{i}, \{x_{j}\}_{j \in N_i}) = 0, \quad \forall i \in N. \notag\\
& g_{i}(x_{i}, \{x_{j}\}_{j \in N_i }) \leq  0, \quad \forall i \in N. \notag\\
& x_i \in X_i, \quad \quad \quad \quad \quad \quad ~\forall i \in N. \notag
\end{align} }
where $f_i: \R^{n_i} \rightarrow \R$ are private objectives of the agents defined on their decision variables $x_i \in \R^{n_i}, i \in N$;  $h_i: \R^{\sum_{j \in N_i} n_j} \rightarrow \R$ and $g_i: \R^{\sum_{j \in N_i} n_j} \rightarrow \R$ characterize the non-linear (often nonconvex) couplings between agent $i$ and its neighbors $N_i$. The subset $X_i$ represents the local bounded convex constraints  of agent $i$. Problem \eqref{pp:p5} models a class of multi-agent optimization where the agents hold private objectives but their  decisions  are coupled in  non-linear (often nonconvex) equality and inequality constraints. Currently, distributed methods to handle problem \eqref{pp:p5} (often nonconvex) efficiently are extremely lacking.

Though ADMM variants for diverse classes of linearly constrained optimization are  available, extending them to non-linearly constrained problem \eqref{pp:p5} (often with nonconvex couplings) is nontrivial due to the difficulty to establish convergence.  Currently, the  existing ADMM variants for solving problem \eqref{pp:p5} are quite limited. To our best knowledge, the \emph{two-level ADMM} proposed in \cite{sun2019two}  has been the main solution with convergence guarantee. We discuss that method below. 

\subsubsection{Two-level ADMM} As a celebrated work, \cite{sun2019two} studied a \emph{two-level ADMM} for solving non-linearly constrained  problem \eqref{pp:p5} in general nonconvex setting (i.e., $f_i$, $h_i$, $g_i$ are nonconvex). The basic  idea is to first convert problem \eqref{pp:p5} into a  linearly constrained optimization and then explore the application of existing ADMM variants.  Though the idea is natural, there are some intrinsic challenges to be overcome, i.e., the two necessary conditions (see the discussions in Section IV-C) to ensure the convergence of ADMM and its variants in nonconvex setting can not be provided simultaneously. To overcome such a challenge, \emph{two-level ADMM} 
proposed the idea of introducing a block of slack variables as the last block. 
The idea of this method should be  clear from below. 

Specifically, by introducing decision copies for the interconnected agents, we have the following equivalent linearly constrained reformulation for problem \eqref{pp:p5}. 
 {	\setlength{\abovedisplayskip}{3pt}
	\setlength{\belowdisplayskip}{3pt}
\begin{align}
\label{pp:p5-v1}  \min_{ \x = (\x_i)_{i=1}^n,  \hx}& F(\x) = \sum_{i=1}^n f_i(\x_i) \tag{$\PP5$-1} \\
\text{s.t.}~ & \sum_{i=1}^n \A_i \x_i + \B \hx = \bm{0}. \notag \\
&  h_i(\x_i) = \bm{0}, g_i(\x_i) \leq \bm{0},  \notag\\
&\x_i \in \X_i, \quad  i \in N. \notag \\
&\hx \in \bar{\mathbf{X}}. \notag
\end{align} }
where $\mathbf{x}_i: = (x_{i}, (x_{j})_{j \in N_i})$ denotes the augmented decision variable held by agent $i$,  which consists of its own decision variable $x_i$ and the estimates $(x_{j})_{j \in N_i}$ for its neighbors indexed by subset $N_i$;  $\bar{\x}: =(x_i)_{i=1}^n$ represents a global copy of all decision variables over the network.
The coupled linear constraints $\sum_{i=1}^n \A_i \x_i + \B \bar{\x} = \bm{0}$ enforce the consistency of the decision copies over the network.  By defining $\A: = (\A_i)_{i=1}^n$ and  $\x: = (\x_i)_{i=1}^n$, the  coupled linear constraints can be  expressed by the compact form of  $\A \x + \B \bar{\x} = \bm{0}$; $\X_i: = \cup_{j \in N_i} X_j$ denotes the augmented local constraints for agent $i$.

Note that problem \eqref{pp:p5-v1} corresponds to a two-block optimization by treating the decision variables  $\x:=(\x_i)_{i=1}^n$ and $\bar{\x}$ as two decision blocks.   This corresponds to the problem structure of most existing ADMM variants. However, they are not applicable  due to the existence of nonconvexity.  As discussed in Section VI-C, we generally require a \emph{well-behaved} last block $\y$ satisfying the two necessary conditions to ensure convergence:  \emph{i)} the last decision block $\y$ is unconstrained and has $\Lips$ differentiable objective, and \emph{ii)} ${\rm Im}(\A)\subseteq {\rm Im}(\B)$, $\B$ has full column rank or the mapping $H(\uu) = \{\arg \min_{\y} \phi(\x, \y): \B \y = \uu\}$ is unique and $\Lips$ smooth. Here  $\phi(\x, \y)$ refers to the objective of problem \eqref{pp:p5-v1},  $\B$ denotes the coefficient matrix of a last block $\y$ and $\A$ represents an augmented coefficient matrix for the decision blocks excluding $\y$.  Despite the direct linearly constrained reformulation \eqref{pp:p5-v1}, it was argued that the two necessary conditions \emph{i)}-\emph{ii)} can not be provided simultaneously. Specifically, to fulfill condition \emph{ii)}, we should have $\x:=(\x_i)_{i=1}^n$ work as the last block (we actually have ${\rm Im } (\A) \supseteq {\rm Im} (\B)$ in \eqref{pp:p5-v1}).  However, the presence of local constraints (i.e., $\X_i, h_i, g_i$) makes the subproblems $\x$ nonsmooth, violating condition \emph{i)}. Oppositely, if we assume $\hx$  as the last block, we do not have condition \emph{ii)}.   To overcome such a challenge, a block of slack variables  $\z$  is introduced as a \emph{well-behaved} last block. This leads to the following equivalent three-block linearly constrained reformulation. 
 {	\setlength{\abovedisplayskip}{3pt}
	\setlength{\belowdisplayskip}{3pt}
\begin{align}
\label{pp:p5-v2}	 \min_{ \x: = (\x_i)_{i=1}^n,  \hx, \z} & F(\x) =\sum_{i=1}^n f_i(\x_i) \tag{$\PP5$-2} \\
\text{s.t.}~& \A \x + \B \hx + \z = \bm{0}. \notag \\
&  h_i(\x_i) = \bm{0}, g_i(\x_i) \leq \bm{0}, \notag\\
& \x_i \in \X_i, \quad  i  \in  N. \notag \\
&\hx \in \bar{\mathbf{X}}. \notag \\
&  \z = \bm{0}. \notag
\end{align} }

On top that,  a  two-level ADMM variant  was proposed for solving \eqref{pp:p5-v2} in which  an ALM is used in the upper level to gradually force the slack variables to \emph{zero} and the classical ADMM is used in the lower level to solve some relaxed multi-block optimization with the slack variables as the last block. 
The method takes the following two-loop iterative scheme. 
\begin{align*}
&\textbf{Two-level ADMM:} ~~\\
&\!\!\! \!\! \text{Outer-loop}  \begin{cases}
	&
	\!\!\! \!\!\!\text{Inner-loop} \begin{cases} 
		& \! \! \! \textbf{Primal update:}~ \\
		&  \! \! \! \x^{k+1} \!=\! \arg \min_{\x} \Lag_{\rho}(\x, \bar{\x}^k, \z^k, \boldlambda^k, \bm{\gamma}^p)\\ 
		& \! \!  \! \bar{\x}^{k + 1} \! =\! \arg \min_{\bar{\x}} \Lag_{\rho}(\x^{k+1}, \bar{\x}, \z^k, \boldlambda^k, \bm{\gamma}^p) \\
		& \! \! \! \z^{k+1} \!=\! \arg \min_{\z}  \Lag_{\rho}(\x^{k + 1}, \bar{\x}^{k + 1}, \z, \boldlambda^k, \bm{\gamma}^p)\\
		& \! \! \! \!\! \!\textbf{Dual update:} \\
		&  \boldlambda^{k+1} \!=\! \boldlambda^k + \rho(\A \x^{k+1} + \B \bar{\x}^{k+1} + \z^{k+1}) \\
	\end{cases} \\
	& \!\!\!\textbf{Dual update:}~~ \bm{\gamma}^{p+1} = \bm{\gamma}^p + \beta \z^{k+1}
\end{cases}
\end{align*}
where $\boldlambda$  and $\bm{\gamma}$  are Lagrangian multipliers associated with the coupled linear constraints $\A \x + \B\bar{\x} +\z = \bm{0}$ and the hard constraints $\z =\bm{0}$; $\rho$ and $\beta$ are penalty parameters  for  the ADMM in  the lower level (Inner-loop)  and the ALM in the upper level (Outer-loop); $k$ and $p$ are related  iteration counters. 
In the primal update of Inner-loop, we have the subproblems $\x:=(x_i)_{i=1}^n$  naturally decomposable across the decision components $\x_i, i \in N$ and thus can be performed in a distributed and parallel manner. 
The convergence of the \emph{two-level ADMM} towards stationary points was established for general nonconvex optimization \cite{sun2019two}.  Further, the efficacy of the method was demonstrated by a concrete  application to   power system control  \cite{sun2021two}.

\emph{Summary:} This section reviewed ADMM variants for solving  non-linearly constrained problem  \eqref{pp:p5} (often nonconvex). Currently, this related methods are quite limited. One solution with convergence guarantee  is the  \emph{two-level ADMM}.   This is mainly caused by the intrinsic challenges that the two necessary conditions to ensure convergence of ADMM variants in nonconvex setting are not satisfied simultaneously. \emph{Two-level ADMM} proposed the idea of introducing  slack variables to fulfill such  two conditions.  Though the efficacy of the method has been demonstrated both theoretically and empirically, the implementation is generally at the cost of high iteration complexity caused by the two-level iterative scheme.

\section{Discussions and future research directions}
Based on the previous survey, this section discusses several important future research directions related to ADMM and its variants for distributed optimization. 
\subsection{Nonconvex extensions}
From this survey,  we see  that ADMM and its variants for convex optimization have  been  studied extensively and broad results are now available for broad classes of problems.  
However, for the more broad nonconvex counterparts,  the related solutions are still quite limited. Moreover, the existing ADMM variants for nonconvex optimization are all restricted to the two necessary conditions  \cite{wang2019global,sun2019two}: \emph{i)}  the last block $\y$ is unconstrained and has $\Lips$ differentiable objective, and \emph{ii)} ${\rm Im}(\A)\subseteq {\rm Im}(\B)$, $\B$ has full column rank or the mapping $H(\uu) = \{\arg \min_{\y} \phi(\x, \y), {\rm s.t.~} \B \y = \uu\}$ is unique and $\Lips$ smooth (see the discussions in Section IV-C).  There exist a wide spectrum of problems  that fail to satisfy such two conditions, such as problems \eqref{pp:p2} and \eqref{pp:p5}. In such situations, it represents a big challenge to develop  an ADMM variant  with  convergence guarantee.

To figure out the solutions, we require to comprehensively understand the general framework of establishing convergence  for an ADMM or its variant in nonconvex setting. Specifically, the key step  is to identify a sufficiently decreasing and lower bounded Lyapunov function, which can indicate the convergent property of  generated sequences. The sufficiently decreasing and lower boundness properties of a Lyapunov  function for an ADMM or its variant  state that \cite{wang2019global, hong2016convergence}
\begin{align*}
	&T(\x^{k+1}, \boldlambda^{k+1}) - T(\x^k, \boldlambda^k) \\
	&\quad \quad \quad \quad \leq -\alpha_{\x}\Vert \x^{k+1} - \x^k\Vert^2 - \alpha_{\boldlambda} \Vert \boldlambda^{k+1} - \boldlambda^k\Vert^2, \\
	& T(\x^k, \boldlambda^k) > -\infty, 
\end{align*}
where  $T(\cdot, \cdot)$ denotes a general Lyapunov function;  $\{\x^k\}_{k=1}^K$  and $\{\boldlambda^k\}_{k=1}^K$ are primal and dual sequences;  $k$ is the iteration counter;  $\alpha_{\x}$ and $\alpha_{\boldlambda}$ are positive scalars. The  AL functions and their variants are often used as the Lyapunov functions (see \cite{wang2019global} for example). 
To identify such a Lyapunov function, we generally require to bound the dual updates  $\Vert \boldlambda^{k+1} - \boldlambda^k\Vert^2$ by the primal updates $\Vert \x^{k+1} - \x^k\Vert^2$. Conditions \emph{i)}-\emph{ii)} are exactly used to achieve such an objective.

In recent years, some interesting and promising efforts have been made to relax such two conditions. 
Particularly, some works have been devoted to overcoming the challenges caused by  the lack of  $\Lips$ differentiable property of  last block. 
For example, \cite{yang2022proximal} proposed to employ a discounted dual update scheme to bound the dual updates manually  when the $\Lips$ differentiable property of last block is lacking. From another perspective,  \cite{zeng2022moreau} proposed the idea of using a smooth and $\Lips$ differentiable \emph{Moreau envelope} to approximate a nonsmooth but weakly convex objective of last block. In terms of the conditions on the coefficient matrices of last block, \cite{sun2019two, yang2020proximal} proposed the idea of introducing slack variables as a \emph{well-behaved} last block.  These works show that the restrictive conditions \emph{i)}-\emph{ii)} are possible to be relaxed by reinforcing the existing methods. This is important to further enhance the capability of ADMM and its variants for more broad distributed optimization and presents one of the important future research directions to be addressed.

%
%
%
%

\subsection{ADMM acceleration}

Though the numerical convergence depends on specific problems, classical ADMM and its many variants only promise an  $\mathcal{O}(1/k)$ convergence rate. For an iterative algorithm, the convergence rate usually characterizes the iteration complexity and communication burden of a method.  We often prefer a faster convergence rate  to approach an appropriate solution with less iterations and less communications. 
To achieve such a goal,  some efforts have been made to accelerate the convergence of classical ADMM. A typical and successful  example is the fast ADMM \cite{goldfarb2013fast,patrinos2014douglas} which  is able to improve the convergence rate of classical ADMM by an order via the integration of Nesterov acceleration technique. However, this method only applies to a special class of problems that are strongly convex and with additional  quadratic structure.   The extension of fast ADMM  to  more general problems still remains an open issue to be addressed. This is nontrivial and faced with some  intrinsic challenges to be overcome. Specifically,  Nesterov acceleration is primarily developed for  first-order descent solvers, however  ALM and its inexact versions (e.g., the ADMM variants based on $\gauss$ and $\Jacobian$ decomposition) are basically not descent solvers. A descent solver states that we have the objective value  decrease along the iterations.  In terms of such issue,  some works argued that the descent property of an ALM or ADMM variant  can  be ensured by adding some monitoring and correction steps \cite{goldfarb2013fast}. This actually sheds some light on generating fast ADMM to more general problems. 

In addition to Nesterov acceleration, Anderson acceleration  also began to draw interest for accelerating ADMM or its variants \cite{fu2020anderson, wang2021asymptotic}. However, only some numerical results are currently available and the solid theoretical results regarding the convergence rate characterizations  remain to be established. 

Overall, the above acceleration techniques follow a  general framework to accelerate ADMM or its variants. The key step  is to twist or modify the generated  primal and dual sequences in the iterative process, which can be uniformly characterized by $\hat{\w}^{k+1} = {\texttt{acc}}\big(\w^{k+1}, \w^k, \w^{k-1}, \cdots\big), k = 1, 2, \cdots$,  where $k$ is the iteration counter,  $\w: = (\x^\top, \boldlambda^\top)^\top$ is the stack of primal and dual variables,  $\w^k$ denotes the generated update by  an ADMM or its variant at iteration $k$,  and $\hat{\w}^k$ denotes the modified update for acceleration at iteration $k$. We use \texttt{acc} to denote an acceleration technique, which often corresponds to a linear  combination. This general acceleration framework for ADMM and its variants has been documented in \cite{buccini2020general}, however the theoretical convergence rate characterizations for such general acceleration framework  remain to be addressed. This represents another important and significant future research direction. 

\subsection{Asynchronous ADMM}
From this survey, we note that most of the existing ADMM variants are synchronous. Specifically, they generally define or assume that the computing agents behave in a synchronous manner. This is in the sense that the agents share the same iteration counter and use the updated information of their interconnected agents to proceed new updates at each step.  This may be problematic considering the practice that one agent usually has no idea about the progress of  another in the iterative process.  Besides, the communications are often not idealized and suffer from communication delays, communication losses and temporary  disconnections.
When a synchronous ADMM is applied to such context, the computation efficiency or convergence property of the method could be largely degraded or disrupted due to the long waiting time or the failure of information synchronization over the networks.
It was demonstrated  by simulations that the asynchronous behaviors of computing agents caused by communication delays and  solver diversity could  lead to the oscillation of a synchronous ADMM \cite{moret2018negotiation}. To address such an issue, the concept of asynchronous ADMM is thus proposed with the idea  to account for the asynchronous behaviors of computing agents while securing the convergence property and computation efficiency of an ADMM or its variant.  

As discussed before, some efforts have been made to develop asynchronous ADMM variants both for convex \cite{zhang2014asynchronous, li2020synchronous} and nonconvex consensus  optimization \cite{chang2016asynchronous, chang2016asynchronous2}. The main idea is that at each step, the coordinator agent only waits for the information of partial computing agents involved in the system to perform a new update with the objective to cut off  waiting time.  
This is often referred to \emph{partial synchronization}.  To ensure  convergence of the method, a critical step is to set a maximal tolerable delay  bound $\tau$ to each computing agent. This enforces that 
the information of all agents used by the coordinator can be at most $\tau$ old. This method has already found many successful applications, such as the peer-to-peer energy trading matching in a deregulated electricity market  \cite{dong2021convergence}.

Though some notable results have been achieved along this direction, the study of  asynchronous implementation of ADMM variants is still at the very early stage. Specifically, most of the exiting asynchronous ADMM variants are specific to consensus optimization that adopts a \emph{master-worker} communication scheme. For the other classes of problems discussed in this paper that also suffer from the asynchronous behaviors of computing agents,  asynchronous solutions are still extremely lacking. In addition, when  a network or a  graph defining the communication scheme across the agents exists, the interactions of computing agents with asynchronous behaviors could be much more complex. In such context, the existing asynchronous ADMM variants are not applicable and we are required to develop other effective solutions. In addition, there often exists a trade-off between the waiting time and the iteration complexity that are determined by the maximal bounded delay configurations and  partial synchronization mechanism. A smaller tolerable delay bound and larger partial synchronization mechanism  often yield low iteration complexity but lead to longer waiting time.   Oppositely, a larger tolerable  delay bound and smaller partial synchronization  often result in higher iteration complexity but shorter waiting time.  In practice, we are often concerned with the time efficiency of a solution method, therefore some theoretical results are required to guide how to make  a trade-off between the tolerable delay bound and partial synchronization to achieve the best time efficiency. These are the interesting topics to be studied in the future.

\subsection{ADMM + Reinforcement Learning}
Learning is one of the most important tools for developing machine intelligence in the big data era. As one of the most important branches of machine learning, reinforcement learning (RL) has been widely studied  and achieved widespread success for sequential decision making under dynamic and uncertain environments for both engineering and computer systems \cite{luong2019applications, mao2018deep}.  RL and ADMM have emerged as the two most important tools for optimization and control, showing complementary benefits. Specifically,  RL is not restricted to system or problem complexity and only entails that  the system performance or problem objective can be evaluated. In contrast, ADMM and its variants provide a powerful and flexible general framework to  achieve reliable coordination of multi-agent computation, thus overcoming  the computation intensity faced by a centralized computation architecture. Considering the dual  benefits, the combination of RL and ADMM  is expected to enable more powerful control and optimization tools and shows wide  prospects.

In  very recent years, the combination of RL and ADMM has began to draw interest. The related works can be divided into two categories by their perspectives. One is to deploy ADMM into an existing RL framework to enable multi-agent reinforcement learning (MARL) or distributed reinforcement learning (DRL), and the other one is to integrate an RL agent into an existing ADMM  variant for handling complex subproblems.  As the first category, \cite{zhao2020distributed} studied the combination of ADMM and inexact ADMM with RL to enable DRL for a class of multi-agent systems where the agents have private local objectives but are coupled through a joint decision making scheme.
The proposed method enables the multiple agents to collaboratively learn a joint control policy in a distributed manner considering their individual objectives while not disclosing their rewards or preferences to the others.  For the same problem set-up, \cite{lei2022adaptive} proposed to combine a  stochastic ADMM with RL  by using the gradient information of system performance for policy updates. As the second category,  \cite{graf2019distributed, zhang2021semi}  studied the integration of RL agents into classical ADMM framework with the objective to overcome the computation challenges of solving complex and non-linear subproblems.

Currently, the combination of ADMM or its variants with RL is at the very beginning. Many critical issues require to be addressed. Typically, how to best integrate the two method to improve computation efficiency  is still under discussion.  In addition, it still remains a big challenge to establish the  theoretical convergence for their combinations due to the feature of RL which tries to improve policies under uncertainties instead of solving optimization problems comprehensively. This represents another important future research directions to be studied.



\subsection{ADMM + Federated Learning}

Federated learning (FL) is an emerging distributed  machine learning setting with the main idea of  empowering massive dispersed  clients to learn models by leveraging locally available data and  relying on a central server to aggregate the models to achieve coordination \cite{mcmahan2017communication, kairouz2021advances}. Generally, FL provides the scope of training machine learning models over an edge-based distributed computation architecture. 
This is in contrast to the traditional centralized learning paradigm where a central server is authorized to learn a comprehensive model independently and thoroughly  by  collecting data from the whole system. 
With the proliferation and penetration of Internet of Things (IoT), data is generated at an unprecedented rate and in a geographically distributed manner.  
Meanwhile, the whole society is raising awareness of data and information privacy. 
This poses the issue of developing communication efficient and privacy-preserving  machine learning tools. FL is thus proposed with the objective to enable  high data privacy, low communication burden, and high computation efficiency. One distinguishing feature of FL is that the clients only communicate  with the server for model parameters and can keep their data private.

FL essentially corresponds to decomposing a mathematical optimization across a number of computing agents. More specifically, by viewing the model parameters as the decision variables, FL generally corresponds to solving a consensus optimization (possibly nonconvex) in a distributed manner. 
As one of the most powerful  distributed methods,  ADMM has emerged  as a  popular tool for enabling FL. A number of works have studied ADMM variants for FL. Typical examples include \cite{zhou2021communication,zhou2022federated, wang2022fedadmm}.  Specifically, to reduce communication and computation burden, \cite{zhou2021communication,zhou2022federated} studied an ADMM variant with a flexible  communication mechanism for FL, i.e.,  the clients only communicate with the central server at fixed time points $k_0, 2k_0, \cdots$ instead of at each iteration, where $k_0$ denotes the interval of communication. To account for communication delays or losses as well as random client ON/OFF behaviors, \cite{wang2022fedadmm} studied  an  ADMM variant with a  flexible client participation mechanism for FL, i.e., at each iteration only a subset  of the clients  are selected to update models \cite{wang2022fedadmm}.  In fact, these ADMM variants are exactly the specific implementation of the ADMM variants for consensus optimization  discussed before.

ADMM presents  an important distributed framework for FL considering its flexibility, robustness and efficiency. Specifically,  ADMM and its variants allow to  activate the computing agents in a flexible manner, such as the random agent activation and the partial agent participation.  Besides, ADMM and its variants rely on quite weak assumptions to ensure convergence and are applicable to broad classes of problems both in convex and nonconvex settings.  In addition, many ADMM variants are compatible with inaccurate or uncertain information as well  communication delays or losses.  Moreover, ADMM  variants generally  show  faster convergence over many other distributed methods.  Last but very important, most of the ADMM variants  allow to solve the subproblems inexactly with quite low per-iteration complexity. These are the favorable features of ADMM and its variants that are expected to be celebrated by FL.  

However, FL is somehow different from mathematical optimization in the sense that  a learning or training agent is generally deployed to train a neural network instead of solving a mathematical subproblem with explicitly available model at each iteration. This poses the challenges to establish the convergence of the methods in the machine learning context. 
In addition, FL often places higher emphasis on the computation and communication efficiency than the solution accuracy. This implies that the existing ADMM variants may have to be adapted or reinforced for FL. These represent some other important and emerging future research directions.

\section{Conclusion}
This paper provided a comprehensive survey on  ADMM and its variants for distributed optimization. We discerned the five major classes of problems that have been mostly concerned in the literature.  For each class of problems, we discussed the related ADMM variants  from the perspectives of main ideas,  main assumptions, decomposition scheme, convergence properties and main features.  Based on the survey, we further identified several important future research directions to be addressed, which include: \emph{i)} the  extension of ADMM variants to  more general nonconex optimization, \emph{ii)} the acceleration of ADMM variants with convergence and convergence rate characterizations. \emph{iii)} the  asynchronous implementation of ADMM variants accounting for the asynchronous behaviors of computing agents, \emph{iv)} the combination of ADMM and reinforcement learning (RL) for enabling more powerful optimization and control tools, \emph{v)} the combination of ADMM and federated learning (FL) for efficient and privacy-preserving machine learning.   This review covered most of the developments of ADMM and its variants in  very recent decades and can work as a tutorial both for developing distributed optimization in broad areas and identifying  existing  research gaps. 


\bibliographystyle{IEEEtran}
\bibliography{reference}

\vfill

\end{document}